\documentclass[a4paper,11pt]{article}
\usepackage{amsmath}
\usepackage{fancyhdr,graphicx}
\usepackage[english]{babel}
\usepackage{hyperref}
\usepackage{natbib}
\usepackage{rotating} 
\usepackage{multirow}
\usepackage{multicol}        
\usepackage{booktabs}
\usepackage{subfig}
\usepackage{relsize}
\usepackage{longtable}
\usepackage{authblk}
\usepackage{color}
\usepackage{array}
\usepackage{amsmath}
    \usepackage[group-separator={,},group-minimum-digits=4]{siunitx} 
\usepackage{makecell}    

\newcolumntype{P}[1]{>{\centering\arraybackslash}p{#1}}

\usepackage[a4paper,
        left=1.4cm,
        right=1.4cm,
        top=3.4cm,
        bottom=3.9cm,
]
{geometry}

\title{Growth, development, and structural change at the firm-level: The example of the PR China}
\author[1,2,3,$\dagger$]{Torsten Heinrich}
\author[2,3,4]{Jangho Yang}
\author[5,6,7]{Shuanping Dai}

\affil[1]{Faculty for Economics and Business Administration, Chemnitz University of Technology, 09111 Chemnitz, Germany}
\affil[2]{Institute for New Economic Thinking at the Oxford Martin School, University of Oxford, Oxford OX1 3UQ, UK}
\affil[3]{Oxford Martin Programme on Technological and Economic Change (OMPTEC), Oxford Martin School, University of Oxford, Oxford OX1 3BD, UK}
\affil[4]{Department of Management Sciences, Faculty of Engineering, University of Waterloo, Waterloo, ON, N2L 3G1}
\affil[5]{School of Economics, Jilin University, 130012, Changchun, China}
\affil[6]{IN-EAST Institute of East Asian Studies, Universit\"{a}t Duisburg-Essen, 47057 Duisburg, Germany}
\affil[7]{China's Public Sector Economy Research Center, Jilin University, 130012, Changchun, China}
\affil[$\dagger$]{\tt torsten.heinrich@uni-bremen.de\authorcr \textcolor{white}{--}}

\begin{document}

\maketitle

\begin{abstract}

Understanding the microeconomic details of technological catch-up processes offers great potential for informing both innovation economics and development policy. We study the economic transition of the PR China from an agrarian country to a high-tech economy as one example for such a case. It is clear from past literature that rapidly rising productivity levels played a crucial role. However, the distribution of labor productivity in Chinese firms has not been comprehensively investigated and it remains an open question if this can be used to guide economic development. We analyze labor productivity and the dynamic change of labor productivity in firm-level data for the years 1998-2013 from the Chinese Industrial Enterprise Database. We demonstrate that both variables are conveniently modeled as L\'{e}vy alpha-stable distributions, provide parameter estimates and analyze dynamic changes to this distribution. We find that the productivity gains were not due to super-star firms, but due to a systematic shift of the entire distribution with otherwise mostly unchanged characteristics. We also found an emerging right-skew in the distribution of labor productivity change. While there are significant differences between the 31 provinces and autonomous regions of the P.R. China, we also show that there are systematic relations between micro-level and province-level variables. We conclude with some implications of these findings for development policy.
\end{abstract}

\tableofcontents

\section{Introduction}

For almost three decades, the PR China has been one of the fastest-growing economies. During this time, it made the transition from a largely agricultural developing country to the world's second-largest industrial economy. Where state-owned enterprises (SOEs) ran the show in the 1980s, the country today is home to a multitude of private corporations of international importance. The PR China used to be a poor country and years behind in technological terms, but today, its development trajectory is of growing importance for the world in science and innovation, in CO2 emissions, and in technological impact on privacy, surveillance, and personal freedom. 
While the development is moderately well-understood in macro-economic terms, many open questions remain with regard to the development of the microstructure of the Chinese economy over the last decades. Which firms were the most productive ones, which were central to the transition process? How was productivity distributed among firms? How did this change over time? Can these processes be observed in all regions? In all sectors? Is it mirrored in profitability and investment rates? 
Can other developing economies achieve the same level of growth and development?

For developed countries, the distributions of firm-level data have been widely investigated and discussed in the literature. Many stylized facts are known although some questions remain contested. \citet{Ijiri/Simon64, Ijiri/Simon77} proposed that firm sizes are highly skewed and follow Pareto distributions, essentially with a process following Gibrat's law\footnote{Gibrat's law with lower bound produces Pareto distributions; without such bound, it generates lognormal distributions \citep{Mitzenmacher04}.} as the root cause of this. The observation was later confirmed with more detailed data sets \citep{Axtell01,Gaffeoetal03}, although some of the literature prefers to model the distribution as a lognormal \citep{Cabral/Mata03} and other generating algorithms have been proposed (\citep{Heinrich/Dai16} offer an overview). It is clear that this has important policy implications for competition law, innovation policy, labor market governance, and the effectiveness of policy interventions in industrial organization. Connections to firm growth, innovation, and technological change \citep{Yuetal15,Li/Rama15} further add to the importance of this distributional approach, as do the later, but equally important investigations of the distributions of firm growth rates \citep{Bottazzi/Secchi06} and productivities \citep{Yangetal19}.

 \begin{figure}[tb!]
 \centering
 \includegraphics[width=0.65\textwidth]{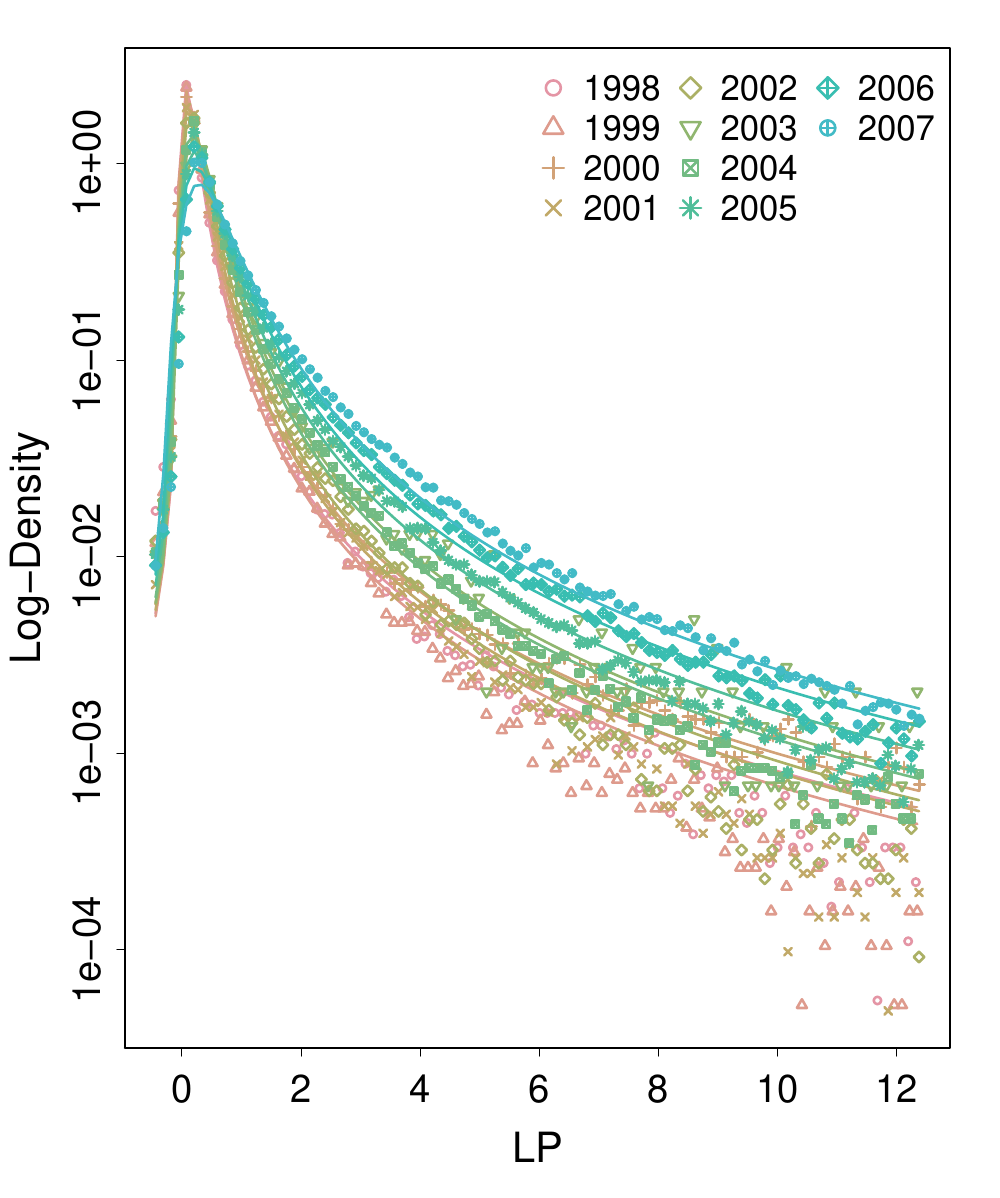}
 \caption{Density of the labor productivity ($LP$) distribution (full sample) by year in semi-log (vertical axis logarithmic). Solid lines indicate Levy alpha stable distribution fits as reported in Table~\ref{tab:patameterfits}.}
 \label{fig:density:year:lp}
 \end{figure}

An important characteristic of firm-level distributions in developed countries is that no significant changes are observed with either time \citep{Yangetal19} or firm age \citep{Cabral/Mata03}. Developing countries, however, may be very different. They are subject to substantial and rapid changes in sectoral structure, technology, economic policy, and social organization. Investigating such distributional changes for developing economies may shed light on the mechanisms driving that development, the effectiveness of policy measures, the microstructural impact of technological change, as well as potentially the history of developed countries. Studying similar historical processes for developed countries would require older data that is almost certainly not available in high resolution. 

For the PR China a look at the data immediately suggests that a systematic shift is underway: Figure~\ref{fig:density:year:lp} shows the distribution (density) function of the labor productivity at the firm-level by year in a semi-log plot (horizontal axis linear, vertical axis logarithmic); the shape of the distribution remains constant, but the right side (positive tail) moves outward and the peak becomes less pronounced. We will discuss other systematic shifts, interpretations, and implications below (see Section~\ref{sect:results}); for now, we emphasize that there are systematic changes in the distributional model during this development phase of the Chinese economy.
 
We use a firm-level data set for the PR China for the years 1998-2013 to investigate changes in the economic microstructure during the years of China's most rapid growth from a distributional model perspective. We will focus on the distribution of labor productivity and of labor productivity change; these are arguably the quantities that are most closely related to economic development. While other distributional models have been suggested for labor productivity (\citet{Yuetal15} consider Asymmetric Exponential Power (AEP) distributions and Gaussian normal distributions), the mounting evidence for heavy tails in both labor productivity and labor productivity change \citep{Yangetal19} suggests the L\'{e}vy alpha-stable distribution \citep{NOLAN1998187,NOLAN2018} as a distributional model. L\'{e}vy alpha-stable distributions generalize Gaussian normal distributions,\footnote{For one particular parameter setting, the L\'{e}vy alpha-stable converges to a Gaussian normal distribution.} but have heavy tails for almost all parameter values.

Important consequences include that the apparent dispersion of labor productivity \citep{Berlingierietal17} depends on how dispersion is measured. If such dispersion exists, it may be an indicator for misallocation of labor, capital, or other resources, a question of considerable relevance for the economy of the PR China \citep{hsieh2009misallocation}. It also relates to the debate on granular origins of aggregate fluctuations \citep{gabaix2011granular,Schwarzkopfetal10} and the question of how characteristics of labor productivity distributions in firm populations should be quantified and interpreted. The characteristics of labor productivity dispersion measures and the consequences for these economic questions are studied comprehensively in \cite{Yangetal19}.

We demonstrate that labor productivity as well as labor productivity change and various other firm-level characteristics in the PR China are indeed fat-tailed with infinite variance. Further, we show that the L\'{e}vy alpha-stable distribution is an excellent fit and discuss how the characteristics of the distribution can be specified and tracked using the parameters of the L\'{e}vy alpha-stable fit. We discuss the emerging temporal and regional patterns as well as the behavior in other subsamples. Finally, we demonstrate connections between the distributions of labor productivity, labor productivity change, profitability, investment rate at the firm-level, as well as between these firm-level patterns and aggregated level data.

The paper is organized as follows: The literature on the development of productivity in the PR China during its period of rapid growth is reviewed in Section~\ref{sect:literature}. Section~\ref{sect:data} describes the data and the variables of interest. Section~\ref{sect:methods} discusses the distributional models that are tested, the L\'{e}vy alpha-stable and the Asymmetric Exponential Power, as well as the fitting procedure, and goodness of fit measures employed. Section~\ref{sect:results} presents the findings and corresponding interpretations. Section~\ref{sect:conclusion} concludes.

\section{Literature}
\label{sect:literature}

The present paper aims to contribute to the study of distributional models of firm-level data, the investigation of the role of firm productivity in economic development, and, more specifically, the analysis of the rapid growth and development of the P.R. China in recent decades. We, therefore, give brief overviews of the literature in these three fields.

\subsection{Distributional models of firm-level data}

For developed countries, it has been established that the distributions of firm sizes, sales, etc. are heavily skewed \citep{Ijiri/Simon64,Axtell01}, with the Pareto \citep{Axtell01} and the lognormal distribution being proposed as distributional models \citep{Cabral/Mata03}. The two-sided distributions of growth rates and productivities equally have much heavier tails than normal distributions; proposed distributional models include Asymmetric Exponential Power distributions for growth rates \citep{Bottazzi/Secchi06,Bottazzietal07,Bottazzi/Secchi11} and L\'{e}vy alpha-stable distributions for productivities \citep{Yangetal19}. 

However, developed economies are relatively static. Very little change has been observed in these distributions over the recent decades for which good data is available (see e.g. \citep{Yangetal19}), and little change would be expected. To understand the development of firm-level distributions, scholars have instead focussed on firms of different age groups, their survival, and their shifts over time \citep{Cabral/Mata03}. Important findings include that the form of the distribution does not change over time \citep{Yangetal19} or with age \citep{Cabral/Mata03,Angelini/Generale08}, that surviving firms are slightly larger \citep{Cabral/Mata03}, and that they increase their productivity (making within-firm productivity gains important at the aggregate level) \citep{Bartelsmanetal13,Li/Rama15}. \citet{Cabral/Mata03} report that surviving firms in a Portuguese data set have less long tails and lower skew; however, if the statistical process which the firm size follows in reality is heavy-tailed with low exponents\footnote{This is indeed indicated by empirical studies \citep{Gaffeoetal03,Fujimotoetal11,Heinrich/Dai16} that find tail exponents around $1.0$ or between $1.0$ and $3.0$ for firm size depending on how firm size is measured.}, these moments may not exist and Cabral and Mata's findings may be statistical artifacts \citep{Yangetal19}. The hypothesis that small firms are more dynamic and account for significant shares of productivity gains and newly created jobs has frequently been proposed, but remains controversial \citep{Li/Rama15}. 

\subsection{Economic development and firm-level productivity}

Comprehensive firm-level data is often not available for developing countries. In turn, many studies have to work with small and potentially biased sample sizes. Notable exceptions are studies of firms in the P.R. China - using the Chinese Industrial Enterprise Database that we also work with - and in India - using government census data. While the general distributional forms found for developed countries are confirmed \citep{hsieh2009misallocation,Maetal08,Zhangetal09,Coad/Tamvada12,Sun/Zhang12,Yuetal15,ding2016determinants,Heinrich/Dai16}, wider dispersions for productivities are reported for India and China specifically \citep{hsieh2009misallocation}. 

In a general equilibrium interpretation, productivities should equalize, as investors should prefer high productivities while low productivity firms should be frozen out. This should be especially true for firms in the same sector and region, since portfolio diversification should not constitute a reason to invest in low productivity establishments. Of course, there may be structural reasons why investment in low productivity firms persists; and it might not be easily observable to investors. However, implications could still be drawn in comparative analyses, if different dispersions are observed. \citet{hsieh2009misallocation} choose to follow this equilibrium interpretation; their contribution has led to the influential interpretation that resources are more misallocated in developing countries \citep{hsieh2009misallocation,song2011growing,Bartelsmanetal13,Li/Rama15,Goyette/Gallipoli15} as well as occasionally strongly worded policy recommendations \citep{Adamopoulos/Restuccia14,Poschke18}. This has been explained with both structural factors, such as constraints on credit availability \citep{Bloometal10,Cabral/Mata03}, and also internal factors of the firm population of developing countries, such as bad management and reluctance to delegate decision-making \citep{Bloometal10,Chaffaietal12}. More recently, it has been found that productivities (both labor productivity and total factor productivity, TFP) are heavy-tailed with tail exponents below $2.0$ such that most dispersion measures, including the ones used in this line of research, are not meaningful \citep{Yangetal19}.

The source of productivity gains is an important question in the study of economic development.  Three main causes are (1) within-firm improvements, (2) selection pressure (unproductive firms do not survive), and (3) changes from distributional differences between entrants and exiting firms. Within-firm improvements were particularly important in developing countries \citep{Li/Rama15} and in successful developing economies such as China \citep{Yuetal15,Yuetal17}. While developed countries show some component from selection (2) and entry/exit (3) \citep{Farinas/Ruano04,Li/Rama15}\footnote{Studies also found a counter-cyclical contribution of entry and exit in developed economies (Spain): In phases of economic growth, when credit is readily available, less productive firms enter, resulting in a negative contribution to productivity growth \citep{Farinas/Ruano04}.}, in some developing countries (Sub-Saharan Africa specifically) the entry-exit-process may come down to churning without any improvements, and firms may survive because they are born larger, not because they learn or improve \citep{vanBiesebroeck05,Li/Rama15,Goyette/Gallipoli15}, resulting in heteroskedastic ''missing middle`` distributions \citep{vanBiesebroeck05}.

Success and growth at the firm-level has been linked to innovation for Argentina \citep{Chudnovskyetal06}, to innovation and technological competence for Indian firms \citep{Coad/Tamvada12}, and export participation for Chilean, Chinese, and European firms \citep{VolpeMartincus/Carballo10,diGiovannietal11,Sun/Zhang12}.

Finally, systematic shifts in distributions have been shown for those developing countries that undergo rapid growth: Both \citet{Yuetal15, Yuetal17} and \citet{ding2016determinants} find a location shift in the productivity distributions (labor productivity and TFP respectively) for China in the 1990s and 2000s, indicating higher productivities across the entire firm population, while the functional form did not change. \citet{Nguyen19} finds a similar location shift in firm-level distributions for Vietnam. \citet{Heinrich/Dai16}, studying the firm size distribution in Chinese provinces, find higher tail exponents in regions with high GDP per capita or high growth.

\subsection{Firm-level productivity and growth in the PR China}

Chinese firm-level distributions follow the same general patterns found elsewhere \citep{Maetal08,Yuetal15,Heinrich/Dai16,Heinrichetal20}. While firm sizes seem to follow power laws \citep{Maetal08,Heinrich/Dai16}, for labor productivities and growth rates, two distributions have been suggested: Asymmetric Exponential Power distributions from the exponential distribution family \citep{Yuetal15} were found to be a much better fit than Gaussians. L\'{e}vy alpha-stable distributions, which have power-law tails on both sides, have been suggested as an alternative since the data seems to be heavy-tailed \citep{Heinrichetal20}. The distinction has important consequences. 

The impressive economic growth of the PR China is reflected in the distributions as a location shift in productivity levels \citep{Yuetal15,Yuetal17,ding2016determinants}; the shift remains present in the gross industrial output per worker (labor productivity per wage), indicating that productivity has grown at a faster pace than labor inputs \citep{Zhang/Liu13}. It is important to note that this is not a changing average, but a shift of the entire distribution which otherwise remains intact in spite of continuing entry and exit processes. The changes in the distribution's parameters have so far not been comprehensively studied.  

We give a brief overview over the contributing factors to China's rapid growth from a historical perspective in Appendix \ref{app:historical}. 

It this worthwhile to note that state-owned enterprises (SOEs) have been found to be in general less productive than privately owned firms, both in terms of average labor productivity and average TFP \citep{song2011growing,Yuetal15,hsieh2015grasp}. While productivities of firms of all types vary widely and follow similar distributional forms as the economy as a whole (see Section~\ref{sect:results}), the mean difference is significant. SOE productivity did improve and has converged towards the productivity levels of private firms until at least 2007, but a gap in average productivity remains \citep{song2011growing,Yuetal15}. \citet{boeing2016china} find that compared to private firms, SOEs are less successful in converting patents into productivity improvements, although a generally positive relationship of productivity and R\&D efforts does exist \citep{hu2004returns,boeing2016china}. SOEs are often seen as a source of misallocation, thereby explaining their lower average productivity and tying the finding to the misallocation hypothesis \citep{hsieh2009misallocation,song2011growing}. However, on the one hand, a closer look at the distributions reveals that the variation is still present within ownership type groups. On the other, firm-level dispersion may not be larger in China than in other countries if the measures employed in achieving these results were misleading for heavy-tailed data \citep{Yangetal19}. The reason why SOEs are catching up and the mean difference between private firms and SOEs is converging is typically seen in the structural transformation of the state sector \citep{hsieh2015grasp, jefferson2000ownership}. Greater flexibility of managers and delegation of decision-making capabilities as well as and the effects of rising incomes and autonomy on employee motivation have also been linked to the productivity improvements in SOEs \citep{groves1994autonomy}. 

Finally, regional disparities in productivity and other variables are well-known and expected for a country of the size of the PR China. Coastal provinces like Shanghai and Guangdong have a better TFP comparing with central and western provinces \citep{ding2016determinants,chen2009analysis}. High productivity firms prefer to concentrate their activities on regions with developed infrastructure, good universities, and related industrial clusters; \citet{zhu2019geography} finds evidence for both sorting and adverse sorting effects. Meanwhile, officials in undeveloped regions are eager to attract investment by providing subsidies. However, government subsidies may attract low-productivity firms, since they have low opportunity costs \citep{zhu2019geography}. Marshall and Jacobs externalities of spatial industrial agglomerations \citep{Beaudry/Schiffauerova09} likely also play a role in creating and maintaining regional disparities, as may the openness of regions towards outside influences, foreign trade, and flexible economic policy \citep{jiang2011understanding}.   

\section{Data}
\label{sect:data}

\subsection{Sources}

We use firm-level data from the \textit{Chinese Industrial Enterprise Database} (CIEDB), which records several hundreds of thousands of firms each year for the time period between 1998 and 2013 and is commonly used by researchers working on firm-level data in China \citep{brandt2012creative,hsieh2015grasp,ding2016determinants,Yuetal15,Yuetal17}. The data ultimately derive from data recorded by the PR China's National Bureau of Statistics. Similar to data provided by the Bureau van Dijk for Europe (ORBIS Europe), the CIEDB records data at the firm-level, not at the level of physical entities (plants). This facilitates investigating structural characteristics such as productivity and profitability at the firm-level, the level at which decision making and management take place. Different from other databases like COMPUSTAT or Bloomberg, but similar to ORBIS Europe, the CIEDB also includes small and medium-sized firms and thus provides better coverage of different types of enterprises. The data set also records the ownership type (state-owned, foreign-owned, private, etc.)\footnote{The database does not include firms from Hong Kong, Macau, and Taiwan, we, therefore, will not cover these three regions in the analysis.}.

There are some notable difficulties with the data, especially for the period after 2008. These difficulties are well-known and recognized in the literature  \citep{Brandtetal14}. \citet{Brandtetal14} qualify the samples after 2008 as unreliable and recommend working with the more reliable date up to 2008 only. We largely follow this strategy. We complement this with later data from the period 2009-2013, where possible, to shed light on some developments after 2008.

Up to 2008, the database includes industrial firms with revenues above $5$ million Yuan. From 2009, only firms with revenues beyond $20$ million Yuan are present in the data set. The set of recorded variables also changes significantly over this time period. For instance, we are unable to compute value-added and productivities for the time period after 2008, as the measures required for their computation are only reported up until 2007.

In addition, we use industry level deflators from and macroeconomic data at province level from China Compendium of Statistics (1949-2008).

\subsection{Data processing}

We extract variables on identity\footnote{Ownership reforms led to continuous legal and structural changes, making it difficult to consistently identify the same firm \citep{jefferson2000ownership}. Using not just the firm ID but also phone number and ZIP code for identification is a typical way to address this \citep{Brandtetal14}.} (ID, phone number, ZIP code), characteristics (founding year, primary sector, ownership type), and structural and financial condition (output, assets, profits, wages, employment, intermediate input). These variables are present in the database throughout the years 1998-2007.\footnote{For 2003, the number of complete observations is very small.} Progressively more variables either missing or reported in substantially different form starting in 2008. (See Table~\ref{tab:observations}.) The monetary variables are deflated using industry level deflators.

We remove duplicates in terms of ID and Year before commencing with the data analysis.

In order to observe productivity changes, we attempt to identify firms that are present over multiple years both directly (using the unique ID) and indirectly, using phone numbers and address details as suggested in \citet{Brandtetal14}.

For the analysis of regional variation, the firms are assigned to the region of their postal address. As the region name is not typically part of the postal address, the ZIP codes were used to identify those regions.

\begin{table}[]
\centering
\begin{tabular}{P{1cm} P{2.2cm} P{2.2cm} P{2.2cm} P{2.2cm} P{2.2cm} P{2.2cm}}
\hline\hline
Year & Labor productivity & Labor productivity change & Labor productivity growth & Labor productivity (imputed) & Profitability & Investment rate \\\hline\hline
1998 & 141,790 & - & - & 141,787 & 149,270 & -\\
1999 & 148,982 & 113,684 & 112,996 & 148,973 & 147,636 & 121,091\\
2000 & 147,196 & 121,977 & 120,882 & 147,188 & 148,088 & 123,492\\
2001 & 158,671 & 116,674 & 116,001 & 158,664 & 157,453 & 119,211\\
2002 & 169,419 & 139,020 & 138,351 & 169,417 & 168,072 & 137,724\\
2003 & 11,404 & 9,503 & 9,490 & 11,404 & 11,369 & 9,461\\
2004 & 267,898 & 120,595 & 120,174 & 267,898 & 263,060 & 118,202\\
2005 & 262,830 & 227,569 & 227,032 & 262,830 & 261,369 & 223,671\\
2006 & 290,762 & 244,918 & 244,406 & 290,762 & 289,566 & 243,448\\
2007 & 324,638 & 268,614 & 268,376 & 324,638 & 323,131 & 267,151\\
2008 & - & - & - & - & 198,945 & 158,114\\
2009 & - & - & - & - & - & 130,155\\
2010 & - & - & - & - & - & 152,979\\
2011 & - & - & - & - & - & -\\
2012 & - & - & - & 42,301 & - & 36,139\\
2013 & - & - & - & 41,625 & 216,817 & 180,634\\\hline\hline
\end{tabular}
\caption{Number of observations per variable and year after cleaning.}
\label{tab:observations}
\end{table}

\subsection{Variables}
\label{sect:data:vars}

In order to investigate structural change at the firm-level, we analyze labor productivity and its dispersion and dynamical change. Labor productivity has been conjectured to hold information about the firm's capabilities, economic potential, and growth prospects. 
Its dynamical change facilitates investigating to what extent this potential is persistent in time. The dispersion of both variables holds information on the structural composition of the firm population of the country, the region, or the industry. 

Labor productivity is defined as \textit{value-added} per employee. 

$$LP_{i,t} = VA_{i,t} / L_{i,t}$$

where $i$ indicates the firms and $t$ is the time. As the value-added is not part of the database, it has to be computed as the difference between output and intermediate input.

$$VA_{i,t} = Q_{i,t} - II_{i,t}$$

where $Q_{i,t}$ is output and $II_{i,t}$ stands for intermediate input of firm $i$ at time $t$. Alternatively, $VA$ can be imputed as the sum of paid wages $W_{i,t}$ and profits $\Pi_{i,t}$

$$\widetilde{VA}_{i,t} = W_{i,t} + \Pi_{i,t}.$$

Imputed value-added differs from the direct computation in that reinvestments cannot be distinguished from negative profits and remain part of the resulting quantity. Reinvestments can be substantial and may occur in systematic patterns across the firm population. 

To observe the dynamic development of labor productivities, we compute the labor productivity change by firm

$$\Delta LP_{i,t} = LP_{i,t} - LP_{i,t-1}.$$

An alternative choice would be labor productivity growth 

$$\Dot{LP}_{i,t} = \frac{LP_{i,t} - LP_{i,t-1}}{LP_{i,t-1}}.$$

However, as this is a growth rate, it has a singularity at $LP_{i,t-1}=0$. Changes in labor productivity in the vicinity of the singularity get grotesquely exaggerated. What is more, $LP_{i,t}$ may be zero (in 1\% of the observations) or negative (3\% of the observations) since stocks and price changes are allowed.\footnote{Typically, output should be larger than intermediate inputs, $Q_{i,t}>II_{i,t}$. However, both are measured in monetary units, so whether $Q_{i,t}>II_{i,t}$ is subject to price changes. Further, the firm may maintain, built up, or reduce stocks intertemporarily.} For this reason, we refrain from using the growth rate and rely on the absolute change $\Delta LP_{i,t}$ as our main indicator of the dynamical change of labor productivities. 

The distributional models for these variables will be investigated in Section~\ref{sect:results:estimation}. It will be shown that this has important consequences for the selection and interpretation of quantitative measures for productivity dispersion. We complement the analysis of labor productivity with the study of the behavior of and dispersion of two more variables: \textit{Return on capital} will serve as an indicator for the firms' \textit{profitability} from the perspective of investors. The \textit{investment rate} is studied to assess investment patterns and growth. These variables are computed as

$$ROC_{i,t}=\frac{\Pi_{i,t}}{K_{i,t}}$$
$$IR_{i,t}=\frac{K_{i,t}-K_{i,t-1}}{K_{i,t-1}}$$

where $\Pi_{i,t}$ are the profits and $K_{i,t}$ is the capital stock (fixed assets) of firm $i$ at time $t$.

The number of observations for all variables by year is given in Table~\ref{tab:observations}.

Additional analyses also use capital intensity, defined as:
$$CI_{i,t} = K_{i,t} / L_{i,t}$$    

\begin{figure}[tb!]
\centering
\includegraphics[width=0.85\textwidth]{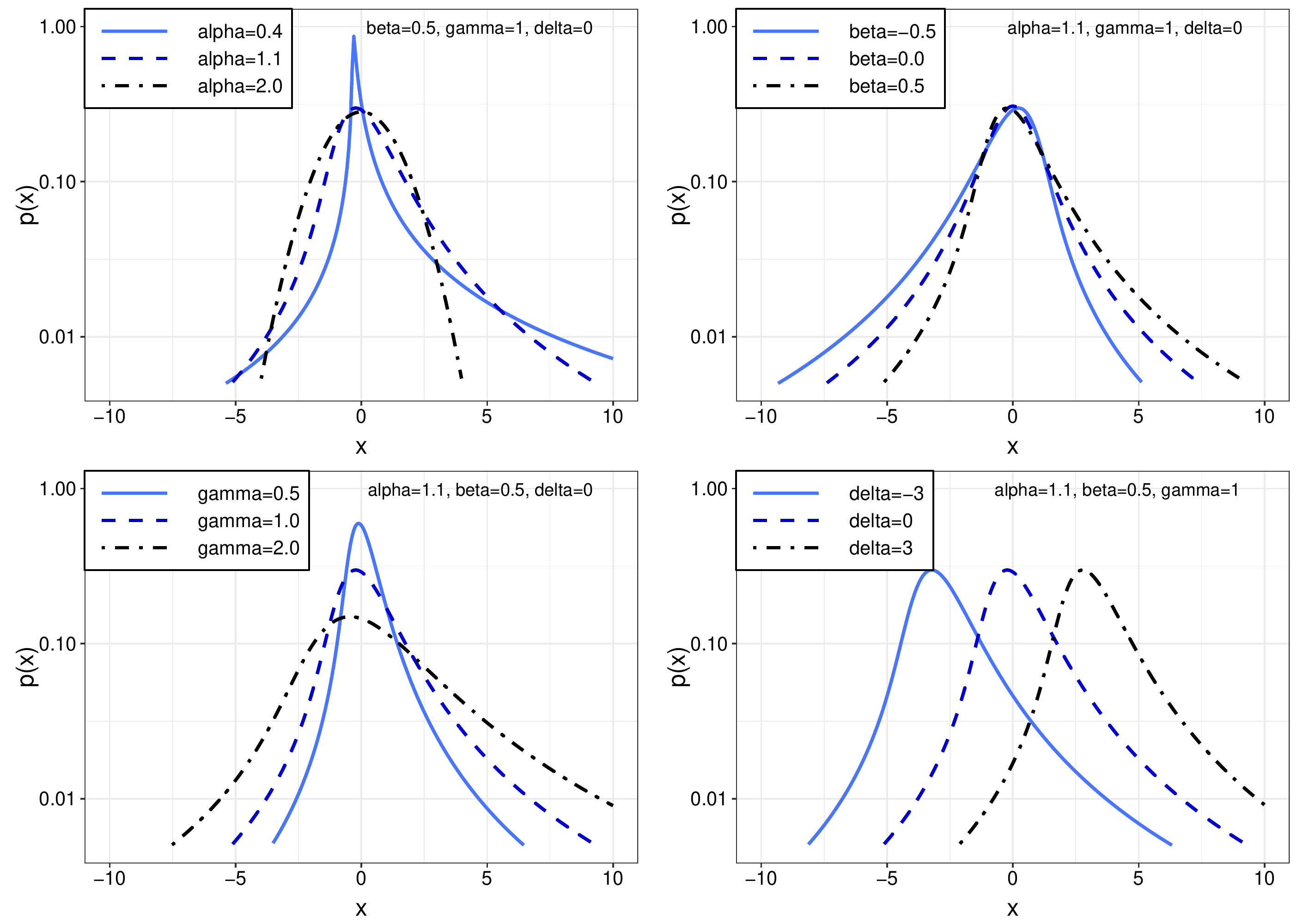}
\caption{Density of the L\'{e}vy alpha-stable distribution for different parameter settings. Upper left: Variation of tail parameter $\alpha$. Upper right: Variation of skew parameter $\beta$. Lower left: Variation of scale parameter $\gamma$. Lower right: Variation of location parameter $\delta$.}
\label{fig:variations:levy}
\end{figure}

\section{Methods}
\label{sect:methods}

\subsection{Distributional models}

Most studies of labor productivity and of firm-level data, in general, are based on generative models. They define, which effects on the measure under investigation are considered under the model; they fix their functional forms; and they establish the resulting distribution. Typically, the approximate form of the distribution to be explained is known, which constrains the variety of candidate models. 

The advantages of the generative approach include that it is illustrative and verifiable by considering other quantities represented in the model. However, specific distributions can frequently be generated by large numbers of different generative models, and matching the correct distribution reveals little information about the correct generating process.

Instead, and in line with much of the modern literature \citep{Frank09,Bottazzi/Secchi06,Yang18,Yangetal19}, 
we consider a different approach: The attractor distribution to which the result of aggregations of (identical, independent) distributions converges. We remain agnostic with regard to the interpretation of the component distributions being aggregated, though temporal aggregation of shocks or aggregation across jobs, processes, or tasks within a firm are natural component separations that suggest themselves. If it is indeed the correct representation of the data, the distribution could be expected to remain stable under a number of changes to the system, as the aggregation continues to converge to this functional form.

In particular, following \citet{Yangetal19}, we use the L\'{e}vy alpha-stable distribution \citep{NOLAN2018,NOLAN1998187} as our main distributional model, although we provide fits to the 4-parameter Asymmetric Exponential Power distribution suggested elsewhere in the literature \citep{Bottazzi/Secchi06,Bottazzietal07,Bottazzi/Secchi11,Yuetal15} as a point of comparison. In the following, we provide a non-technical explanation and some intuition why the L\'{e}vy alpha-stable distribution may be a good distributional model. A technical description is given in the Appendix~\ref{app:dist}.

Random variables distributions can be aggregated in convolutions (i.e. summation of the variables), which yields a different distribution of the results (for technical details, see Appendix~\ref{app:dist:agg}). Aggregation leads to a loss of information; it washes out less strong signals and only a dominant pattern remains. As the convoluted distributions are independent, this pattern is the one that carries the least information (highest entropy), the one that is the most likely one without additional information, the one that constitutes the maximum entropy distribution under constraints that depend on the component distributions $X$. 

The maximum entropy perspective may be helpful in that it allows computing the resulting distribution and understanding the type of its constraints in a concise way. The resulting distributional form is determined by the constraints in the maximum entropy perspective, or equivalently by the type of convolution and the characteristics of the component distributions in the convolution perspective. For instance, a single constraint on the mean of the distribution will yield an exponential or Laplacian (two-sided exponential) maximum entropy distribution. The Asymmetric Exponential Power distributions that are often used for the distributional models for firm growth \citep{Bottazzi/Secchi06,Bottazzietal07,Bottazzi/Secchi11} or productivity \citep{Yuetal15}, belong to this family, albeit with a modification that allows for asymmetry\footnote{The maximum entropy constraint includes a sign function under this modification to distinguish the two tails and account for different shapes of both sides.} (for technical details, see Appendix~\ref{app:dist:aep}). A single constraint on the mean of the distribution under logarithmic transformation will yield a Pareto maximum entropy distribution, typically considered for distributional models of firm size distributions. A constraint on the variance of a distribution (implying a second constraint on the mean) will yield a Gaussian normal maximum entropy distribution. 

Almost all maximum entropy distributions do not constitute attractors under further aggregation. If the resulting distribution is further convoluted, it continues to change. Those that do constitute attractors, i.e. those that yield an identical distribution under convolution are known as L\'{e}vy alpha-stable distributions (for technical details, see Appendix~\ref{app:dist:levy-as}).  The  L\'{e}vy alpha-stable is a generalization of several families of distributions, including Gaussian normal distributions, Cauchy, distributions and L\'{e}vy distributions. The generalized central limit theorem (GCLT) states that any sum of independent, identical distributions will converge to a L\'{e}vy alpha-stable distribution. Specifically, if the convoluted distributions have a finite variance, the sum will converge to a Gaussian normal, a member of the family of L\'{e}vy alpha-stable distributions (for technical details, see Appendix~\ref{app:dist:clt}). If not, it will yield a different member of this family with a heavy tail and a tail parameter $<2$.

L\'{e}vy alpha-stable distributions do not have a closed-form representation as a function in the frequency domain, except for special parameter sets.\footnote{For $\alpha=2$ it yields a Gaussian normal distribution, for $\alpha=1$, it yields a Cauchy distribution, and for $\alpha=0.5$ it yields a  L\'{e}vy distribution.} The functional form in the Fourier domain (the \textit{characteristic function}, for technical details, see Appendix~\ref{app:dist:Fourier}) is 

\begin{eqnarray}
\varphi(s)=\operatorname{E}[e^{(isx)}]  ={\begin{cases}
	e^{(-\gamma^\alpha|s|^\alpha[1+i\beta \text{tan}\left({\tfrac {\pi \alpha }{2}}\right) \operatorname {sgn}(s)\left((\gamma|s|)^{1-\alpha}-1 \right))] + i\delta s)}
	&\alpha \neq 1
	\\e^{(-\gamma|s|[1+i\beta{\tfrac {2}{\pi}} \operatorname {sgn}(s)\log(\gamma|s|)] + i\delta s)}&\alpha =1
	\end{cases}}
\label{eq:solution:levy:fourierdomain}
\end{eqnarray}

Figure~\ref{fig:variations:levy} shows the bahaviour of the four parameters of the L\'{e}vy alpha-stable distribution in semi-log scale (y-axis logarithmic). The upper left panel contrasts the Gaussian case (black curve) with skewed fat-tailed cases for different tail indices $\alpha$. Note that the curve bends outward for the two fat-tailed cases, indicating that the tails are heavier than in an exponential distribution, which would be linear in a semi-log scale. This is a tell-tale sign of fat-tailedness. The lower left panel shows variations of the scale or width of the distribution. The scale, $\gamma$ is another measure of dispersion besides the tail index and is independent from it. In the Gaussian case ($\alpha=2$), the scale is simply the standard deviation. For fat-tailed variants ($\alpha<2$) such as the ones depicted in this panel, this is not the case, as the standard deviation is infinite. The right panels demonstrate different skew values and a location shift respectively.

More technical details on L\'{e}vy alpha-stable distributions can be found in \citet{NOLAN1998187, NOLAN2018}; a comprehensive discussion of maximum entropy, aggregation of distributions, and characteristic equations in the Fourier domain is offered in \citet{Frank09}.

\subsection{Fitting}

\subsubsection{L\'{e}vy alpha-stable distributions}

We use Nolan's \citep{NOLAN1998187,NOLAN2018} parametrization $0$ for the L\'{e}vy alpha-stable distribution as given in equation~\ref{eq:solution:levy:fourierdomain}. Common methods to fit the distribution include maximum likelihood, the general method of moments (GMM), and McCulloch's \citep{McCulloch86} quantile based estimation. Maximum likelihood is generally considered the most reliable, but requires much more computation power than the alternatives and is, for the data sizes considered here, not practical. A direct comparison of McCulloch's method with GMM\footnote{For this comparison, we used Hansen's \citeyearpar{Hansen82} two-step algorithm with Carrasco et al.'s \citeyearpar{Carrascoetal07} spectral cut-off regularization.} showed that for the relevant data sizes of at least $1000$, $5000$, and $10000$ (depending on the type of the subsample, see Section~\ref{sect:data}) observations each as considered here, McCulloch's quantile-based estimation is more accurate and gives generally better Soofi ID scores (see Section\ref{sect:methods:soofi}).

We use the R package \verb|StableEstim| \citep{STABLEESTIM}.

\subsubsection{Asymmetric exponential power (AEP) distributions}

Contrasting our distributional model to AEP distributions is expedient not only because it has been suggested as a distributional model in the literature \citep{Bottazzi/Secchi06,Bottazzietal07,Bottazzi/Secchi11,Yuetal15}, but also because AEP distributions show radically different tail behavior compared to L\'{e}vy alpha-stable distributions, which are heavy-tailed and have infinite variance for $\alpha<2$. While finite samples always have a finite variance, it will diverge in the sample size if the underlying distribution of the sample is heavy-tailed. As a result, measuring the variance of a sample from a heavy-tailed distribution will yield misleading results \citep{NOLAN2018,Emberchts97,Yangetal19}, as they are tainted by other quantities such as the sample size. Similar problems exist for other dispersion measures \citep{Yangetal19}. Performing OLS correlations on variables with hevy tails will likely also fail, since the error distribution will likely inherit the heavy tails and OLS requires errors with finite variance. For technical details, see Appendix~\ref{app:dist:tails}.

We use a 4-parameter AEP distribution with the functional form given in equation~\ref{eq:AEP:density} as an alternative distributional model for comparison. Fitting relies on the L-moments method as discussed in \cite{ASQUITH2014955} and implemented in the R package \verb|lmomco| \citep{lmonco}. 

\subsection{Goodness of fit}

Two measures for model selection and validation are used, both based on information theory considerations. Additional techniques, such as the Kolmogorov-Smirnov test\footnote{The KS test is known to have low precision and to lead to many false negatives. 
} or cross-validation are possible, but were not applied in the present study.

\subsubsection{Soofi ID index}
\label{sect:methods:soofi}

Our main goodness-of-fit metric is based on Soofi et al.'s \citep{Soofietal95} information distinguishability ($ID$) concept, which gives the distinguishability of two distributions based on their information content. We can use this measure to assess to what extent a fitted model $p(\theta| x)$ with parameters $\theta$ is distinguishable from the entropy maximizing distribution $q(\theta| x)$ given a set of observations $x$. 

Formally, $ID$ is based on the Kullback-Leibler divergence between the two distributions

$$ D_{\mathrm {KL} }(p\|q)=\sum _{i}p(x_i)\,\log {\frac {p(x_i)}{q(x_i)}}, $$

where $\|$ is the divergence operator.\footnote{I.e., for any concept of divergence, $p\|q$ is the divergence of $p$ and $q$; $p\|q|\theta$ is the divergence of $p$ and $q$ given $\theta$.} Information distinguishability is defined as

\begin{equation}
ID(p\|q|\theta) = \exp[-D_{\mathrm {KL} }(p\|q|\theta)],       
\end{equation}

and has support $ID \in [0,1]$. $ID=0$ indicates that the distributions are indistinguishable, while the differences are more pronounced the higher $ID$. For convenience, we construct a Soofi ID score $SIDS$ as previously used by \citet{Yang18,Yangetal19} by rescaling $ID$

\begin{equation}
SIDS = 100 \times (1-ID),    
\end{equation}

with support $SIDS \in [0,100]$ such that $SIDS=100$ indicates a perfect match while low values indicate that the distributional model under investigation is probably not correct for the sample in question.

\subsubsection{Akaike information criterion (AIC)}

Akaike's \citep{Akaike73} information criterion (AIC) is based on the likelihood of a distributional model fit while accounting for the number of parameters. Formally,

\begin{equation}
\operatorname{AIC} = 2k - 2\log\mathcal{L}(\theta|x) 	
\end{equation}

where $\mathcal{L}(\theta|x)$ is the likelihood function of parameters $\theta$ given data $x$ and $k$ is the number of estimated parameters $\theta$. 

The AIC relies on the same concepts as the $SIDS$, namely minimizing Kullback-Leibler divergence, but rescales differently and applies a correction for $k$. It offers a measure for model comparison, but is difficult to interpret in illustrative terms. $SIDS$, on the other hand, has a straight forward interpretation as the extent of similarity with an entropy maximizing model given the data. 

 \begin{figure}[tb!]
 \centering
 \includegraphics[width=0.65\textwidth]{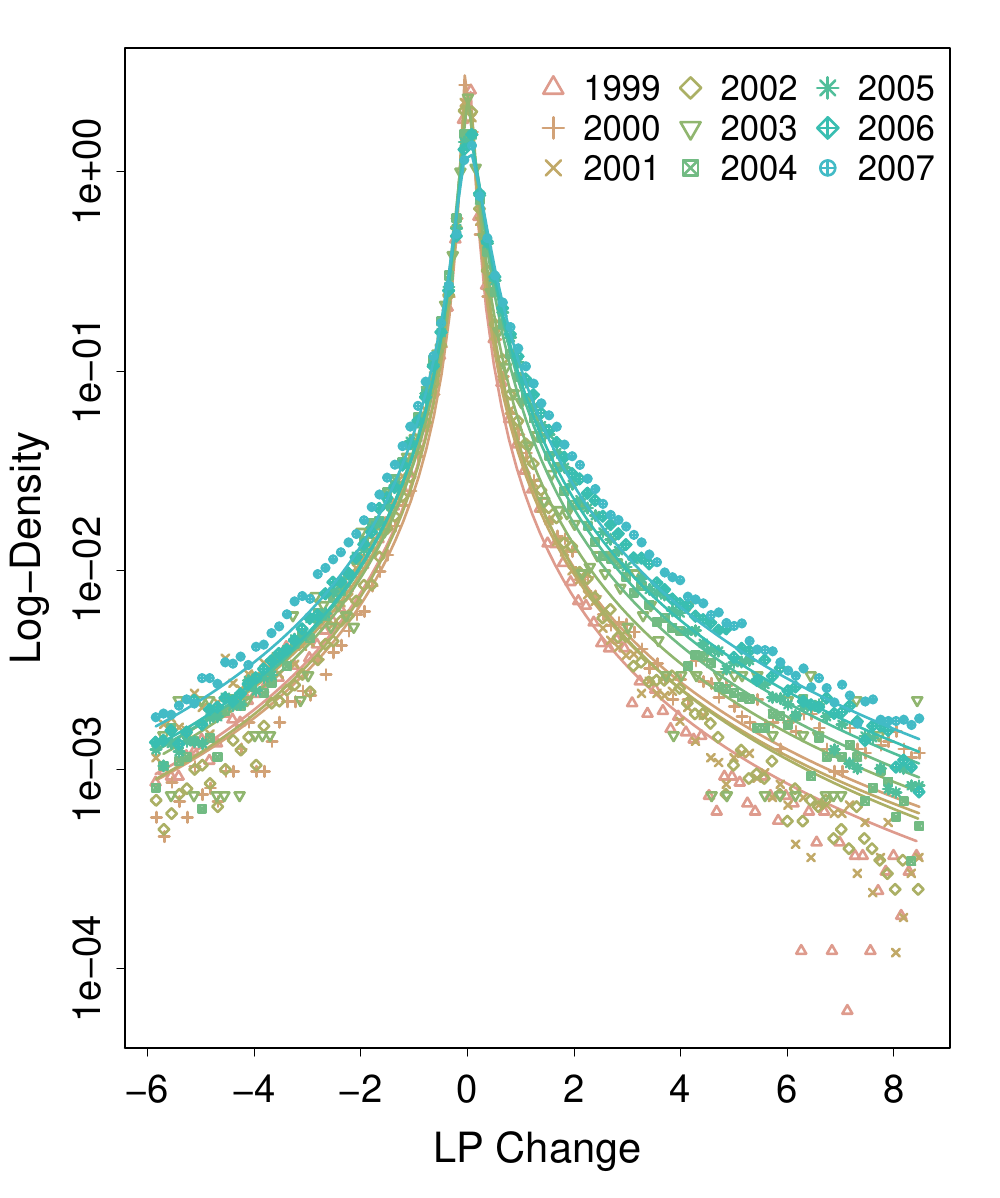}
 \caption{Density of the labor productivity change ($\Delta LP$) distribution (full sample) by year in semi-log (vertical axis logarithmic). Solid lines indicate Levy alpha stable distribution fits as reported in Table~\ref{tab:patameterfits}.}
 \label{fig:density:year:lp_change}
 \end{figure}

\begin{table}[ht]
\centering
\small
\begin{tabular}{llrrrrrrrrr}
  \hline\hline
       & \multicolumn{1}{c}\textbf{Year} & \multicolumn{1}{c}\textbf{$\#$Obs.} & \multicolumn{4}{c}{\textbf{L\'{e}vy alpha-stable fit}} & \multicolumn{4}{c}{\textbf{AEP fit}} \\ 
       \cmidrule(lr){2-2}    \cmidrule(lr){3-3}  \cmidrule(lr){4-7} \cmidrule(lr){8-11}

       &      &  & $\alpha$ & $\beta$ & $\gamma$ & $\delta$ & $\kappa$ & $h$ & $\sigma$ & $\xi$ \\
        \cmidrule(lr){4-7} \cmidrule(lr){8-11} 
        
	\multirow{10}{*}{LP}&   1998 & 140,372 & \textbf{1.00} & \textbf{0.95} & \textbf{0.11} & \textbf{0.11} & 0.47 & 0.40 & 0.01 & 0.06 \\ 
 & 1999 & 147,492 & \textbf{1.06} & \textbf{0.95} & \textbf{0.12} & \textbf{0.13} & 0.45 & 0.46 & 0.02 & 0.07 \\ 
 & 2000 & 145,724 & \textbf{0.97} & \textbf{0.95} & \textbf{0.14} & \textbf{0.14} & 0.42 & 0.27 & 0.00 & 0.10 \\ 
 & 2001 & 157,083 & \textbf{1.08} & \textbf{0.95} & \textbf{0.15} & \textbf{0.18} & 0.42 & 0.48 & 0.02 & 0.10 \\ 
 & 2002 & 167,723 & \textbf{1.08} & \textbf{0.95} & \textbf{0.17} & \textbf{0.20} & 0.44 & 0.48 & 0.03 & 0.12 \\ 
 & 2003 & 11,288 & \textbf{1.04} & \textbf{0.95} & \textbf{0.21} & \textbf{0.28} & 0.43 & 0.40 & 0.02 & 0.18 \\ 
 & 2004 & 265,218 & \textbf{1.06} & \textbf{0.95} & \textbf{0.20} & \textbf{0.25} & 0.42 & 0.45 & 0.03 & 0.15 \\ 
 & 2005 & 260,200 & \textbf{1.03} & \textbf{0.95} & \textbf{0.25} & \textbf{0.30} & 0.40 & 0.45 & 0.03 & 0.17 \\ 
 & 2006 & 287,854 & \textbf{1.00} & \textbf{0.95} & \textbf{0.30} & \textbf{0.36} & 0.37 & 0.43 & 0.03 & 0.21 \\ 
 & 2007 & 321,390 & \textbf{0.99} & \textbf{0.95} & \textbf{0.36} & \textbf{0.43} & 0.36 & 0.43 & 0.03 & 0.24 \\ \hline
  
  \multirow{9}{*}{$\Delta LP$}  & 1999 & 112,546 & \textbf{0.96} & \textbf{-0.05} & \textbf{0.08} & \textbf{0.01} & 1.07 & 0.34 & 0.00 & 0.02 \\ 
  & 2000 & 120,757 & 0.90 & 0.25 & 0.09 & -0.00 & 0.72 & 0.22 & 0.00 & -0.04 \\ 
  & 2001 & 115,506 & \textbf{0.93} & \textbf{-0.11} & \textbf{0.10} & \textbf{0.01} & 1.28 & 0.24 & 0.00 & 0.07 \\ 
  & 2002 & 137,628 & \textbf{0.99} & \textbf{0.12} & \textbf{0.11} & \textbf{0.01} & 0.92 & 0.41 & 0.02 & 0.01 \\ 
  & 2003 & 9,407 & \textbf{0.97} & \textbf{0.14} & \textbf{0.13} & \textbf{0.01} & 0.90 & 0.38 & 0.01 & 0.00 \\ 
  & 2004 & 119,389 & \textbf{1.03} & \textbf{0.20} & \textbf{0.18} & \textbf{0.02} & 0.86 & 0.42 & 0.03 & 0.00 \\ 
  & 2005 & 225,293 & \textbf{1.01} & \textbf{0.26} & \textbf{0.19} & \textbf{0.04} & 0.84 & 0.42 & 0.03 & 0.02 \\ 
  & 2006 & 242,468 & \textbf{0.99} & \textbf{0.29} & \textbf{0.20} & \textbf{0.05} & 0.81 & 0.40 & 0.03 & 0.02 \\ 
  &2007 & 265,926 & \textbf{0.99} & \textbf{0.25} & \textbf{0.25} & \textbf{0.06} & 0.84 & 0.40 & 0.03 & 0.04 \\\hline 
   \multirow{12}{*}{$ROC$}   & 1998 & 136,435 & 0.91 & 0.37 & 0.08 & 0.00 & \textbf{0.73} & \textbf{0.34} & \textbf{0.00} & \textbf{-0.01} \\ 
  & 1999 & 145,513 & \textbf{0.91} & \textbf{0.44} & \textbf{0.07} & \textbf{0.00} & 0.69 & 0.34 & 0.00 & -0.01 \\ 
  & 2000 & 141,570 & \textbf{0.93} & \textbf{0.51} & \textbf{0.08} & \textbf{0.01} & \textbf{0.67} & \textbf{0.34} & \textbf{0.00} & \textbf{-0.00} \\ 
  & 2001 & 155,527 & \textbf{0.93} & \textbf{0.55} & \textbf{0.09} & \textbf{0.02} & 0.65 & 0.32 & 0.00 & 0.00 \\ 
  & 2002 & 166,246 & \textbf{0.98} & \textbf{0.61} & \textbf{0.10} & \textbf{0.03} & \textbf{0.64} & \textbf{0.34} & \textbf{0.00} & \textbf{0.01} \\ 
  & 2003 & 11,250 & \textbf{1.01} & \textbf{0.59} & \textbf{0.11} & \textbf{0.05} & 0.65 & 0.36 & 0.01 & 0.03 \\ 
  & 2004 & 260,311 & \textbf{0.96} & \textbf{0.61} & \textbf{0.12} & \textbf{0.05} & 0.63 & 0.31 & 0.00 & 0.02 \\ 
  & 2005 & 258,741 & \textbf{1.00} & \textbf{0.69} & \textbf{0.14} & \textbf{0.06} & \textbf{0.59} & \textbf{0.35} & \textbf{0.01} & \textbf{0.02} \\ 
  & 2006 & 286,602 & \textbf{1.01} & \textbf{0.78} & \textbf{0.14} & \textbf{0.06} & \textbf{0.56} & \textbf{0.35} & \textbf{0.01} & \textbf{0.03} \\ 
  & 2007 & 319,847 & \textbf{1.02} & \textbf{0.92} & \textbf{0.15} & \textbf{0.07} & \textbf{0.52} & \textbf{0.35} & \textbf{0.01} & \textbf{0.03} \\ 
  & 2012 & 40,840 & \textbf{1.00} & \textbf{0.75} & \textbf{0.18} & \textbf{0.09} & \textbf{0.55} & \textbf{0.36} & \textbf{0.01} & \textbf{0.03} \\ 
  & 2013 & 40,090 & \textbf{0.96} & \textbf{0.74} & \textbf{0.18} & \textbf{0.09} & 0.55 & 0.33 & 0.01 & 0.04 \\ \hline
 \multirow{14}{*}{$IR$}   & 1999 & 119,401 & \textbf{0.86} & \textbf{0.40} & \textbf{0.09} & \textbf{-0.04} & 0.61 & 0.24 & 0.00 & -0.09 \\ 
& 2000 & 118,108 & \textbf{0.86} & \textbf{0.40} & \textbf{0.09} & \textbf{-0.06} & 0.61 & 0.25 & 0.00 & -0.12 \\ 
& 2001 & 117,732 & \textbf{0.86} & \textbf{0.41} & \textbf{0.10} & \textbf{-0.02} & 0.60 & 0.25 & 0.00 & -0.09 \\ 
& 2002 & 136,224 & \textbf{0.84} & \textbf{0.46} & \textbf{0.11} & \textbf{-0.02} & 0.57 & 0.24 & 0.00 & -0.08 \\ 
& 2003 & 9,362 & \textbf{0.93} & \textbf{0.58} & \textbf{0.13} & \textbf{-0.05} & 0.53 & 0.23 & 0.00 & -0.10 \\ 
& 2004 & 116,966 & \textbf{0.92} & \textbf{0.66} & \textbf{0.24} & \textbf{-0.12} & 0.49 & 0.26 & 0.00 & -0.22 \\ 
& 2005 & 221,420 & 0.82 & 0.60 & 0.15 & -0.02 & 0.49 & 0.23 & 0.00 & -0.09 \\ 
& 2006 & 240,962 & \textbf{0.87} & \textbf{0.58} & \textbf{0.14} & \textbf{-0.03} & 0.53 & 0.25 & 0.00 & -0.09 \\ 
& 2007 & 264,428 & \textbf{0.86} & \textbf{0.59} & \textbf{0.14} & \textbf{-0.04} & 0.52 & 0.25 & 0.00 & -0.11 \\ 
& 2008 & 156,516 & \textbf{0.78} & \textbf{0.59} & \textbf{0.15} & \textbf{-0.07} & 0.48 & 0.22 & 0.00 & -0.16 \\ 
& 2009 & 128,846 & \textbf{0.87} & \textbf{0.56} & \textbf{0.18} & \textbf{0.02} & 0.53 & 0.25 & 0.00 & -0.07 \\ 
& 2010 & 151,441 & 0.63 & 0.20 & 0.03 & -0.03 & 0.51 & 0.10 & 0.00 & -0.10 \\ 
& 2012 & 8,746 & \textbf{0.88} & \textbf{0.95} & \textbf{0.58} & \textbf{-0.03} & 0.36 & 0.25 & 0.00 & -0.22 \\ 
& 2013 & 35,452 & \textbf{0.86} & \textbf{0.55} & \textbf{0.11} & \textbf{-0.05} & 0.54 & 0.22 & 0.00 & -0.10 \\ 
   \hline\hline
\end{tabular}
\caption{L\'{e}vy alpha-stable and AEP parameter fits for labor productivity $LP$, labor productivity change $\Delta LP$, profitability $ROC$, and investment rate $IR$ by year. Fits with comparatively better goodness in either $SIDS$ or $AIC$ in bold, provided $SIDS>95$. Details of the associated goodness of the fits are given in Table~\ref{tab:goodness-of-fits}.}
\label{tab:patameterfits}
\end{table}

\begin{table}[ht]
\centering
\small
\begin{tabular}{llrrrrrrc}
  \hline\hline
        & Year & \multicolumn{2}{c}{L\'{e}vy alpha-stable} & \multicolumn{2}{c}{AEP} & $\Delta SIDS$ & $\Delta AIC$ & Preferred model\\ 
         \cmidrule(lr){2-2}    \cmidrule(lr){3-4}  \cmidrule(lr){5-6} \cmidrule(lr){7-8} \cmidrule(lr){9-9}
       
        &      & $SIDS$ & $AIC$ & $SIDS$ & $AIC$ &  &  & \\ 
 \cmidrule(lr){3-4}  \cmidrule(lr){5-6}
\multirow{10}{*}{$LP$} & 1998 & 98.78 & 880.11 & 88.71 & 913.49 & 10.07 & -33.38 &  L\'{e}vy $\alpha$-s. \\ 
& 1999 & 98.10 & 803.04 & 93.86 & 845.21 & 4.24 & -42.17 &  L\'{e}vy $\alpha$-s. \\ 
& 2000 & 99.76 & 1,249.25 & 99.14 & 1,320.87 & 0.62 & -71.61 &  L\'{e}vy $\alpha$-s. \\ 
& 2001 & 98.26 & 797.22 & 94.86 & 843.04 & 3.40 & -45.82 &  L\'{e}vy $\alpha$-s. \\ 
& 2002 & 98.43 & 826.50 & 94.88 & 859.14 & 3.55 & -32.64 &  L\'{e}vy $\alpha$-s. \\ 
& 2003 & 98.78 & 979.66 & 83.90 & 1,023.38 & 14.88 & -43.72 &  L\'{e}vy $\alpha$-s. \\ 
& 2004 & 98.45 & 884.45 & 95.01 & 927.27 & 3.44 & -42.82 &  L\'{e}vy $\alpha$-s. \\ 
& 2005 & 97.88 & 923.83 & 94.10 & 975.49 & 3.78 & -51.66 &  L\'{e}vy $\alpha$-s. \\ 
& 2006 & 97.49 & 972.51 & 93.41 & 1,025.50 & 4.08 & -52.99 &  L\'{e}vy $\alpha$-s. \\ 
& 2007 & 96.75 & 1,012.76 & 95.21 & 1,071.37 & 1.54 & -58.61 &  L\'{e}vy $\alpha$-s. \\ \hline
  \multirow{9}{*}{$\Delta LP$}  &1999 & 99.35 & 914.08 & 95.89 & 937.44 & 3.46 & -23.36 &  L\'{e}vy $\alpha$-s. \\ 
  &2000 & 94.48 & 1,322.04 & 94.48 & 1,357.34 & 0.00 & -35.30 & - \\ 
  &2001 & 97.92 & 1,301.04 & 84.58 & 1,337.90 & 13.34 & -36.87 &  L\'{e}vy $\alpha$-s. \\ 
  &2002 & 99.22 & 824.87 & 98.95 & 852.95 & 0.27 & -28.09 &  L\'{e}vy $\alpha$-s. \\ 
  &2003 & 98.96 & 901.36 & 98.85 & 931.11 & 0.11 & -29.75 &  L\'{e}vy $\alpha$-s. \\ 
  &2004 & 99.12 & 908.20 & 96.55 & 929.40 & 2.57 & -21.20 &  L\'{e}vy $\alpha$-s. \\ 
  &2005 & 99.22 & 925.94 & 99.56 & 948.58 & -0.34 & -22.64 &  L\'{e}vy $\alpha$-s. \\ 
  &2006 & 99.14 & 965.16 & 99.53 & 989.38 & -0.39 & -24.22 &  L\'{e}vy $\alpha$-s. \\ 
  &2007 & 99.24 & 1,003.31 & 99.33 & 1,024.51 & -0.09 & -21.20 &  L\'{e}vy $\alpha$-s. \\\hline 
  \multirow{12}{*}{$ROC$}  & 1998 & 92.99 & 919.04 & 96.53 & 960.82 & -3.54 & -41.78 & AEP \\ 
  & 1999 & 98.53 & 899.66 & 91.12 & 934.49 & 7.41 & -34.83 &  L\'{e}vy $\alpha$-s. \\ 
  & 2000 & 98.86 & 920.60 & 99.79 & 961.19 & -0.93 & -40.60 & even \\ 
  & 2001 & 98.38 & 978.17 & 96.66 & 1,018.65 & 1.72 & -40.48 &  L\'{e}vy $\alpha$-s. \\ 
  & 2002 & 99.06 & 973.09 & 99.11 & 1,017.00 & -0.05 & -43.92 & even \\ 
  & 2003 & 98.85 & 911.05 & 96.70 & 939.00 & 2.15 & -27.95 &  L\'{e}vy $\alpha$-s. \\ 
  & 2004 & 99.55 & 1,067.32 & 97.29 & 1,110.23 & 2.26 & -42.91 &  L\'{e}vy $\alpha$-s. \\ 
  & 2005 & 98.16 & 1,003.81 & 99.50 & 1,038.32 & -1.34 & -34.52 & even \\ 
  & 2006 & 97.26 & 1,016.92 & 99.15 & 1,047.34 & -1.89 & -30.41 & even \\ 
  & 2007 & 95.28 & 1,062.73 & 97.86 & 1,072.13 & -2.58 & -9.41 & even \\ 
  & 2012 & 95.20 & 1,031.04 & 97.74 & 1,060.46 & -2.54 & -29.43 & even \\ 
  & 2013 & 96.63 & 1,063.87 & 94.74 & 1,095.59 & 1.89 & -31.72 &  L\'{e}vy $\alpha$-s. \\ \hline
\multirow{14}{*}{$IR$}   & 1999 & 98.54 & 1,230.72 & 94.70 & 1,277.72 & 3.84 & -47.00 &  L\'{e}vy $\alpha$-s. \\ 
& 2000 & 98.27 & 1,221.62 & 94.69 & 1,283.48 & 3.58 & -61.86 &  L\'{e}vy $\alpha$-s. \\ 
& 2001 & 95.30 & 1,243.37 & 90.81 & 1,293.24 & 4.49 & -49.88 &  L\'{e}vy $\alpha$-s. \\ 
& 2002 & 96.57 & 1,243.78 & 96.28 & 1,296.99 & 0.29 & -53.21 &  L\'{e}vy $\alpha$-s. \\ 
& 2003 & 96.10 & 1,419.75 & 96.05 & 1,481.73 & 0.05 & -61.98 &  L\'{e}vy $\alpha$-s. \\ 
& 2004 & 95.63 & 1,408.22 & 93.21 & 1,475.98 & 2.42 & -67.76 &  L\'{e}vy $\alpha$-s. \\ 
& 2005 & 94.48 & 1,350.69 & 85.73 & 1,408.57 & 8.75 & -57.88 & - \\ 
& 2006 & 97.43 & 1,263.57 & 90.15 & 1,308.09 & 7.28 & -44.52 &  L\'{e}vy $\alpha$-s. \\ 
& 2007 & 97.06 & 1,239.16 & 79.77 & 1,283.18 & 17.29 & -44.02 &  L\'{e}vy $\alpha$-s. \\ 
& 2008 & 96.17 & 1,373.84 & 92.12 & 1,428.20 & 4.05 & -54.35 &  L\'{e}vy $\alpha$-s. \\ 
& 2009 & 97.73 & 1,302.20 & 94.31 & 1,351.16 & 3.42 & -48.96 &  L\'{e}vy $\alpha$-s. \\ 
& 2010 & 88.00 & 1,786.06 & 77.35 & 1,810.14 & 10.65 & -24.09 & - \\ 
& 2012 & 98.39 & 1,508.08 & 81.76 & 1,565.72 & 16.63 & -57.64 &  L\'{e}vy $\alpha$-s. \\ 
& 2013 & 95.42 & 1,297.17 & 90.06 & 1,342.75 & 5.36 & -45.57 &  L\'{e}vy $\alpha$-s. \\ 
   \hline\hline
\end{tabular}
\centering
\caption{Goodness of fit measures for L\'{e}vy alpha-stable and AEP parameter fits for labor productivity $LP$, labor productivity change $\Delta LP$, profitability $ROC$, and investment rate $IR$ by year as reported in Table~\ref{tab:patameterfits}. The last column notes the fit with comparatively better goodness in either $SIDS$ or $AIC$ in bold, provided $SIDS>95$.}
\label{tab:goodness-of-fits}
\end{table}

\begin{figure}[tb!]
 \centering
 \includegraphics[width=0.85\textwidth]{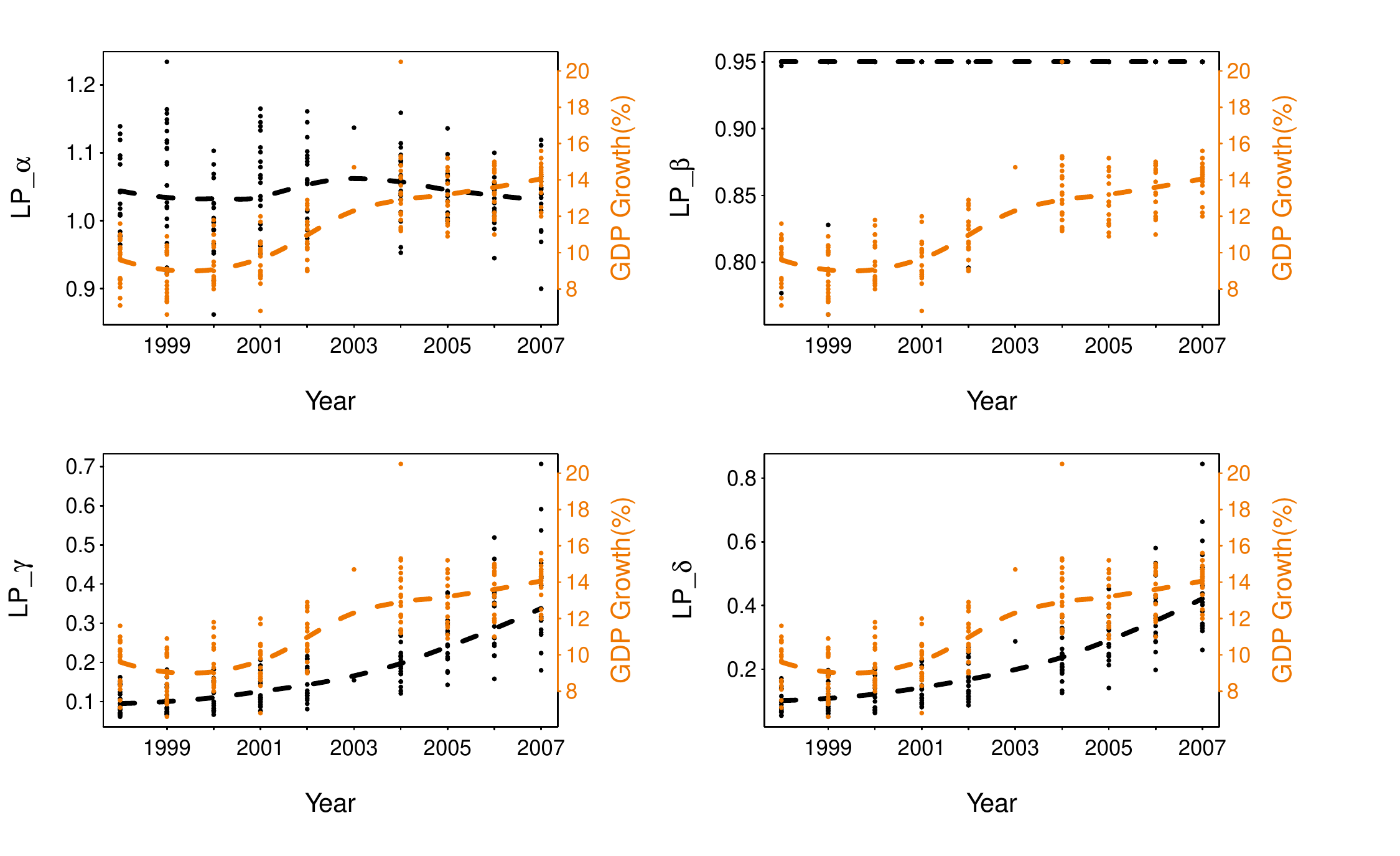}
 \caption{$LP$ (labor productivity) by region and year (black) in comparison to GDP growth (orange).}
 \label{fig:timedev:province:lp}
\end{figure}

\begin{figure}[tb!]
\centering
\includegraphics[width=0.85\textwidth]{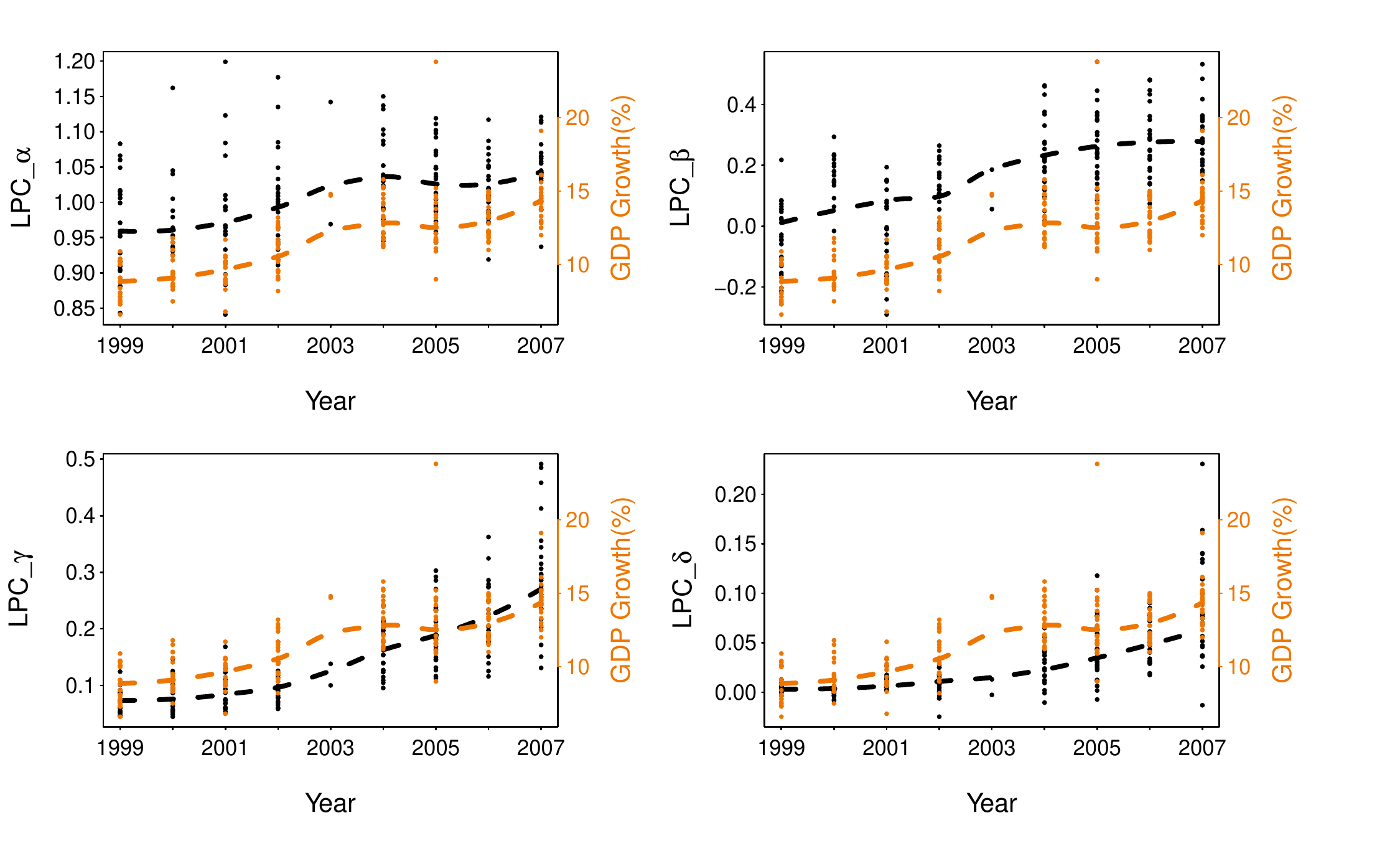}
\caption{$\Delta LP$ (labor productivity change) by region and year (black) in comparison to GDP growth (orange).}
\label{fig:timedev:province:lpc}
\end{figure}

\section{Results}
\label{sect:results}

In this section, we investigate questions that are of importance for understanding developing economies in general and the Chinese case during the decade of rapid catch-up (1999-2013, the period for which we have data) in particular. 
\begin{enumerate}
  \item What distributional model should be used for productivity microdata for developing countries (here, the P.R. China)? Do they differ from developed countries? (Section \ref{sect:results:estimation})
  \item Do the parameters of these distributions change with advancing development level? (Section \ref{sect:results:development})
  \item Are there persistent differences between regions (or countries)? (Section \ref{sect:results:regional})
  \item If there are any systematic differences or developments, how do they relate to other characteristics at the micro- or macro-level (firm age, GDP growth, capital intensity, employment)? (Section \ref{sect:results:determinants})
\end{enumerate}

Although our analysis is limited to the P.R. China, we can draw comparisons to the distributions of firm-level productivity data for developed economies \citep{Yangetal19} and conjecture that other developing economies may show similar patterns in phases of rapid economic catch-up. We also leverage the considerable diversity between Chinese provinces to assess regional differences, which may be an indicator of how different developing countries should be expected to be from one another in this regard.

\subsection{Fitting productivity distributions}
\label{sect:results:estimation}

We performed parameter fits with the L\'{e}vy alpha-stable model and, as a point of comparison, for the AEP model for all variables listed in section \ref{sect:data:vars}. However, we concentrate our analysis on the labor productivity $LP$, and the labor productivity change $\Delta LP$, while the other variables ($ROC$, $IR$) serve as a point of comparison and to show that the functional forms of the distributions are connected. 
L\'{e}vy alpha-stable fit lines as well as empirical density by year are shown in Figures~\ref{fig:density:year:lp} ($LP$) and \ref{fig:density:year:lp_change} ($\Delta LP$). The parameter values for the fits are given in the upper two sections of Table~\ref{tab:patameterfits}, while the goodness of fit measures are listed in Table~\ref{tab:goodness-of-fits}. 

The distributions of both variables ($LP$, $\Delta LP$) have striking and regular characteristics. They are (i) unimodal (one pronounced peak), (ii) heavy-tailed (bent outwards in semi-log), (iii) have wide support over both negative and positive numbers, and (iv) are highly stable over time. 
The L\'{e}vy alpha-stable model is an excellent fit of the distribution and the data, better than the alternative AEP in all cases.\footnote{This is confirmed by both goodness of fit measures employed here, the Soofi ID score ($SIDS$) and the Akaike information criterion ($AIC$). Table~\ref{tab:goodness-of-fits} explicitly lists which model provides a better fit for which sample in either criterion ($SIDS$ or $AIC$). AEP performs systematically worse than L\'{e}vy alpha-stable in the labor productivity, the labor productivity change, and the investment rate. Both models appear to be good fits for the profitability ($ROC$). Only in the case of the labor productivity change in 1999, both models resulted in a Soofi ID score $SIDS<95$ which indicates a less perfect fit. It is, however, a marginal case with $SIDS>94$ for both L\'{e}vy alpha-stable and AEP.} This is confirmed both in the goodness of fit measures in Table~\ref{tab:goodness-of-fits} and in the fit lines in Figures~\ref{fig:density:year:lp} and \ref{fig:density:year:lp_change}. 

\subsection{Economic development and systematic changes to productivity distributions}
\label{sect:results:development}

The fundings in Section~\ref{sect:results:estimation} show that the productivity distribution is found for the P.R. China is consistent with those identified for a wide range of developed economies \citep{Yangetal19}. However, while \citet{Yangetal19} find that there is no systematic change for developed economies over a period of 10 years (2006-2015), there is a persistent shift in several parameters in the case of China. This is evidenced by the densities and fit lines in Figures~\ref{fig:density:year:lp} ($LP$) and \ref{fig:density:year:lp_change} ($\Delta LP$) not overlaying each other but being neatly aligned next to one another in the exact order of years. The parameter values for the complete sample are given in Table~\ref{tab:patameterfits}; further, the black line in Figures~\ref{fig:timedev:province:lp} and \ref{fig:timedev:province:lpc} illustrate the development of the average of these parameter fits for each of the P.R. China's 31 provinces and autonomous regions.\footnote{This does not include the Special Administrative Regions Hong Kong and Macao, which are not represented in the database.}

For both distributions, the modal value became less pronounced, while the wings (not necessarily the tails) were pushed out, especially the one to the positive side (higher labor productivity, higher intertemporal gains in labor productivity). For both $LP$ and $\Delta LP$, the location parameter $\delta$ increases substantially from 1998 to 2008 and the scale parameter $\gamma$ increases in concert. This reflects the increase in labor productivity over the period of study with yearly changes and variation growing proportionally. While tail index $\alpha$ and the skew $\beta$ remain almost unchanged for $LP$, they increase systematically for $\Delta LP$. This has several implications:
\begin{itemize}
  \item Super-star firms do not become more prevalent with China's development push. The tail index of $LP$ remains approximately the same. Not even the skew of the $LP$ distribution changes. Instead the entire distribution shifts.
  \item The tails of the labor productivity change ($\Delta LP$) distribution grow shorter (higher $\alpha$, indicating that large productivity changes in one and the same firm become less common. Instead, the productivity change remains in the body of the distribution, therefore becoming more uniform across the economy.
  \item A right-skew emerges in the labor productivity change ($\Delta LP$) distribution. The body of the distribution stretches to the right (higher positive, but not excessively large productivity gains).
\end{itemize}

\clearpage

\begin{figure}[hbt!]
\centering
\subfloat[$LP$]{\includegraphics[width=0.4\textwidth]{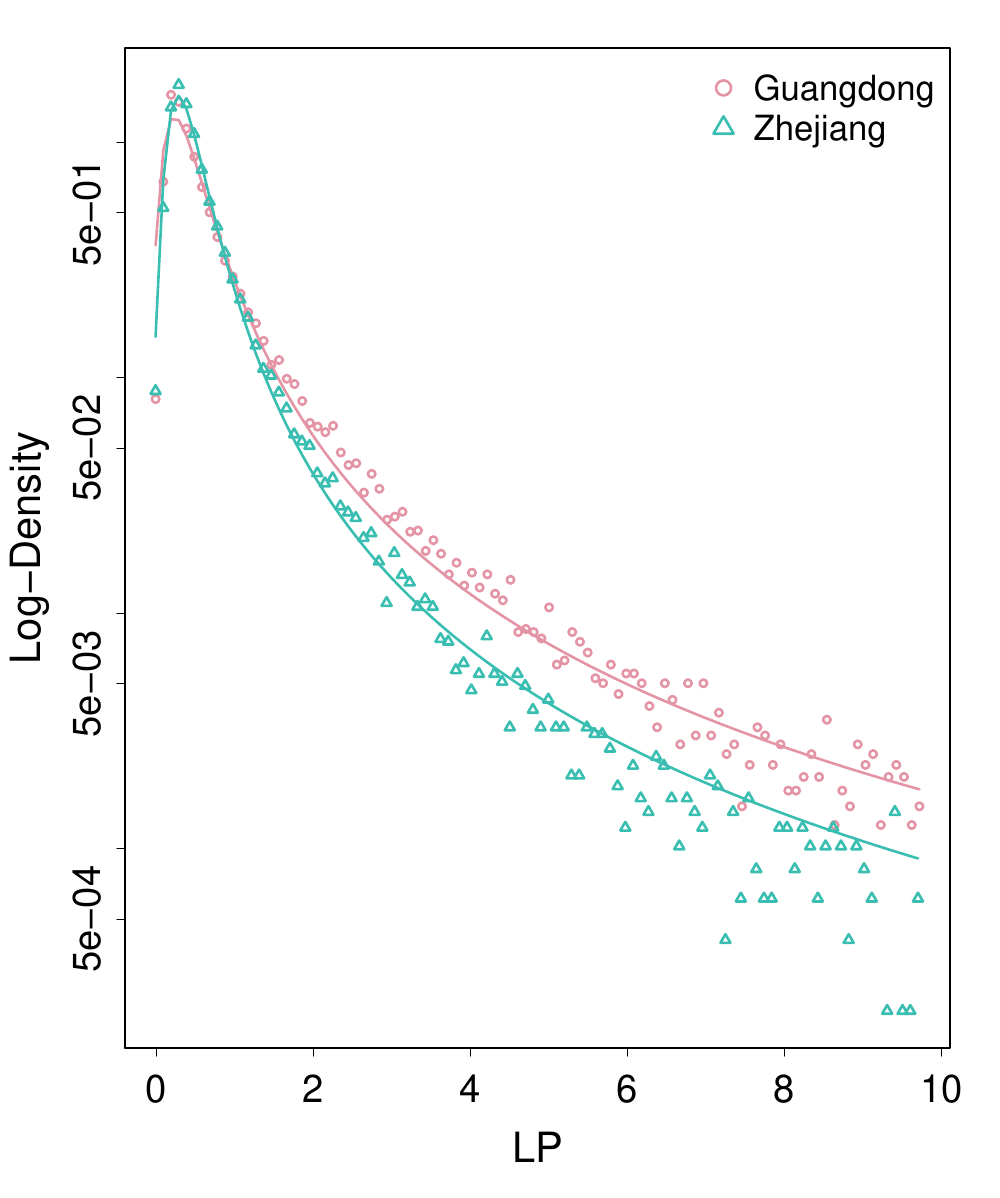}}
\subfloat[$\Delta LP$]{\includegraphics[width=0.4\textwidth]{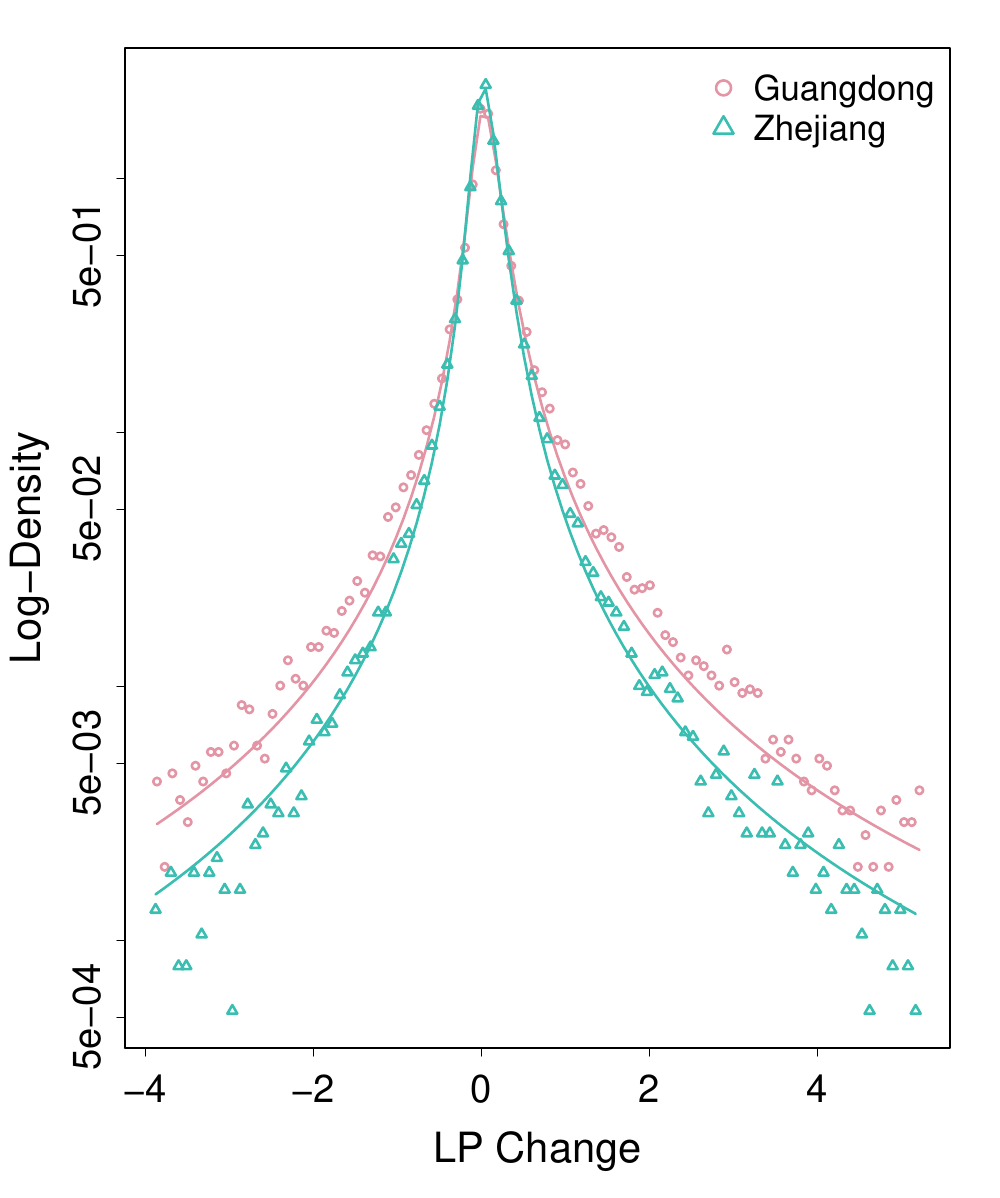}}\\
\caption{Density of $LP$ and $\Delta LP$ for regions Guangdong and Zhejiang in 2007}
\label{fig:2provinces:LP-LPC}
\end{figure}

\begin{figure}[hbtp!]
\centering
\subfloat[1998]{\includegraphics[width=0.9\textwidth]{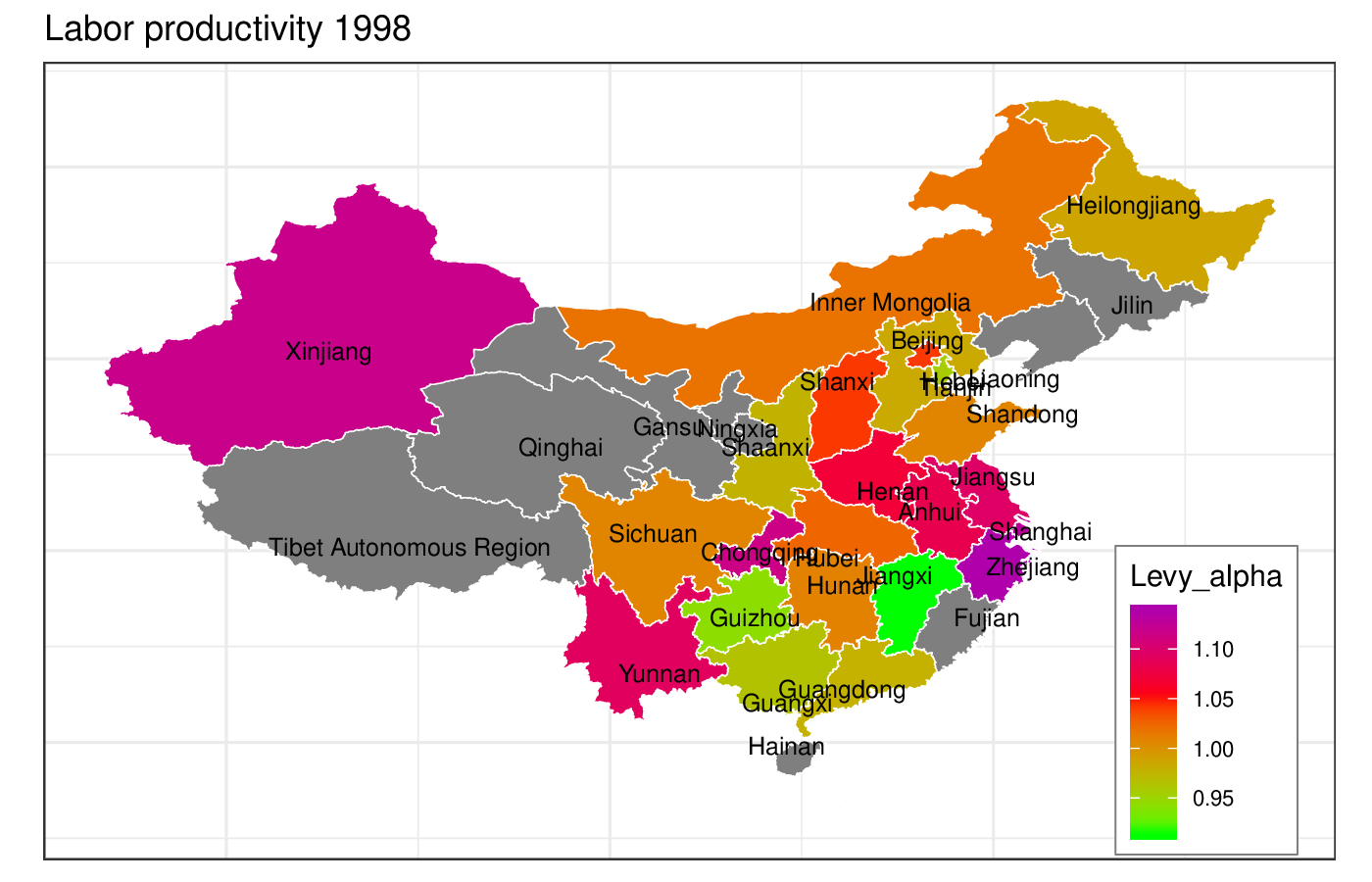}}\\
\subfloat[1999]{\includegraphics[width=0.33\textwidth]{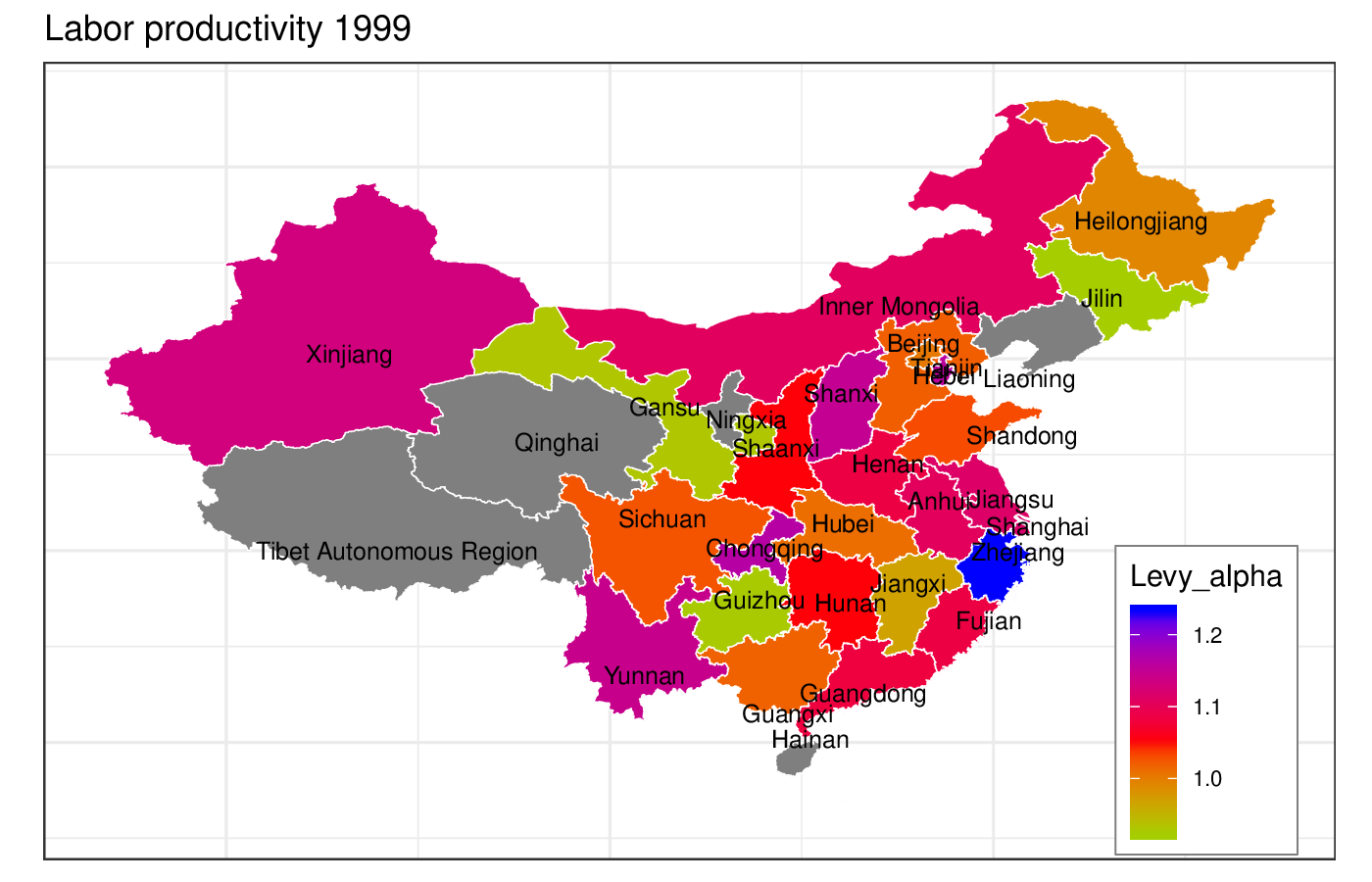}}
\subfloat[2000]{\includegraphics[width=0.33\textwidth]{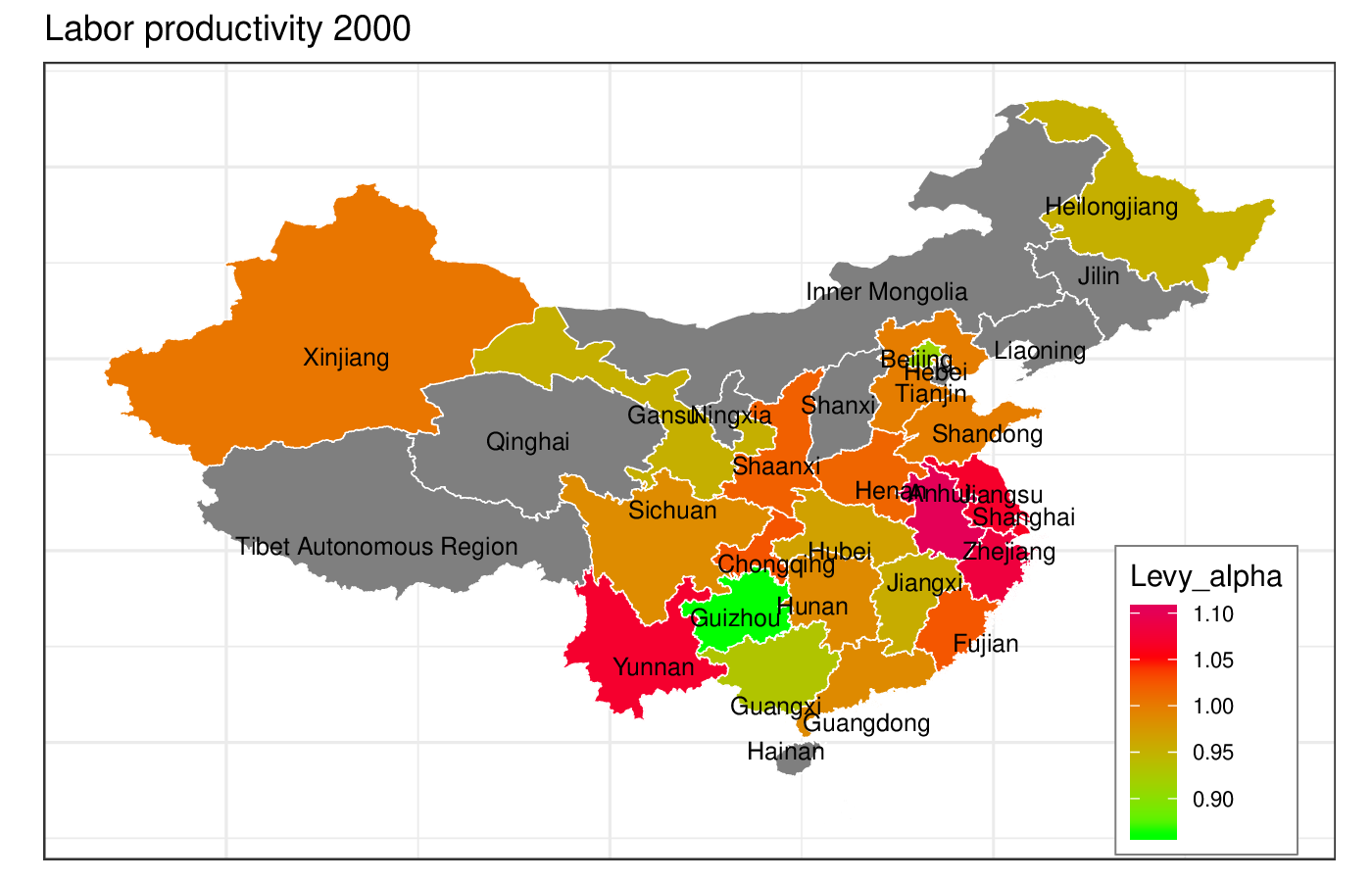}}
\subfloat[2001]{\includegraphics[width=0.33\textwidth]{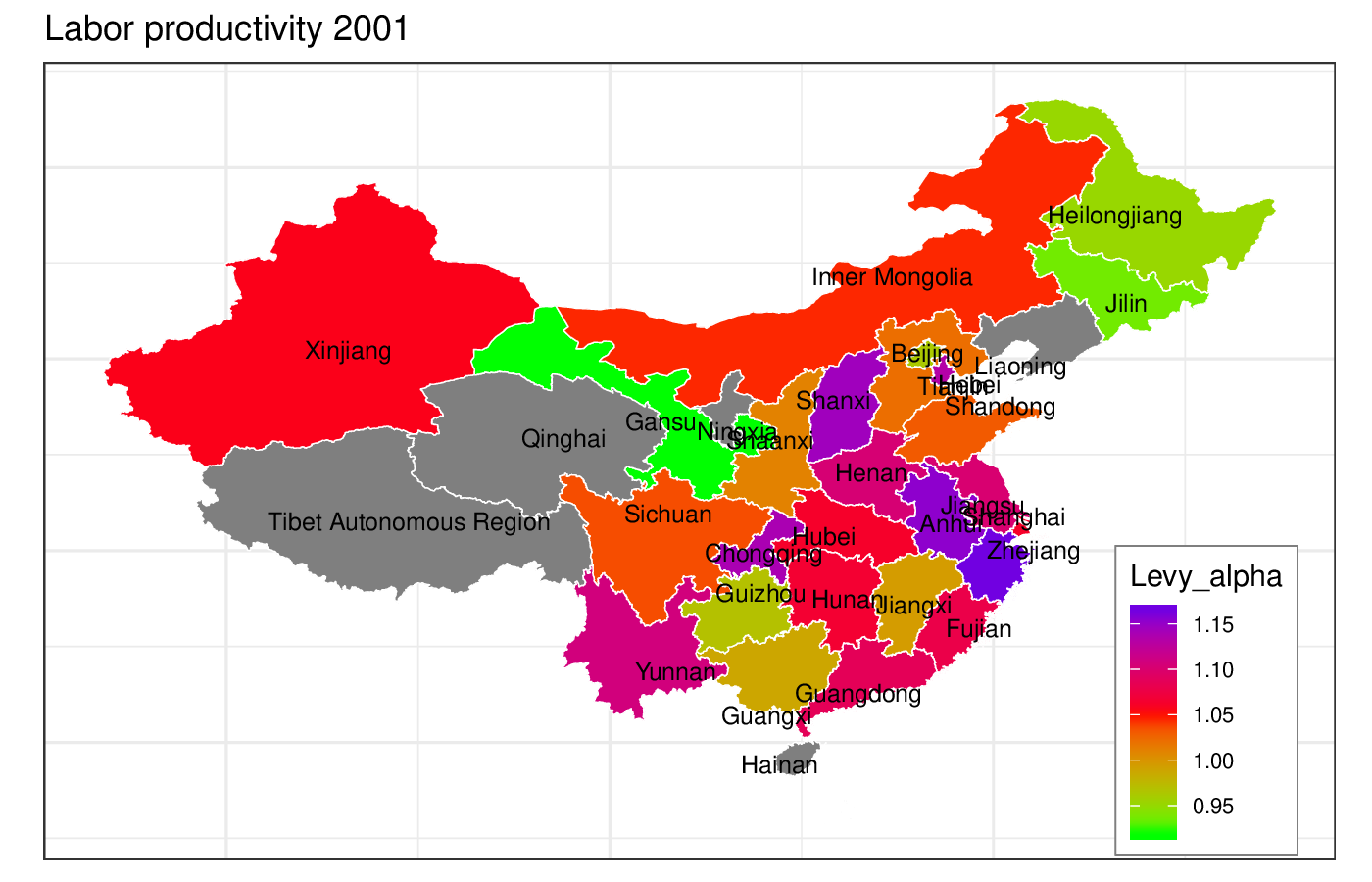}}\\
\subfloat[2002]{\includegraphics[width=0.33\textwidth]{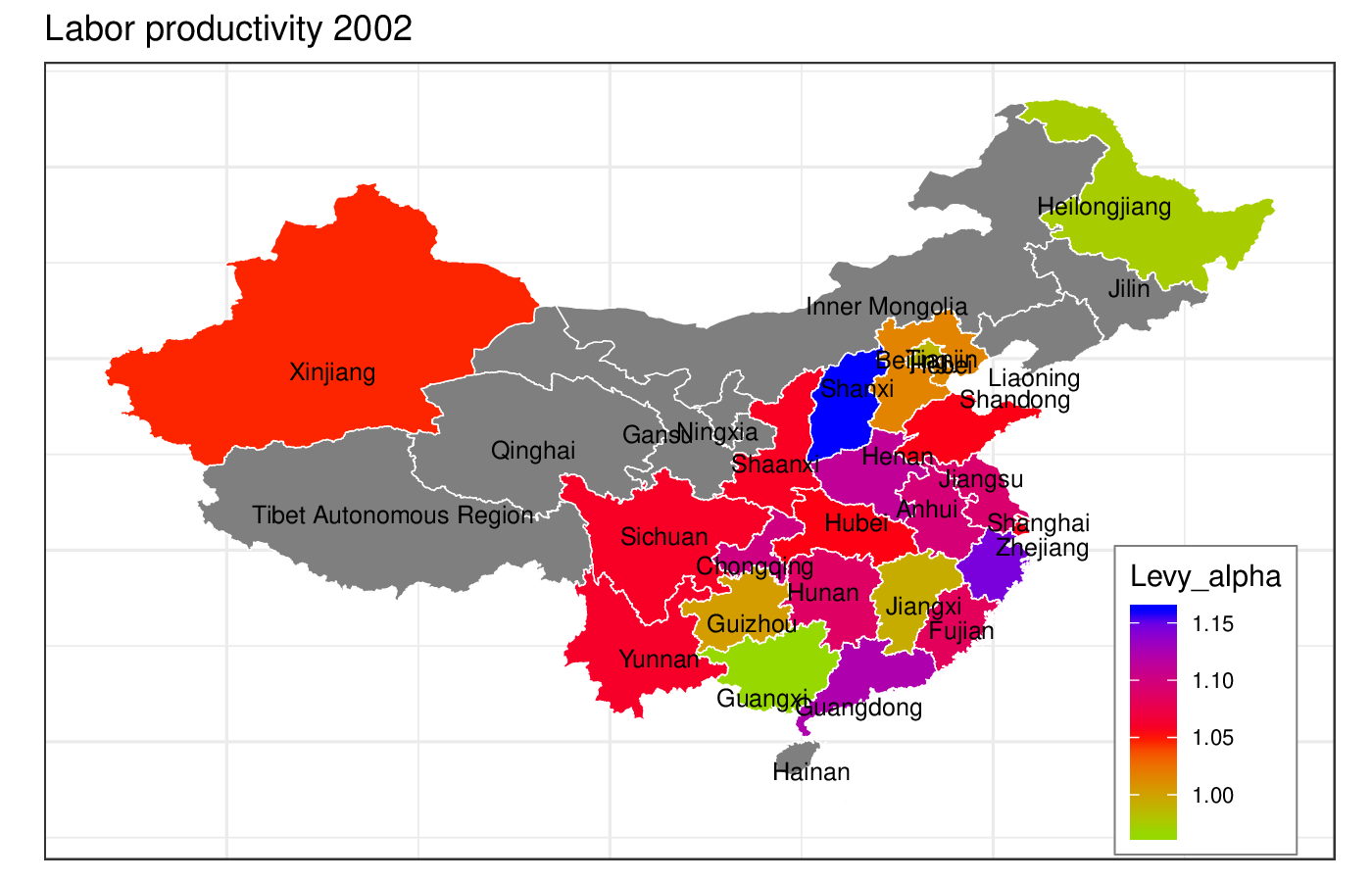}}
\subfloat[2004]{\includegraphics[width=0.33\textwidth]{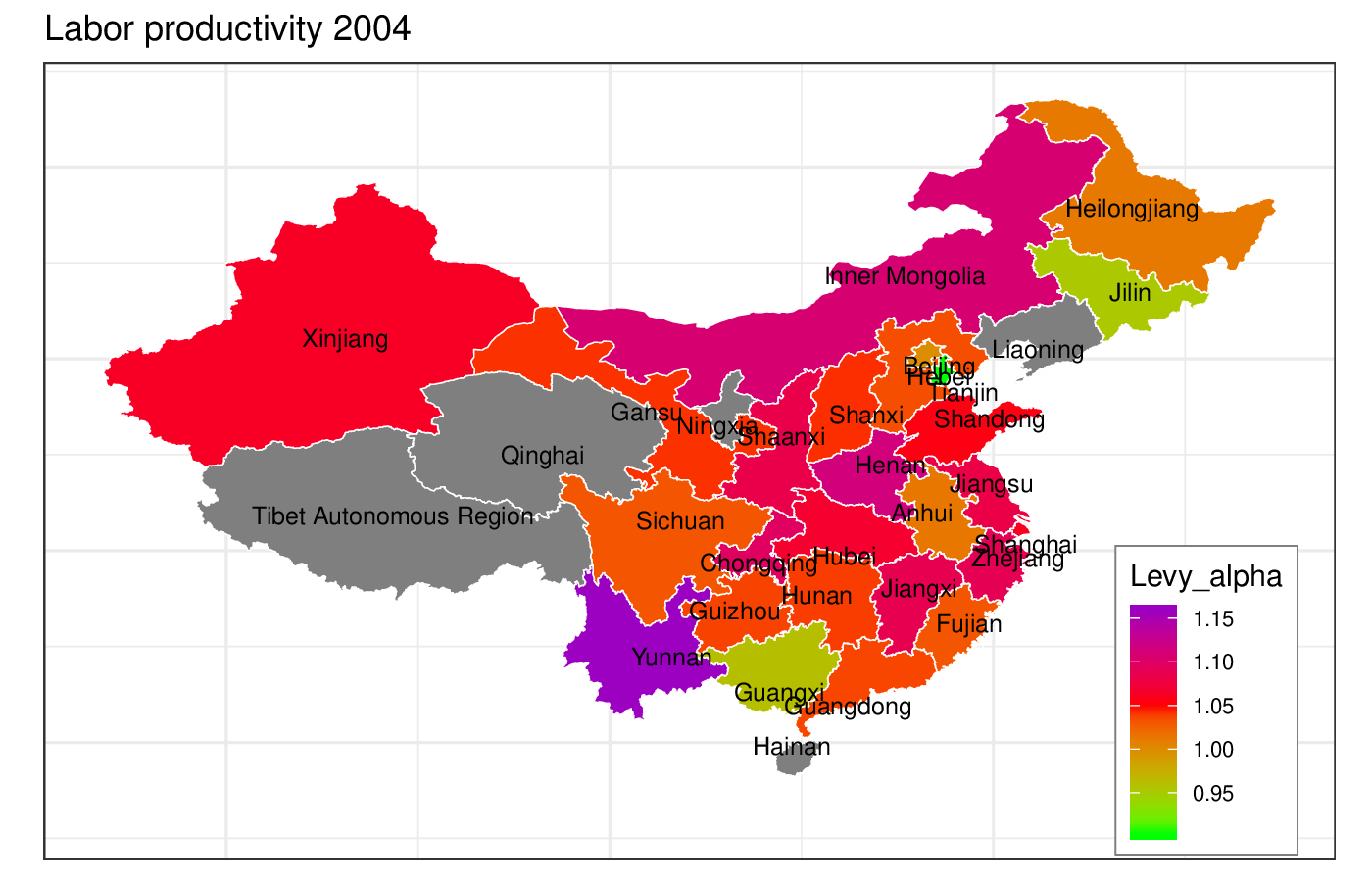}}
\subfloat[2005]{\includegraphics[width=0.33\textwidth]{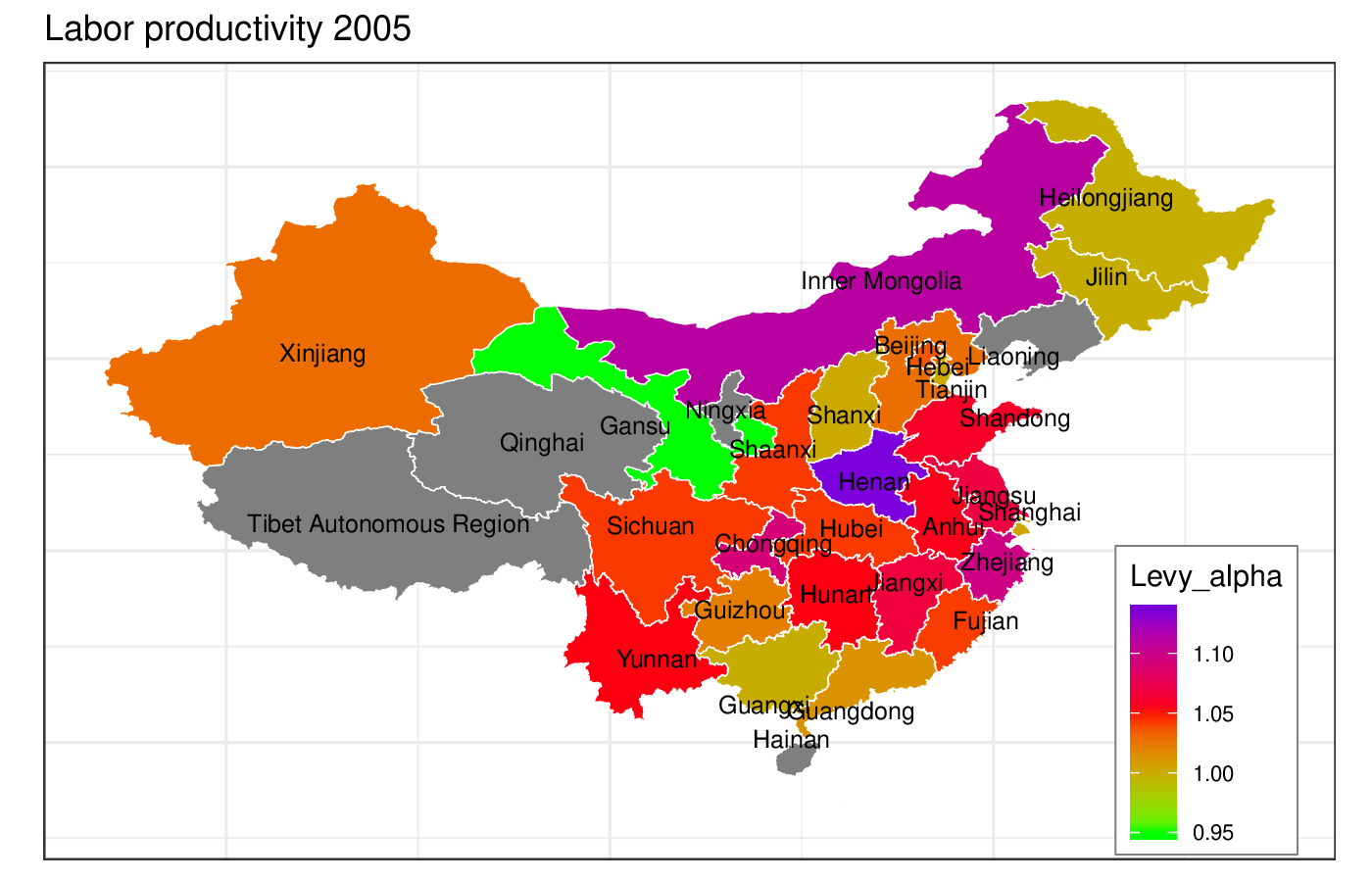}}\\
\subfloat[2006]{\includegraphics[width=0.33\textwidth]{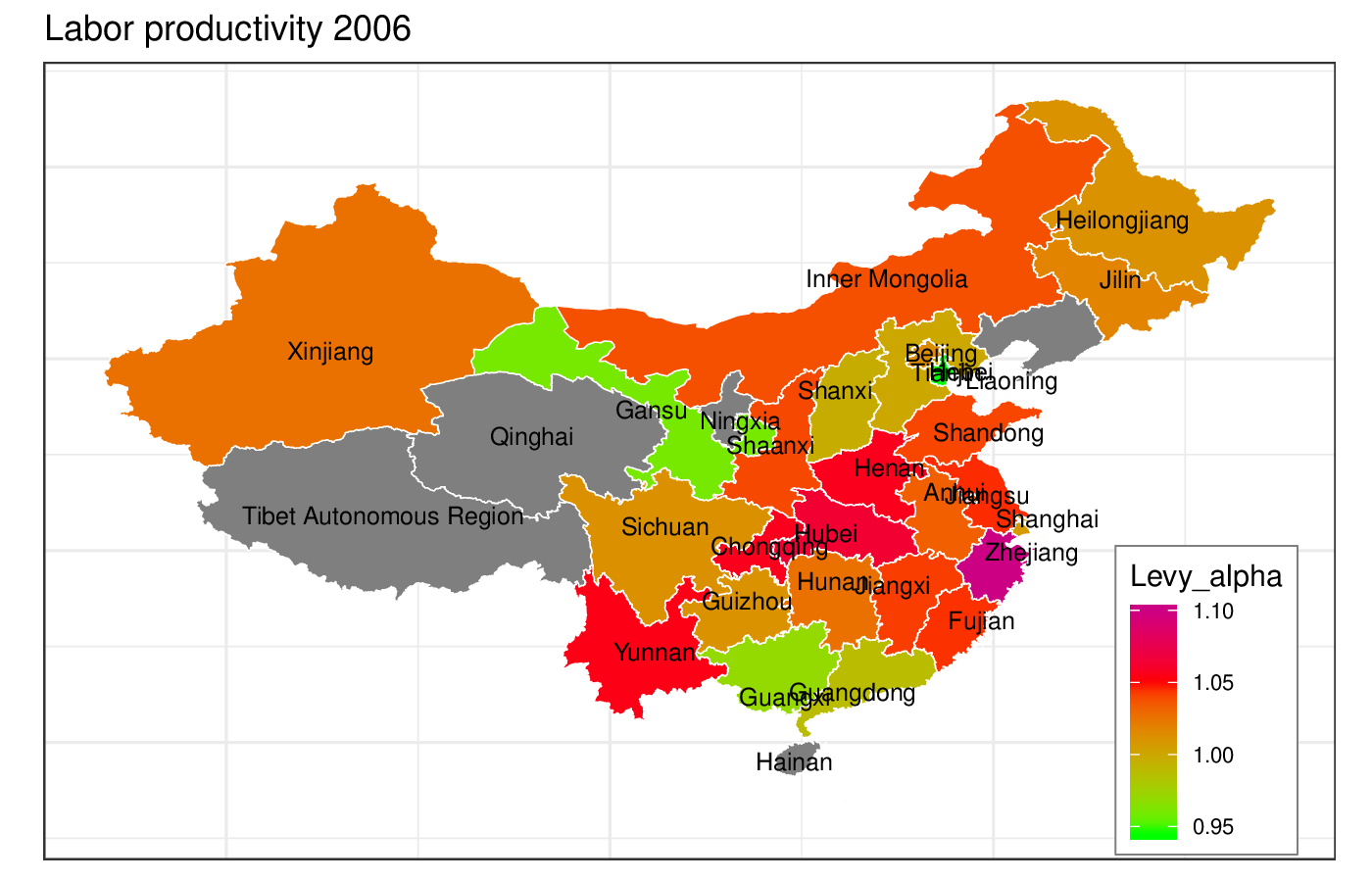}}
\subfloat[2007]{\includegraphics[width=0.33\textwidth]{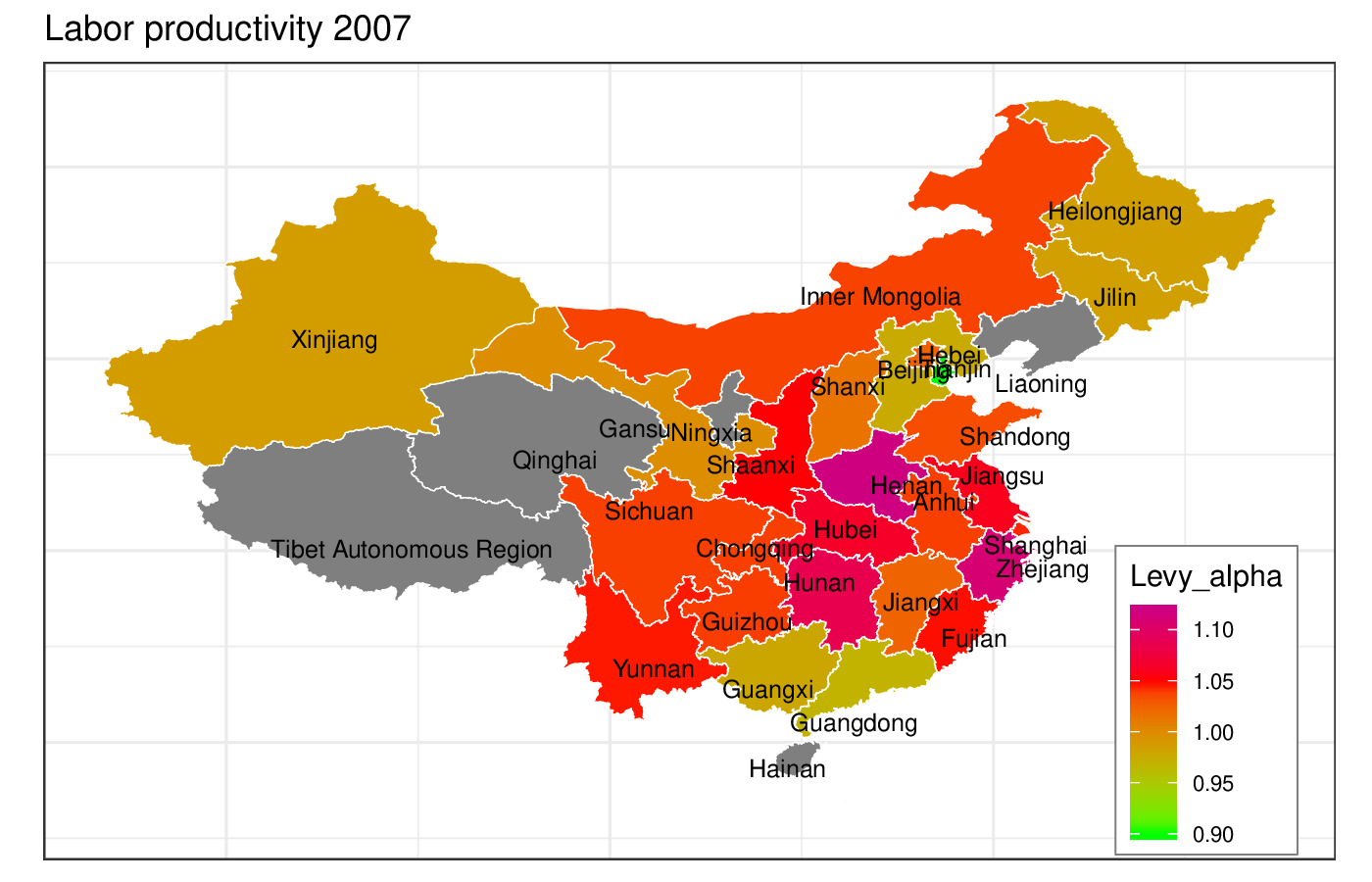}}
\caption{Levy $\alpha$ parameter fits for $LP$ (labor productivity) by Region}
\label{fig:maps:lp}
\end{figure}

\begin{figure}[hbtp!]
\centering
\subfloat[1999]{\includegraphics[width=0.33\textwidth]{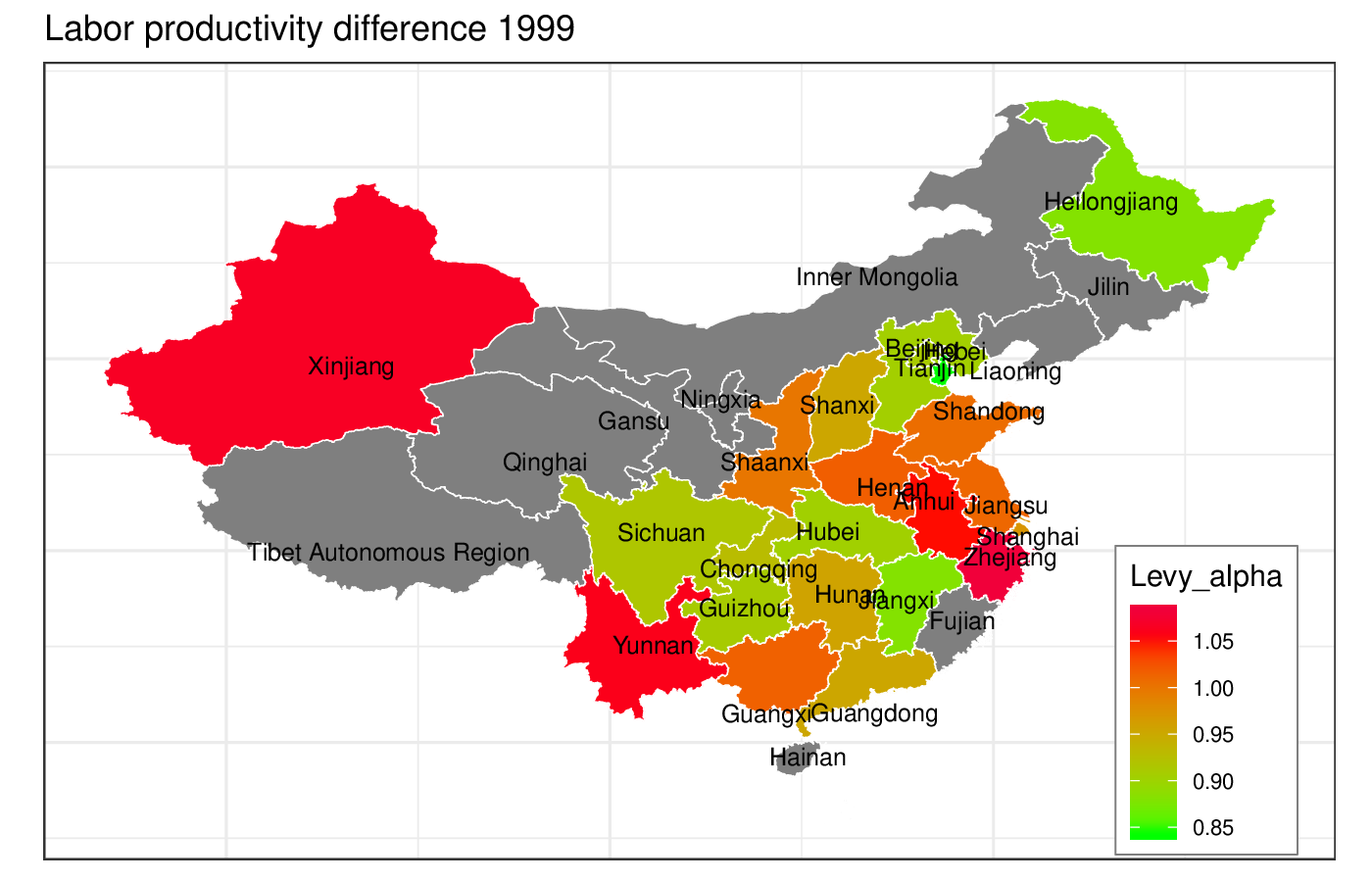}}
\subfloat[2000]{\includegraphics[width=0.33\textwidth]{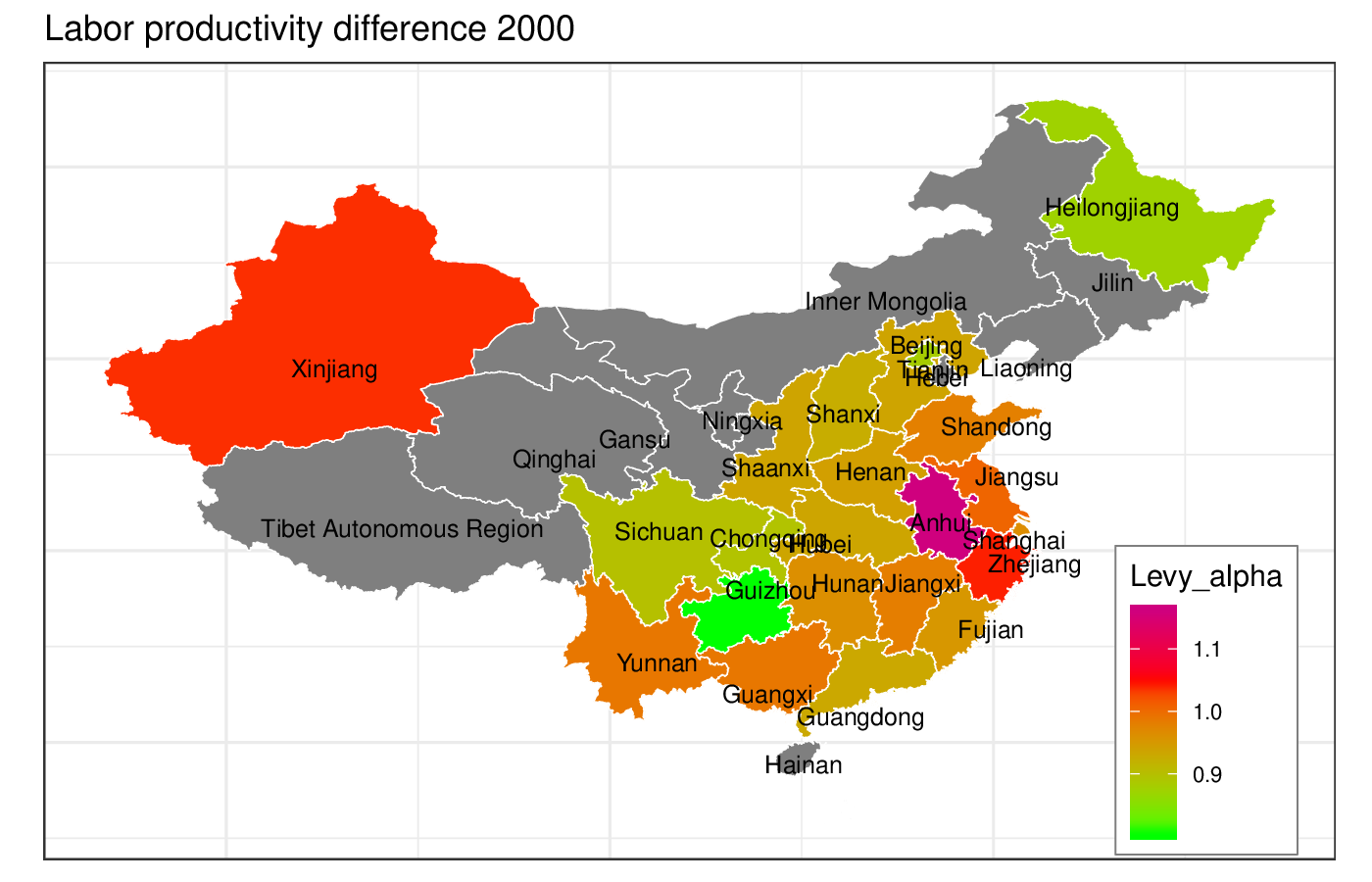}}
\subfloat[2001]{\includegraphics[width=0.33\textwidth]{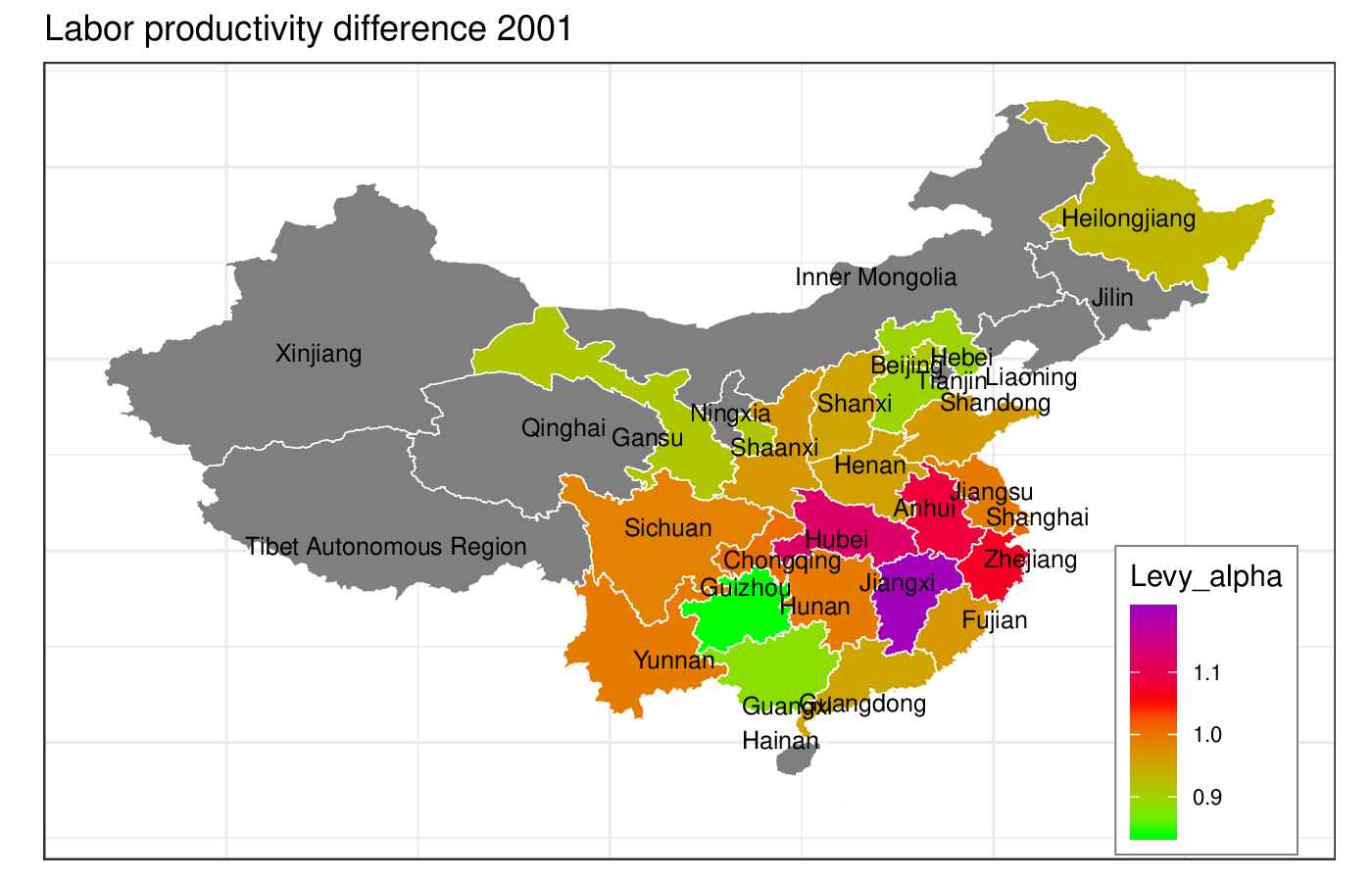}}\\
\subfloat[2002]{\includegraphics[width=0.33\textwidth]{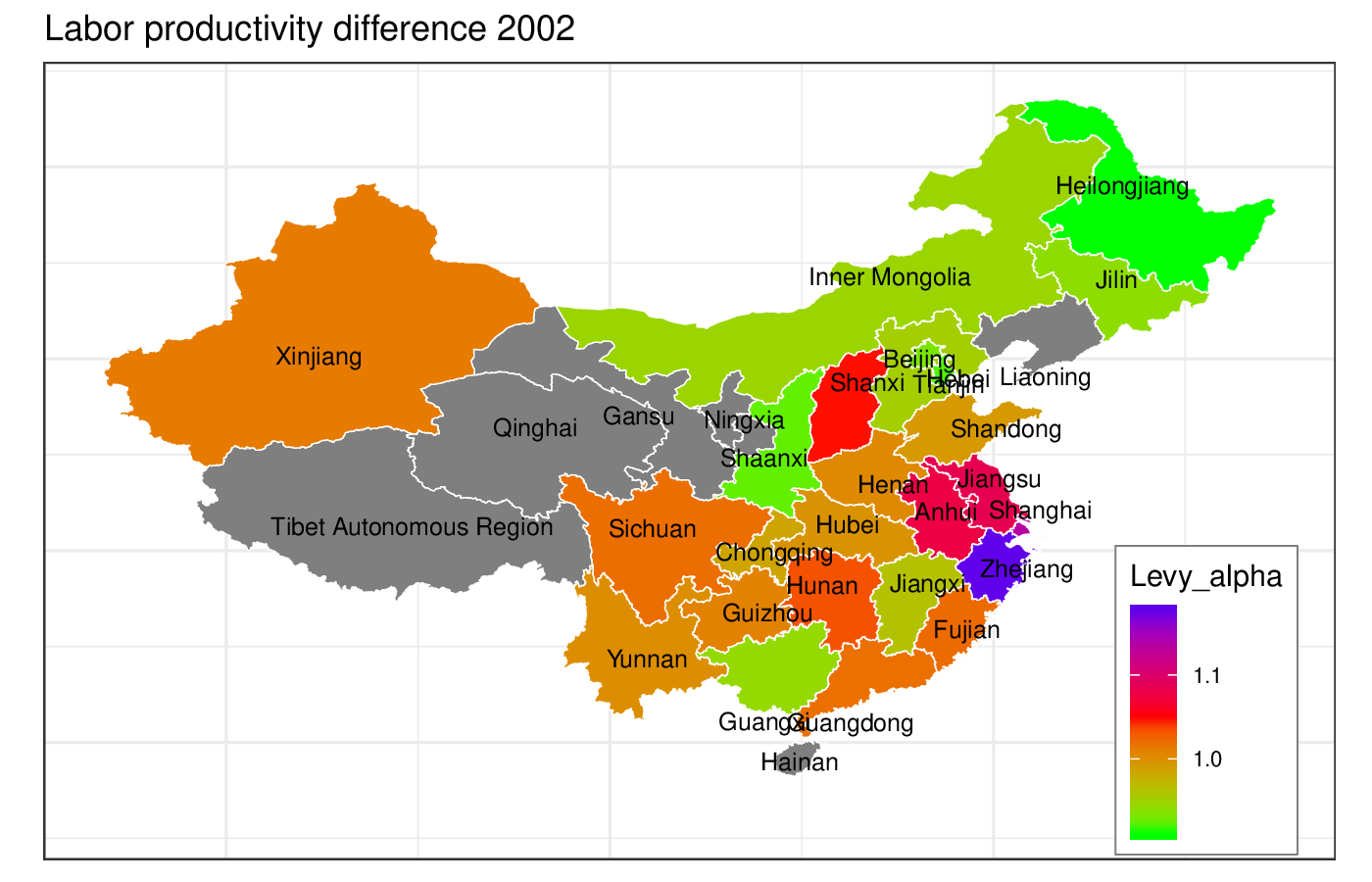}}
\subfloat[2004]{\includegraphics[width=0.33\textwidth]{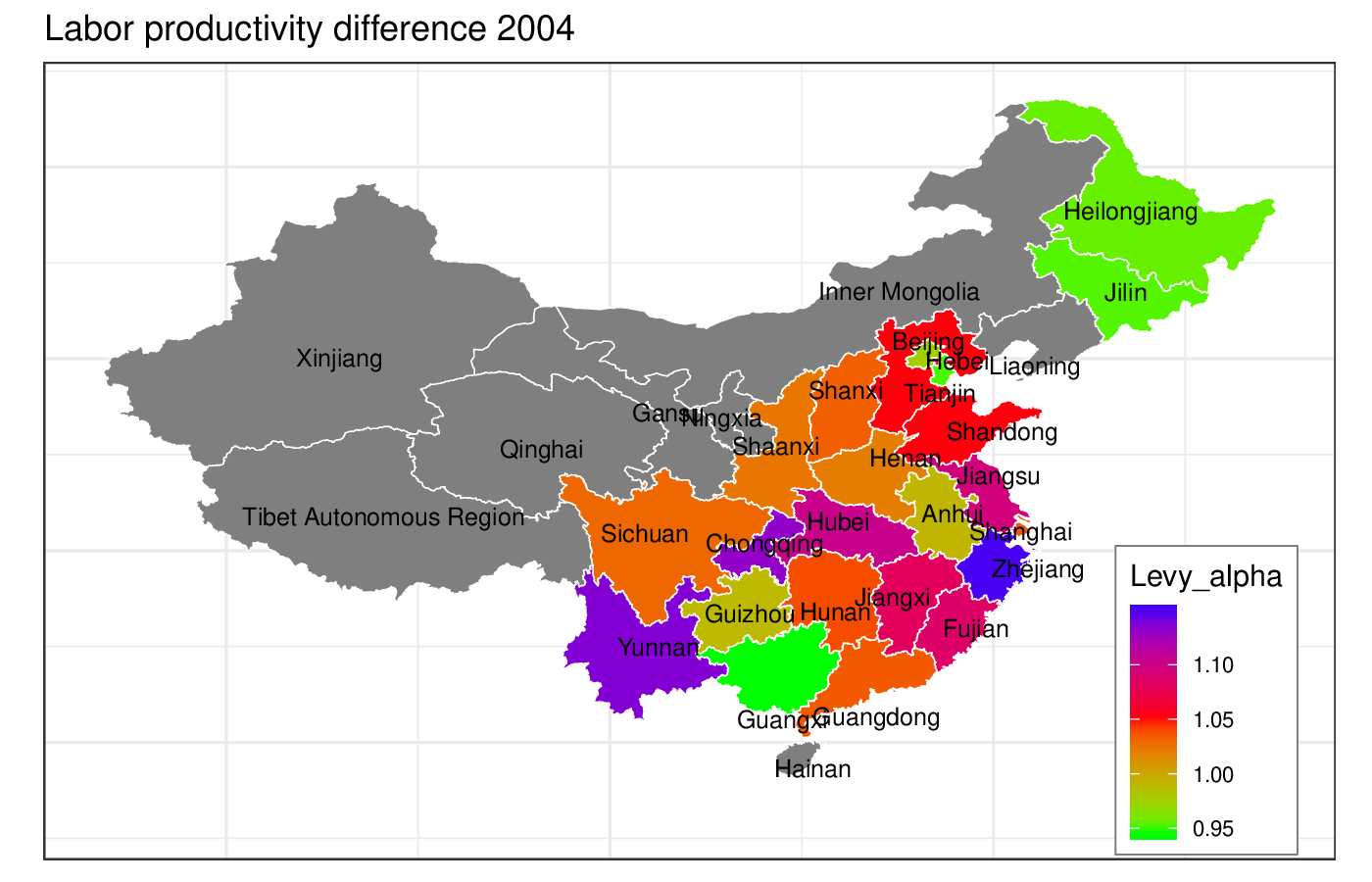}}
\subfloat[2005]{\includegraphics[width=0.33\textwidth]{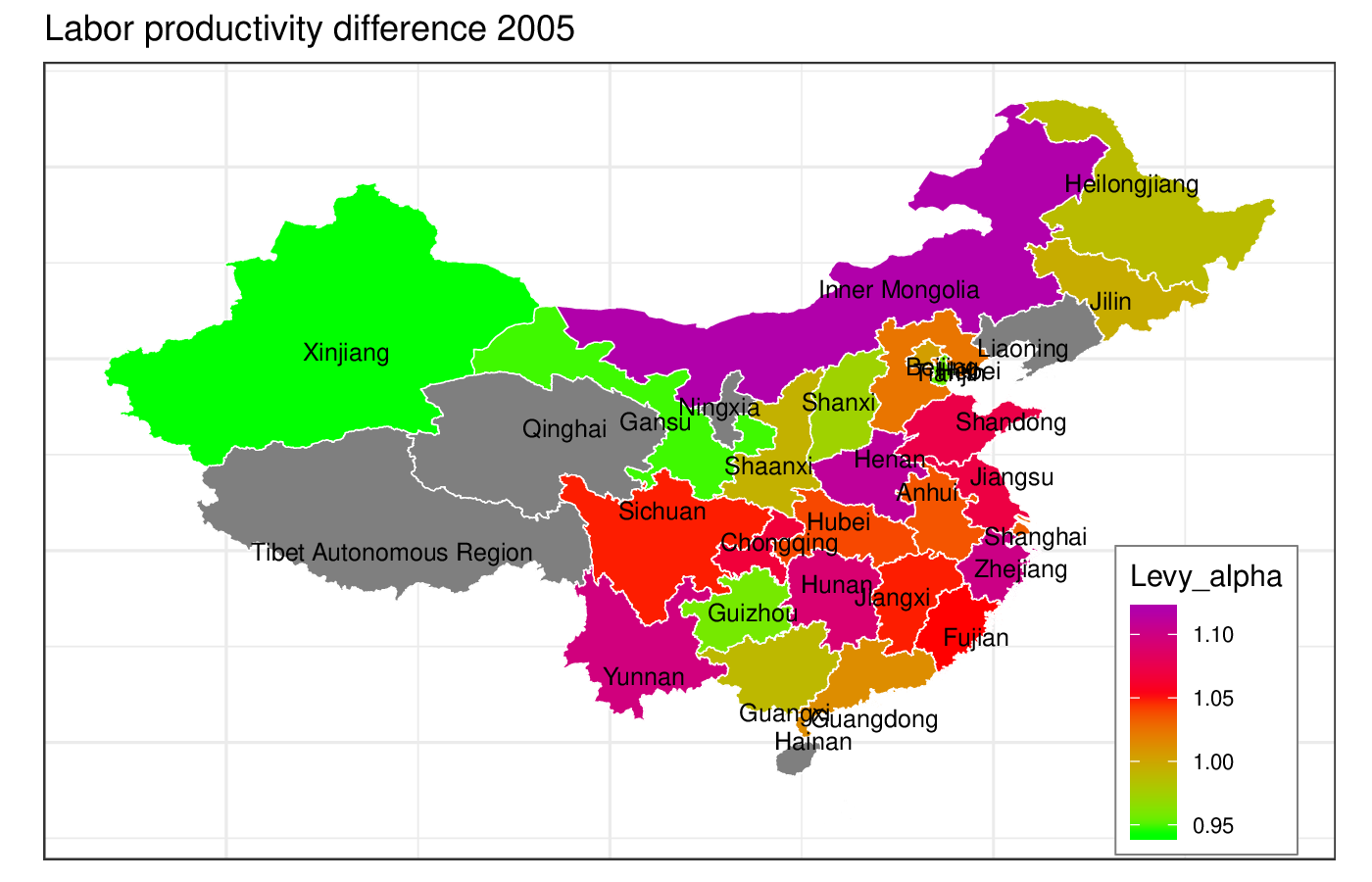}}\\
\subfloat[2006]{\includegraphics[width=0.5\textwidth]{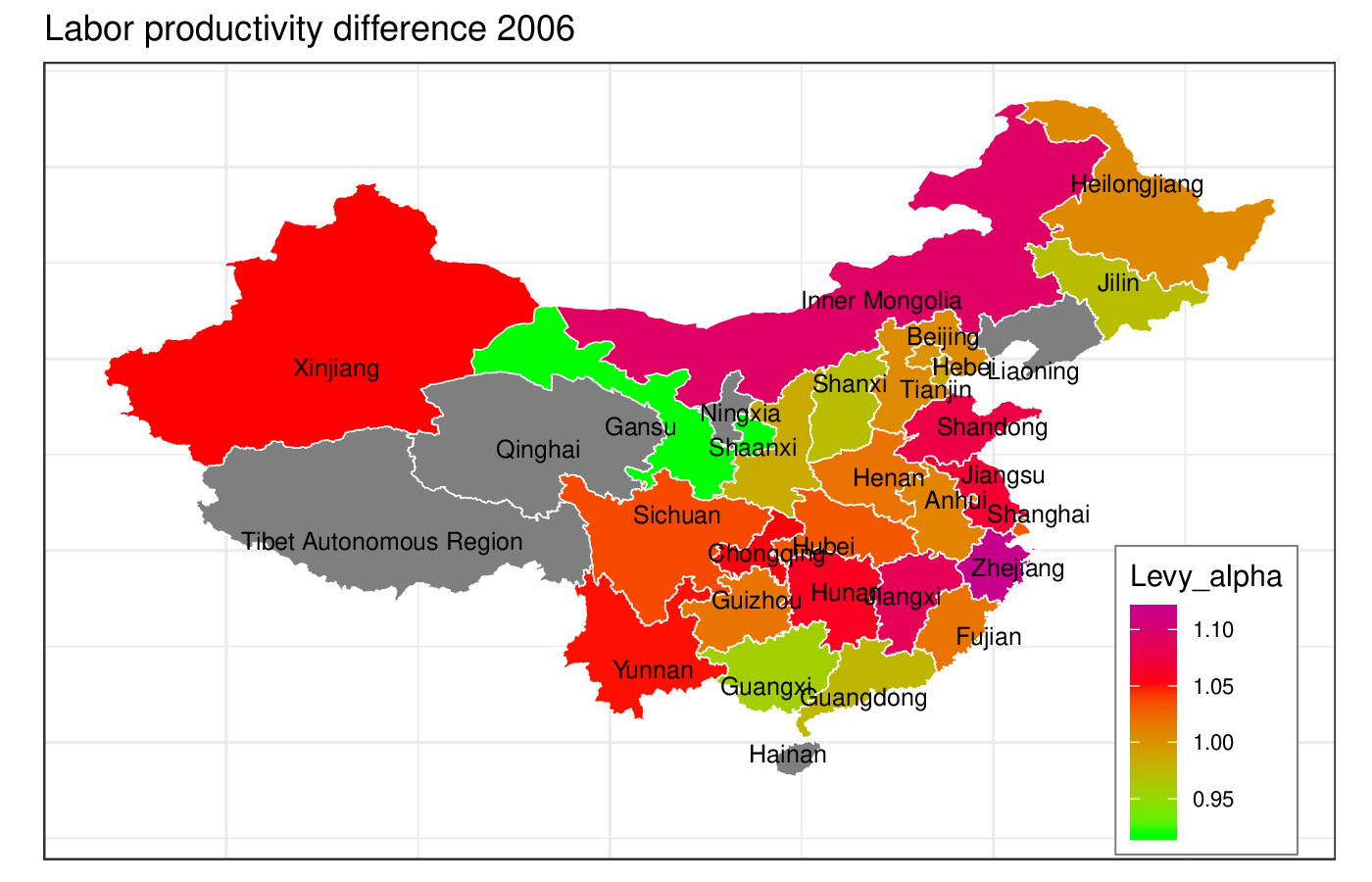}}
\subfloat[2007]{\includegraphics[width=0.5\textwidth]{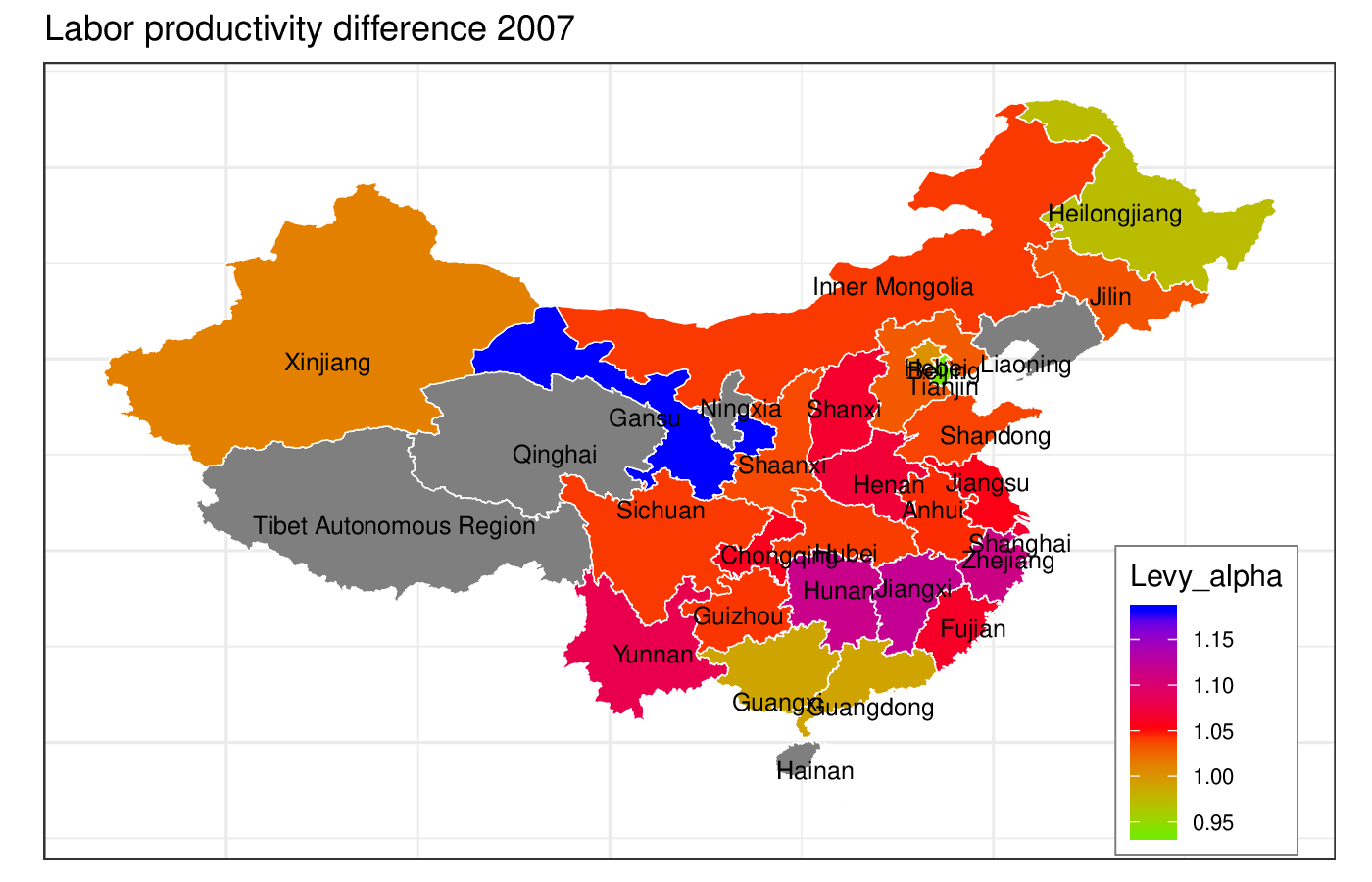}}
\caption{Levy $\alpha$ parameter fits for $\Delta LP$ (labor productivity change) by Region}
\label{fig:maps:lp_diff}
\end{figure}

\subsection{Regional variation}
\label{sect:results:regional}

While we only have data for one country, the P.R. China, we can investigate regional variation by considering subsamples of the data set for the 31 Chinese provinces and autonomous regions. In cross-country studies, where the cases are not subject to the same policy decision and idiosyncratic influences, any systematic differences should be more pronounced.

Two of the most dynamic and economically strongest regions of the PR China are Guangdong and Zhejiang. While Zhejiang is one of the most active regions supporting private economic sectors, and has benefited from the early economic boom in Yangtze River Delta, Guangdong is in South China, north of Hong Kong, and driven significantly by the opening-up endeavors. Both are coastal provinces and have the potential to develop the same industries. Yet, the regional distributions of $LP$ and $\Delta LP$ differ sharply and persistently. Guangdong maintains one of the lowest tail indices among all provinces, i.e. a very high probability weight in the tails ($\alpha_{\Delta LP}=0.93$ to $\alpha_{\Delta LP}=1.03$). Zhejiang is at the other end of the provincial dispersion, with a moderately high tail index between $\alpha_{\Delta LP}=1.05$ and $\alpha_{\Delta LP}=1.18$. 
Figure~\ref{fig:2provinces:LP-LPC} shows the density of labor productivity $LP$ and labor productivity change $\Delta LP$ in both regions.

Figures~\ref{fig:maps:lp} and \ref{fig:maps:lp_diff} show the development of the tail index parameter $\alpha$ of the L\'{e}vy alpha-stable fit of labor productivity $LP$ and labor productivity change $\Delta LP$ at the region level. Indeed it can be seen that Guangdong and neighboring regions (Guangxi, Jiangxi) and Zhejiang and its neighbors (Shanghai, Anhui) persistently find themselves at opposing ends of the variation. The Guangdong area tends to have longer tails (lower exponents) than the Shanghai/Zhejiang area.

Other regions with longer tails include the Northeast (Heilongjiang), the Beijing region, Inner Mongolia and, 
in the earlier years of the period of study, the central region. As pointed out in \citet{Heinrich/Dai16}, this subtle change over time may reflect the transition of various regions to a standard market economy system that occurred at different time periods (see details in \cite{Heinrich/Dai16}). Some regions, such as Inner Mongolia, likely stand out because of regional specificities; in Inner Mongolia a domination of the mining sector with rather large firms and certain volatility of empirical labor productivity depending on world market prices for metals etc. Some regions with a smaller population of firms (Xinjiang, Guangsu) are more volatile with less confidence being warranted for these fits.

\begin{figure}[tb!]
\centering
\includegraphics[width=0.85\textwidth]{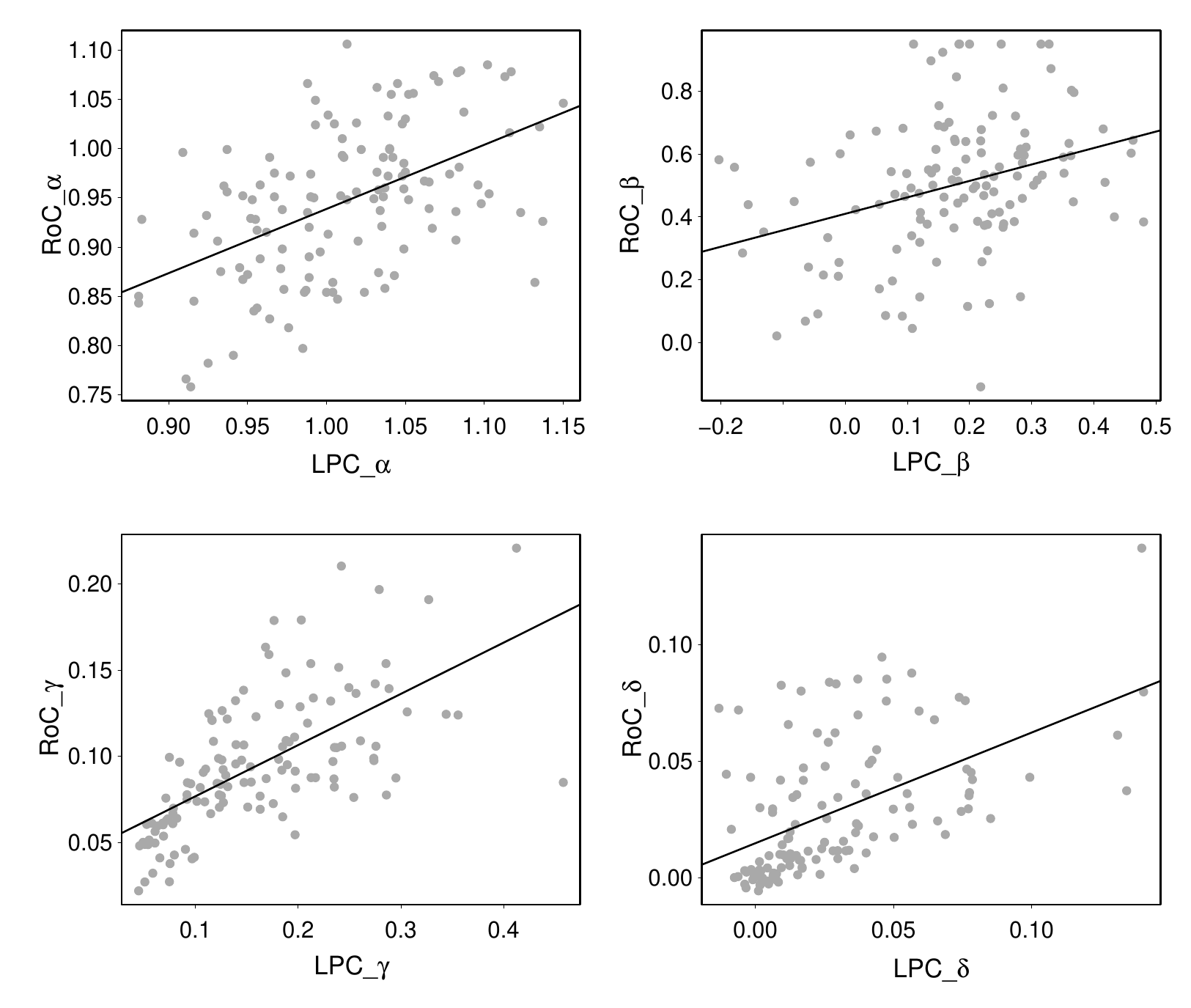}
\caption{Scatter plot of L\'{e}vy parameters of Returns on Capital ($ROC$) and Labor Productivity Change ($\Delta LP$). All four parameters tend to be positively correlated.}
\label{fig:lpc_inv}
\end{figure}

\begin{figure}[tb!]
\centering
\includegraphics[width=0.85\textwidth]{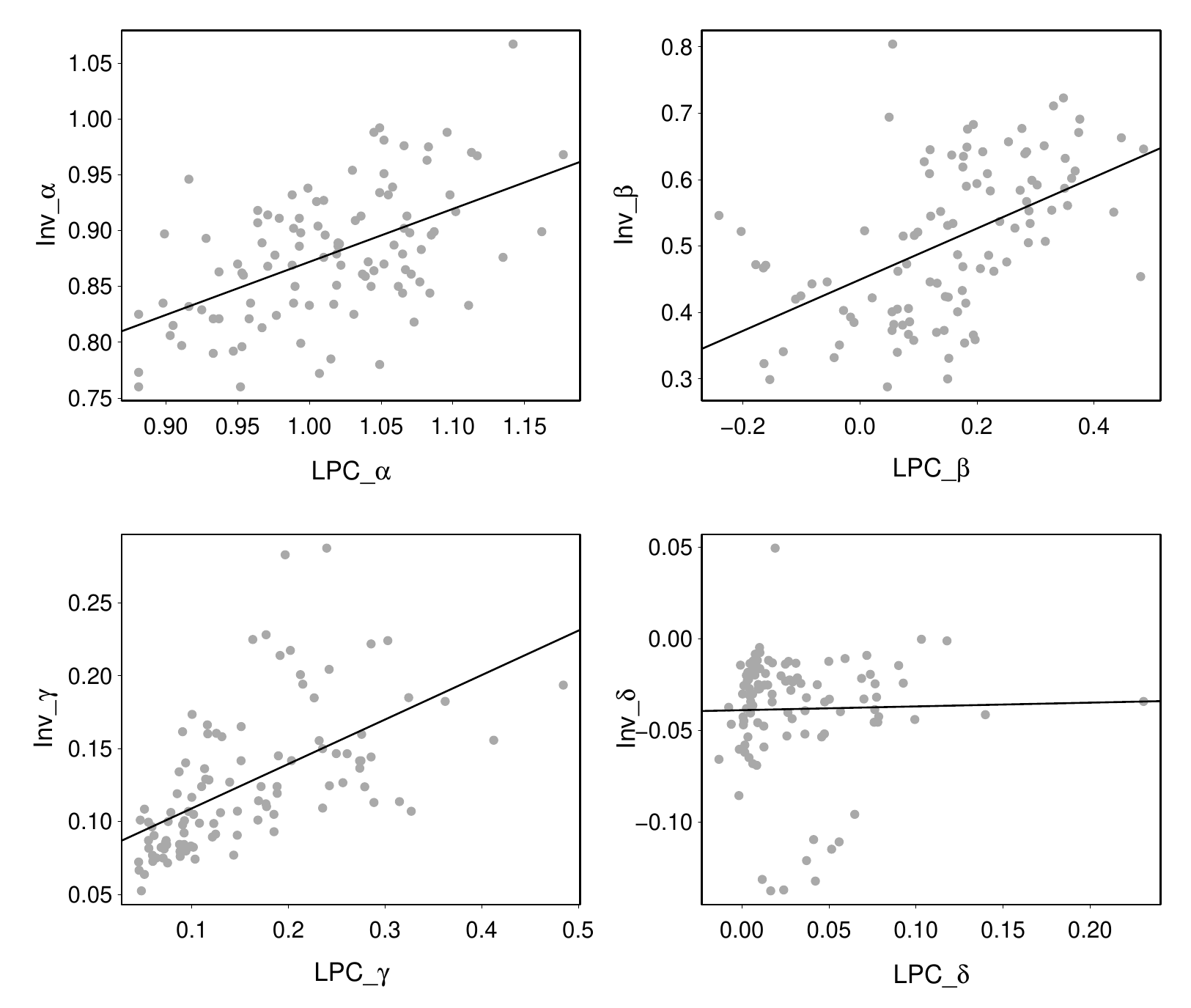}
\caption{Scatter plot of L\'{e}vy parameters of Investment Rate ($IR$) and Labor Productivity Change ($\Delta LP$). $\alpha, \beta, \gamma$  tend to be positively correlated, while $\delta$ has no relationship. }
\label{fig:lpc_roc}
\end{figure}

So far, we have shown that there are strong and persistent differences in the characteristics of the productivity distributions among the regions of China. This indicates that this would very likely also hold between countries. Beyond that, we can show in the Chinese case that micro-level variables are strongly interconnected: For instance, there is a cluster of regions with a shorter tail (higher tail index $\alpha$) around Shanghai and Zhejiang in both $LP$ (Figure~\ref{fig:maps:lp}) and $\Delta LP$ (Figure~\ref{fig:maps:lp_diff}). This regional pattern is also present in the figures for profitability and the investment rate shown in the Appendix~\ref{app:results} (Figures~\ref{fig:maps:roc} and \ref{fig:maps:fias_g}). Indeed, it can be shown that at the regional level, the distributions of the key firm-level variables are interconnected. The correlations of parameters fitted for the labor productivity change ($\Delta LP$) distribution and investment rate ($IR$)   distribution are shown in Figure~\ref{fig:lpc_inv}; those of the parameters for ($\Delta LP$) and the profitability ($ROC$) are plotted in Figure~\ref{fig:lpc_roc}. For each pair of distributions, the same parameters are highly correlated. I.e. a high tail index $\alpha$ for the labor productivity change distribution is associated with a high tail index for the distributions of profitability and investment rate. The same is true for the skew ($\beta$) and the scale ($\gamma$) of all distributions and the location of the modal value ($\delta$) of $\Delta LP$ and $ROC$ distributions.\footnote{The only exception is the $\delta$ parameter of the distribution of the investment rate, which is not associated with any of the other fits and appears to fall into clusters. The reason is that there is a strong time signal in this variable (L\'{e}vy parameter $\delta$ of $IR$) with left shifts in the investment rate distribution in 2004 and in 2008, which may be caused by idiosyncratic shocks (such as the financial crisis hitting in 2008), by tax policy or by accounting intricacies.}

These systematic regional differences evident in concert in a number of variables in the Chinese case indicate that such patterns should also be expected in cross-country comparisons. The following section will furthermore investigate relations between productivity distributions and various other macro- and micro-level variables using a Bayesian multi-level regression approach.

\subsection{Relation of productivity distributions and their parameters to other economic measures}
\label{sect:results:determinants}

To understand how L\'{e}vy parameters of each variable are associated with the geographical predictors (in this case, Chinese regions), we run a simple mixed linear regression model. We use a Bayesian multi-level approach to properly account for group-level similarities and differences through a partial pooling. The likelihood function is written as follows:

\begin{eqnarray}
\operatorname{Parameter_{i}} &\sim& \operatorname{Normal}(\mu_{i}, \sigma)  \nonumber \\
\mu_{i} &=& \alpha + \alpha_{_{j[i]}} +  \alpha_{_{t[i]}} + \beta_1\operatorname{GDP\_r} + \beta_2\operatorname{Firm\_Age} + \beta_3\operatorname{Emp} + \beta_4\operatorname{Cap\_Intensity} 
\label{eq:regressionmodel}
\end{eqnarray}

where the subscript $t[i]$ and $j[i]$ indicate the year and province index which will be used as the main group effect variable. See Appendix \ref{app:prior} for a detailed discussion on prior specification. We assume that the outcome variable $\operatorname{Parameter_i}$ is distributed according to the Gaussian likelihood function around a mean $\mu_i$ and the standard deviation $\sigma$. The independent variables (predictors) include GDP growth (GDP\_r), firm age, employment (Emp), as well as capital intensity ($\operatorname{Cap\_Intensity}$) at the regional level. The firm-level variables among these (not the GDP growth) are computed using the regional averages in the data set. The intercept has two varying components by year and province along with the overall intercept. 

\begin{itemize}
\item $\alpha$: The overall intercept. The expected value of $\operatorname{Parameter_{i}}$ when all other explanatory variables are zero. 
\item $\alpha_{_{t[i]}}$: The varying intercept effect coming from the year index. The deviation in the intercept for year $t$ from the overall intercept $\alpha$. 
\item $\alpha_{_{j[i]}}$: The varying intercept effect coming from the province index. The deviation in the intercept for province $j$ from the overall intercept $\alpha$. 
\item $\beta_1$: The coefficient of GDP Growth. The expected change in $\operatorname{GDP\_Growth}$ across all $t$ and $j$. 
\item $\beta_2$: The coefficient of Firm Age. The expected change in $\operatorname{Firm\_Age}$ across all $t$ and $j$. 
\item $\beta_3$: The coefficient of Employment. The expected change in $\operatorname{Employment}$ across all $t$ and $j$. 
\item $\beta_4$: The coefficient of Capital Intensity. The expected change in $\operatorname{Capital\_Intensity}$ across all $t$ and $j$. 
\end{itemize}


We use the \textit{Hamiltonian Monte Carlo} (HMC) to obtain a sequence of random samples from a posterior probability distribution. We use the Bayesian programming language Stan and the R-package  \verb|brms| that operationalize the HMC algorithm to efficiently compute posterior distributions \citep{rstan-software2015, brms_paul}. For a more detailed discussion on HMC and Stan, see \citep{gelman2014bayesian, carpenter2017stan}. 

Table ~\ref{tab:regsummary} summarizes the estimation results for each variable and parameter. We only report the mean and standard deviation (in parenthesis) of the posterior distribution of each parameter. A full summary table can be found in the Appendix~\ref{app:regression}.

\begin{table}[htb!]
	\centering
	\scriptsize
	\begin{tabular}{lcccccc}
		\hline\hline
		\multicolumn{1}{c}{
		Variable} & \multicolumn{1}{c}{Parameters}  &\multicolumn{1}{c}{Intercept} & \multicolumn{1}{c}{GDP Growth}    &\multicolumn{1}{c}{Firm Age} & \multicolumn{1}{c}{Employment}  & \multicolumn{1}{c}{Capital Intensity}   \\ 
    \midrule
		\multirow{4}{*}{LP Change}  & $\alpha$ & 0.9841 (0.0462) & 0.0083 (0.0027) & -0.0012 (0.0006) & -0.0001 (0.0001) & -0.0001 (0.0001) \\ 
  &$\beta$ & -0.0564 (0.1025) & 0.0256 (0.0065) & -0.0035 (0.0014) & 0 (0.0001) & -0.0001 (0.0002) \\ 
  &$\gamma$ & 0.0555 (0.0466) & 0.0094 (0.0026) & 0 (0.0005) & -0.0002 (0.0001) & 0.0004 (0.0001) \\ 
  &$\delta$ & -0.0105 (0.024) & 0.0054 (0.0014) & -0.0001 (0.0003) & -0.0001 (0) & 0 (0) \\ 
		\midrule
	\multirow{4}{*}{LP}  &	$\alpha$ & 1.0823 (0.0509) & 0.0007 (0.0031) & -0.0001 (0.0002) & -0.0001 (0.0001) & -0.0001 (0.0001) \\ 
 & $\gamma$ & 0.2304 (0.063) & 0.0054 (0.0032) & 0 (0.0002) & -0.0005 (0.0001) & 0.0004 (0.0001) \\ 
 & $\delta$ & 0.2574 (0.0681) & 0.009 (0.0035) & 0 (0.0002) & -0.0005 (0.0001) & 0.0003 (0.0001) \\ 
  		\midrule
		\multirow{4}{*}{RoC} &  $\alpha$  & 1.1235 (0.0564) & -0.0033 (0.0033) & -0.0008 (0.0002) & -0.0003 (0.0001) & -0.0001 (0.0001) \\ 
  &$\beta$& 0.167 (0.1337) & 0.0333 (0.0081) & 0 (0.0004) & -0.0003 (0.0002) & -0.0001 (0.0002) \\ 
  &$\gamma$ & 0.0713 (0.0208) & 0.0034 (0.0012) & -0.0001 (0.0001) & 0 (0) & -0.0001 (0) \\ 
  &$\delta$ & 0.0341 (0.0181) & 0.0018 (0.0011) & 0 (0.0001) & -0.0001 (0) & -0.0001 (0) \\ 
		\midrule
		\multirow{4}{*}{Inv Rate}  &   $\alpha$  & 0.7954 (0.0486) & 0.0004 (0.0025) & -0.0014 (0.0008) & 0.0002 (0.0001) & 0.0002 (0.0001) \\ 
  &$\beta$ & 0.4201 (0.0955) & 0.0047 (0.0044) & 0.0005 (0.0015) & -0.0001 (0.0001) & 0.0001 (0.0002) \\ 
  &$\gamma$ & 0.122 (0.0286) & 0.0028 (0.0016) & -0.0007 (0.0005) & 0 (0) & 0 (0.0001) \\ 
  &$\delta$  & -0.0687 (0.0191) & -0.0001 (0.0009) & -0.0001 (0.0003) & 0.0001 (0) & 0 (0) \\ 
	\hline\hline
	\end{tabular}
	\caption{Summary statistics of a mixed linear regression model of L\'{e}vy parameters with four predictors: GDP Growth, Firm Age, Employment, and Capital Intensity. }
\label{tab:regsummary}
\end{table}

For each variable, we show the regression results for L\'{e}vy parameters. Note that we do not report on the regression of $\beta$ in the labor productivity variable since it is always very close to 1 (maximal skewness) and doesn't have much variation. 

\paragraph{GDP Growth}: 
\begin{figure}[tb!]
\includegraphics[width=1\textwidth]{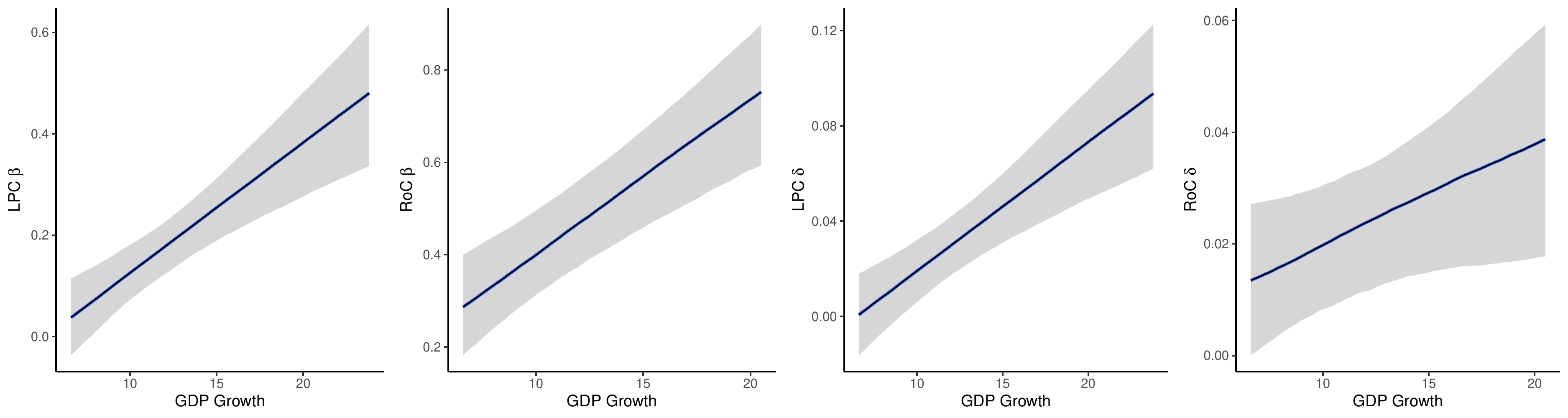}
\caption{Marginal effects of GDP Growth on $\beta$ and $\delta$ of labor productivity change ($\Delta LP$) and profitability ($ROC$). The blue line is the mean estimate and the grey shade area is the 90\% uncertainty interval.}
\label{fig:reg_gdp}
\end{figure}
A higher regional economic growth tends to be associated with a higher $\beta, \gamma$, and $\delta$ in $\Delta LP$ and  profitability. The GDP coefficients in labor productivity and investment rate tend to be rather noisy except for $\gamma$. $\alpha$ parameter is only informative in the $\Delta LP$ case: the higher the provincial GDP growth the higher $\alpha$ and thus the thinner the tails. $\alpha$ in other variables has neither enough variation nor a clear pattern in relation to regional GDP growth. Figure~\ref{fig:reg_gdp} shows the marginal effects of GDP Growth on key parameters.  From this, we can infer that the economic growth in China is characterized by four distinctive patterns. As the economy grows in China, 1) firms become more technologically dynamic and more profitable (a high $\delta$ in $\Delta LP$ and profitability), 2) the economy has an increasing number of highly innovative and profitable firms (a high $\beta$ in $\Delta LP$ and profitability), 3) firms become more diverse in their performance (a high $\gamma$ in  $\Delta LP$, $LP$, profitability and investment rate), and 4) the technological competition among firms gets more fierce over time (a high $\alpha$ in $\Delta LP$).

\paragraph{Firm Age}: 
\begin{figure}[tb!]
\includegraphics[width=1\textwidth]{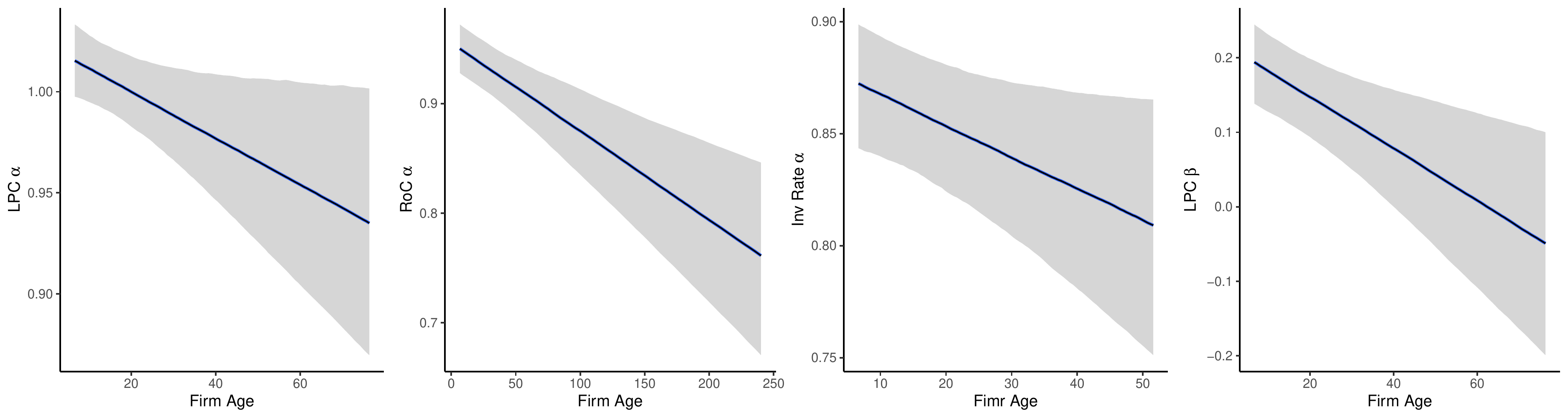}
\caption{Marginal effects of Firm Age on $\alpha$ of labor productivity change ($\Delta LP$), profitability ($ROC$), and investment rate $IR$, and $\beta$ of $\Delta LP$.  The blue line is the mean estimate and the grey shade area is the 90\% uncertainty interval. }
\label{fig:reg_age}
\end{figure}
The higher the average firm age, the lower $\alpha$ and thus heavier tails in $\Delta LP$, profitability, and Investment rate, and the lower $\beta$ in $\Delta LP$. $\gamma$ and $\delta$ are very noisy in all parameters except for $\gamma$ in profitability and investment rate. Figure~\ref{fig:reg_age} shows the marginal effects of firm age on key parameters.  From this, we can infer that, when the province has a higher average firm age, 1) the market tends to be less competitive for technological change, profitability, and firm growth (a low $\alpha$ in $\Delta LP$, profitability, and investment rate), 2) the economy has an increasing number of less innovative firms (a low $\beta$ in $\Delta LP$), and 3) the firm performance in terms of profitability and investment tends to be diverse but with a relatively high degree of uncertainty.

\paragraph{Employment}: 
\begin{figure}[tb!]
\includegraphics[width=1\textwidth]{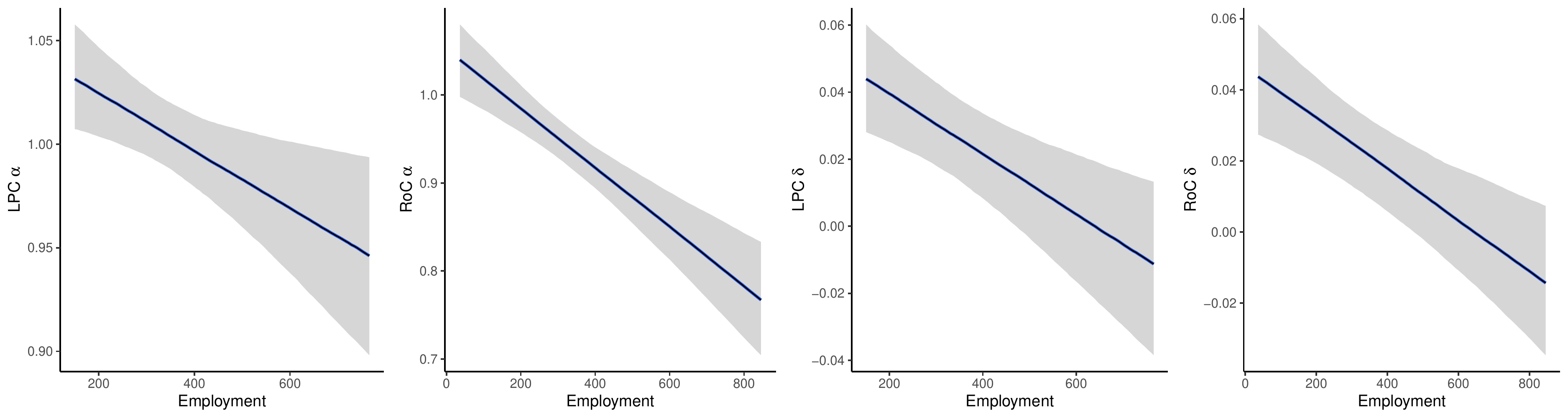}
\caption{Marginal effects of Employment on $\alpha$ and $\delta$ of labor productivity change ($\Delta LP$) and profitability ($ROC$). The blue line is the mean estimate and the grey shade area is the 90\% uncertainty interval.}
\label{fig:reg_emp}
\end{figure}
A larger average employment size is associated with a lower $\alpha$ in $\Delta LP$, $LP$, profitability, a lower $\beta$ in profitability, a lower $\gamma$ in $\Delta LP$, $LP$, and a lower $\delta$ in $\Delta LP$ and Profitability. Investment Rate has a somewhat different pattern and has a positive relationship between employment size and $\alpha$ and $\delta$. Figure~\ref{fig:reg_emp} shows the marginal effects of employment on key parameters. From this, we can infer that, when the province has larger size firms with a high number of employees, 1) the market tends to become less competitive overall (a low $\alpha$ in $\Delta LP$, $LP$, profitability), 2) firms become less technologically dynamic and more profitable (a low $\delta$ in $\Delta LP$ and profitability), and 3) firms become less diverse in their performance in technological change  (a low $\gamma$ in  $\Delta LP$ and $LP$). 

\paragraph{Capital Intensity}: 
\begin{figure}[tb!]
\includegraphics[width=1\textwidth]{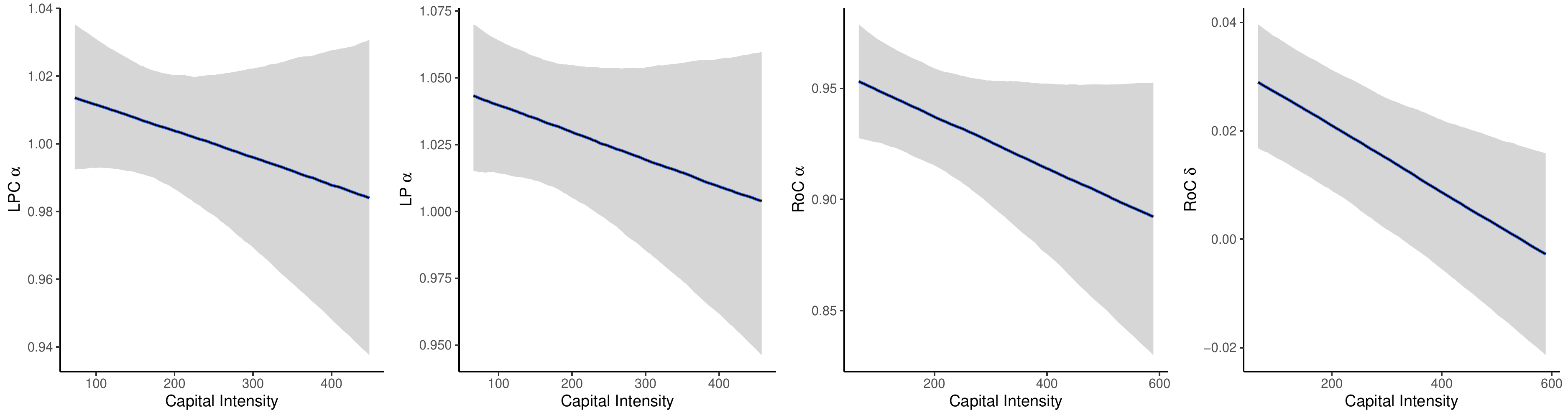}
\caption{Marginal effects of Capital Intensity on $\alpha$ of labor productivity change ($\Delta LP$), $LP$, Returns on Capital, and $\delta$ of  Returns on Capital. The blue line is the mean estimate and the grey shade area is the 90\% uncertainty interval.}
\label{fig:reg_cap}
\end{figure}
Higher average capital intensity is associated with a lower $\alpha$ in $\Delta LP$, $LP$, and Profitability but a higher $\alpha$ in the Investment rate, a higher $\gamma$ in $\Delta LP$ and $LP$, and a lower $\delta$ in profitability. Figure~\ref{fig:reg_cap} shows the marginal effects of capital intensity on key parameters. From this, we can infer that, as the province has more firms with a higher capital intensity (a higher degree of mechanization), 1) the market tends to become less competitive overall (a low $\alpha$ in $\Delta LP$, $LP$, profitability), 2) firms become more diverse in their performance in technological change  (a high $\gamma$ in  $\Delta LP$ and $LP$), and 3) the firms tend to be less profitable overall (a low  $\delta$ in profitability).

\section{Conclusion}
\label{sect:conclusion}

The distribution of productivity at the firm level has, like other economic quantities, been thoroughly investigated in recent years. For the developed economies, many stylized facts are known now: The distribution is unimodal, strongly right-skewed, has heavy tails, and is persistent in time. \citet{Yangetal19}, who also summarize the state of the art, report no systematic changes in their study covering millions of observations for European developed countries over a period of 10 years. 

It is a fair question to ask if we might have systematic changes in developing countries, since these countries experience more rapid structural, demographic, and technological changes. Is the distribution the same? Do the parameter estimates show any trends? If so, what does that tell us about the development process and about development policy? Can and should productivity distributions be managed?

While we cannot provide direct evidence for the entire developing world, we did offer evidence for one country, the PR China as an example in this paper. Our study covers a crucial period of Chinese history, 1998-2007 (and further to 2013 with less reliable data), a period in which the country experienced the highest growth rates; when the economy and the technology sector took off; when the Chinese converged to the consumption and lifestyle habits of the developed world. 

We demonstrated that the distributions of a wide range of quantities at the firm-level are heavy-tailed with a L\'{e}vy alpha-stable distribution being an excellent distributional model and clearly superior to the alternative AEP distribution that was also tested. This includes labor productivity ($LP$) and labor productivity change ($\Delta LP$). Consistent with the developed economies, both the shape of the distribution and the parameter values were remarkably consistent over time. 

However, we did find a systematic shift: The location parameter of both productivity and productivity change were steadily increasing over the period of the study. The scale parameter $\gamma$ followed suit. Tail index ($\alpha$) and skew ($\beta$) were stable for the productivity level, but for the productivity change, the tail grew shorter ($\alpha$ increasing) and the distribution developed a right skew ($\beta$ increasing). What this means is that the productivity gains China experienced in the period of study do not come from super-star firms of exceptionally high productivity: The tail weight of $LP$ remained unchanged and that of $\Delta LP$ decreased. Instead, labor productivity increases became more consistent, concentrated, and uniform across the economy (decreasing tail weight of $\Delta LP$) and the body of labor productivity change extended to the right (emerging right-skew). 

Further, we showed that there are significant and systematic differences across the regions of China that persist in time over the period of the study. This is even the case for differences for superficially similar regions such as the technology centers of Guangdong and Zhejiang/Shanghai. If this can be shown for the regions of China, that are subject to similar policies, environmental factors, and idiosyncratic shocks, differences in cross-country studies with multiple developing countries would be expected to be more pronounced. 

Nevertheless, there were systematic relations between the parameters of the productivity distribution, a range of other distributions of micro-level variables (profitability, capital intensity, firm age), as well as macro-level characteristics (GDP growth, employment) of the respective regions, as shown in our Bayesian multi-level regression in Section~\ref{sect:results:determinants}.

The tail indices of the distributions were found to be between $\alpha=0.9$ and $\alpha=1.2$ in most cases, implying infinite variance (since $\alpha<2$) and very slow convergence to the theoretical mean, if the mean even exists (only for $\alpha\geq 1$). As a consequence, characteristics of the labor productivity distribution in the form of moments would be avoided. Such characteristics are, however, commonly given as variants of direct moments, the mean for the location, the variance, standard deviation, or Olley-Pakes gap for the dispersion. \citet{hsieh2009misallocation} for instance, in his otherwise exemplary study of misallocation in China and India, uses sample variances.

What does this mean for development policy? First, concentration on super-star firms - be that domestic ones or branches of foreign groups - might be the wrong approach for successful technological catch-up. It would certainly be a different approach from that taken in China. Second, it instead seems important to ensure that productivity gains can also be realized by other firms. The most direct approach for this is encouraging technology transfer and providing incentives for sourcing intermediate products locally (which would also lead to cooperation and technology transfer). Technologies are arguably the most important factor in determining productivity at the firm level. Third, other factors such as instrumental institutions and a comprehensive education system could support this process. Fourth, significant differences would be expected between different countries. Direct comparisons of the parameters of productivity distributions in isolation are likely of only limited value. Instead, such comparisons should be done with a range of measures for the economic micro-structure at the firm level (productivity, age, capital intensity, etc.) while also taking the intertemporal development of these variables into account. Fifth, particular skepticism is advised with respect to measures that rely on moments of the productivity distribution (or similar quantities), as these may not exist. An example is the use of variance as a dispersion indicator, which will certainly fail.

While our findings are encouraging, more research is needed to confirm our findings for other developing countries. Do firm-level data in India, Vietnam, Nigeria, and other rapidly developing countries have the same characteristics? Can catch-up processes like the one showcased here be repeated in still other developing economies in the future? What impact might the Covid-19 pandemic have, that changed the face of the world economy by hitting many developed economies, but also some individual developing countries (like Tanzania) very hard?

Finally, can our example teach us something about the development history of European and other developed countries, the USA, Japan, South Korea? These countries' catch-up phases were much longer ago in history, at a time, when economic microdata were not collected to the same extent as today. We may be unable to reconstruct the microdata, but it may still be possible to infer how the development process unfolded.

\bibliography{main}
\bibliographystyle{apa}

\newpage
\appendix

\section{Technical explanation of aggregation, maximum entropy, and distributional models}
\label{app:dist}

\subsection{Aggregation and maximum entropy distributions}
\label{app:dist:agg}

The form of aggregation of random variable distributions that is typically considered is convolution, i.e. the summation of random processes.\footnote{Summation of random processes is different from summation of scalars, since the resulting densities and probabilities have to be obtained through convolution.} For a defense why convolution is a suitable form of aggregation, see \citet{Frank09}. 

The sum of $n$ independent and identically distributed (i.i.d.) random variables $X_1+X_2+\ldots+X_n$. The distribution of (all) $X$ is an attractor distribution, if the aggregation converges to a distribution of the same form $X$. Formally \citep{NOLAN1998187,NOLAN2018,Frank09},
\begin{equation}
    X_1+X_2+\ldots+X_n \sim c_n X + d_n
\end{equation}
where $c_n$ and $d_n$ are scalars dependent on the number of convoluted distributions $n$. Such a distribution is called a {\it stable distribution}. The general form of stable distributions, {\it L\'{e}vy alpha-stable distributions}, are a generalization of Gaussian, Cauchy, and other specific distributions. We will consequently work with L\'{e}vy alpha-stable distributions as our distributional model for labor productivities and other firm-level variables in this paper. 

Before we discuss the form, properties, and parametrizations of the L\'{e}vy alpha-stable distribution, we will give some background on aggregation and convergence to the entropy maximizing distribution under aggregation. Intuitively, aggregation leads to a loss of information; it washes out less strong signals and only a dominant pattern remains. As the convoluted distributions are independent, this pattern is the one that carries the least information (highest entropy), the one that is the most likely one without additional information, the one that constitutes the maximum entropy distribution under constraints that depend on the component distributions $X$.

For instance, constraints requiring a constant mean $\int_{-\infty}^{\infty}p(x)xdx=\mu$ and variance $\int_{-\infty}^{\infty}p(x)(x-\mu)dx=\sigma^2$ lead to a Gaussian as the entropy maximizing distribution. A constraint imposing a constant first moment (e.g., the mean) without constraints on the variance yields an exponential (one sided\footnote{Note that support for the exponential distribution is $(0, \infty)$, while it is $(-\infty, \infty)$ for the Laplacian.}, constraint on mean, $\int_{0}^{\infty}p(x)xdx=\mu$) or Laplace distribution (two-sided with discontinuity, constraint on absolute deviation, $\int_{-\infty}^{\infty}p(x)(x-\mu)dx=a$). We will return to the constraint corresponding to the more general L\'{e}vy alpha-stable distribution in Section~\ref{sect:methods:levy-as}.

It should be noted that not every maximum entropy distribution is a stable distribution, since the convolution of distributions $X$ may have an entropy maximizing distribution with a different functional form, which, in turn, may again aggregate to a distribution with another, different functional form.

\subsection{The classical central limit theorem}
\label{app:dist:clt}

The most well-known stable distribution is the Gaussian. This is known as the classical central limit theorem: Any sum over $n$ i.i.d. random variables with fixed mean and variance will converge to a Gaussian. The Gaussian is the solution to the entropy maximizing problem under this constraint\footnote{It has been shown that the Gaussian can be obtained with different component variables that do not have to be i.i.d. (Lindeberg condition). However, the condition of fixed variance remains as does the fact that the Gaussian is the stable distribution to which aggregates of i.i.d. random variable distributions with fixed variance converge. Cf. \cite{Frank09}.} (fixed variance, $(x-\mu)^2=\sigma^2$):

\begin{equation}
\Lambda = -\int_{-\infty}^{\infty}p(x)\log\left(\frac{p(x)}{m(x)}\right)dx - \lambda_1\left(\int_{-\infty}^{\infty}p(x)dx -1\right) -\lambda_2(p(x)(x-\mu)^2-\sigma^2)
\label{eq:gaussian:maxent}
\end{equation}

where the first term ($-\int_{-\infty}^{\infty}-p(x)\log(\frac{p(x)}{m(x)})dx$) is the entropy ($m(x)$ being the invariance measure), the second term the normalization constraint (probability must sum to $1$) and the last term the maximum entropy constraint.\footnote{Fixed variance is not included as a separate constraint as it is implied by the constraints in equation~\ref{eq:gaussian:maxent}.} We obtain first-order conditions 

$$\frac{\partial \Lambda}{\partial p(x)} = 0 = -\log\left(\frac{p(x)}{m(x)}\right)-1-\lambda_1-\lambda_2(x-\mu)^2$$
$$\frac{\partial \Lambda}{\partial \lambda_1} =0= \int_{-\infty}^{\infty}p(x)dx -1$$
$$\frac{\partial \Lambda}{\partial \lambda_2} =0= \int_{-\infty}^{\infty}p(x)(x-\mu)^2dx -\sigma^2$$

the first one of which directly yields the functional form of the density function $p(x)$ as

\begin{equation}
p(x) = me^{-\lambda_1}e^{-\lambda_2(x-\mu)^2}=ke^{-\lambda_2(x-\mu)^2}
\end{equation}
where $k=me^{-\lambda_1}$ and $\lambda_2$ are constants. The values of these constants can be determined by substituting the function of $p(x)$ into the other two first-order conditions and solving.\footnote{Note that the function contains an exponent with a quadratic function of the variable of integration $x$, so the solution can be expressed in terms of the Gaussian error function and $\pi$. The normalization to $1$ in the second first-order condition forces us to use this form.} We then obtain $k=\frac{1}{\sigma\sqrt{2\pi}}$ and $\lambda_2=\frac{1}{2\sigma^2}$ and finally the Gaussian

\begin{equation}
p(x) = \sqrt{\frac{1}{2\pi\sigma^2}} e^{-\frac{1}{2\sigma^2}(x-\mu)^2}.
\end{equation}

\subsection{Fourier domain representations of random variable distributions}
\label{app:dist:Fourier}

Distributions can also be represented as characteristic functions $\varphi(t)$ in the Fourier domain. While density $p(x)$ in the direct domain $x$ gives probabilities of realizations of values $x$, $\varphi(s)$ gives the intensity of fluctuations of frequencies $s$ in frequency space. If both functions, $p(x)$ and $\varphi(s)$, exist in functional form, there is a bijective mapping (a unique, invertible, one-to-one mapping) between the two, the Fourier transform

\begin{equation}
    \varphi(s) = \operatorname{E}[e^{(isx)}] = \int_{-\infty}^{\infty}e^{(isx)}p(x)dx
\end{equation}

where $\operatorname{E}$ represents the expectation. The inverse operation (inverse Fourier transform) is

\begin{equation}
    p(x) = \frac{1}{2\pi}\int_{-\infty}^{\infty}e^{(-isx)}\varphi(s)ds.
\end{equation}

However, in some cases, as for L\'{e}vy alpha-stable distributions, there is no functional representation of the density in the direct domain $p(x)$ and only the characteristic function $\varphi(s)$ in the Fourier domain exists.

Since the two representations are absolutely equivalent, the maximum entropy distribution can equivalently be obtained in Fourier domain.

Let $\varphi'(s)$ be the normalized characteristic function.\footnote{That is, $\varphi(s)$ is normalized as 
$$\varphi'(s)=\frac{\varphi(s)}{\int_{-\infty}^{\infty}\varphi(s)ds}$$
so that the area sums to one, $\int_{-\infty}^{\infty}\varphi'(s)ds=1$ and $\varphi'(s)$ thus constitutes a probability distribution.} We apply entropy $S$ and entropy constraints $f_i(s)$ just like in the direct domain and maximize\footnote{The normalization constraint is unnecessary since the function is already normalized with the transformation to $\varphi'(s)$.}

\begin{equation}
\Lambda = S - \sum_i\lambda_i\int_{-\infty}^{\infty}f_i(s)\varphi'(s)ds = - \int_{-\infty}^{\infty}\varphi'(s)\log\left(\frac{\varphi'(s)}{M(s)}\right)ds  - \sum_i\lambda_i\int_{-\infty}^{\infty}f_i(s)\varphi'(s)ds
\label{eq:lagrange:fourierdomain}
\end{equation}

with first-order conditions

$$\frac{\partial \Lambda}{\partial \varphi(s)} = 0 = - \log\left(\frac{\varphi'(s)}{M(s)}\right) - 1 - \sum_i \lambda_if_i(s)$$
$$\frac{\partial \Lambda}{\partial \lambda_i} = 0 = \int_{-\infty}^{\infty}f_i(s)\varphi'(s)ds.$$

From the first condition, we obtain the general functional form of the maximum entropy distribution in the Fourier domain

\begin{equation}
\varphi'(s) = M(s)e^{-1}e^{-\sum_i\lambda_if_i(s)}=ke^{-\sum_i\lambda_if_i(s)}.
\label{eq:solution:fourierdomain}
\end{equation}

where $k$ and $\lambda_1$ are factors that must be fixed by solving the other first-order conditions.

Returning to the example of the Gaussian above, we obtain the characteristic function of the same distribution (that we would also get by taking the Fourier transform of the density function in the direct domain) if we set the same constraint. Recall that for the Gaussian, this constraint is to fix the variance. In the Fourier domain, this is $s^2=\chi$. Hence, we substitute $f=s^2-\chi$ in equation~\ref{eq:lagrange:fourierdomain} and after $\chi$ drops out in the derivative we obtain 

\begin{equation}
\varphi'(s) = ke^{-\lambda s^2},
\label{eq:solution:gaussian:fourierdomain}
\end{equation}

or, more generally,

\begin{equation}
\varphi'(s) = ke^{-\lambda (s^{\alpha}-\chi)} \quad\quad\quad \textup{ with } \alpha=2.
\end{equation}

\subsection{L\'{e}vy alpha-stable distributions}
\label{sect:methods:levy-as}
\label{app:dist:levy-as}

Instead of fixing the second moment, the variance, $\alpha=2$, as $\chi=s^{\alpha}=s^2$, the constraint can fix lower-order moments as the highest finite moments of the distribution. These moments do not need to be integer moments (fractional lower-order moments). They will, in fact, not be integers, except in case $\alpha=1$ (the Cauchy distribution). The computation of the maximum entropy distribution is equivalent, as long as we impose that the distribution is symmetric, $\beta=0$, and centered around zero, $\delta=0$. Analogous to equation~\ref{eq:solution:gaussian:fourierdomain}, the distribution is now

\begin{equation}
\varphi'(s) = ke^{-\lambda s^{\alpha}}.
\label{eq:solution:symmetriclevy:fourierdomain}
\end{equation}

For $\alpha < 2$, $\varphi'(s)$ does not have a functional representation in the direct domain any longer (except again for the Cauchy distribution, $\alpha=1$). The tails of the distribution do, however, asymptotically approach the power law 

\begin{equation}
p_{Tail}(x) = C|x|^{-(\alpha+1)}.
\label{eq:tails:symmetriclevy:fourierdomain}
\end{equation}

and the distribution consequently has fat tails instead of quickly dropping to zero as in the Gaussian case with $\alpha=2$.

For the general case, without imposing symmetry and central location at zero, the characteristic equation becomes more complex and has four parameters, interpreted as the tail index ($\alpha$), the skew ($\beta$), the scale ($\gamma$), and the location ($\delta$). The functional form is\footnote{Observe that for $\beta=0$, $\delta=0$, the function reduces to 
$$\varphi(s)=\operatorname{E}[e^{(isx)}]  =e^{(-\gamma^{\alpha}|s|^{\alpha}},$$
the solution obtained above (with $k=1$, $\lambda=\gamma^{\alpha}$).} 

\begin{eqnarray}
\varphi(s)=\operatorname{E}[e^{(isx)}]  ={\begin{cases}
	e^{(-\gamma^\alpha|s|^\alpha[1+i\beta \text{tan}\left({\tfrac {\pi \alpha }{2}}\right) \operatorname {sgn}(s)\left((\gamma|s|)^{1-\alpha}-1 \right))] + i\delta s)}
	&\alpha \neq 1
	\\e^{(-\gamma|s|[1+i\beta{\tfrac {2}{\pi}} \operatorname {sgn}(s)\log(\gamma|s|)] + i\delta s)}&\alpha =1
	\end{cases}}
\end{eqnarray}

More technical details on L\'{e}vy alpha-stable distributions can be found in \citet{NOLAN1998187, NOLAN2018}; a comprehensive discussion of maximum entropy, aggregation of distributions, and characteristic equations in the Fourier domain is offered in \citet{Frank09}.

\begin{figure}[tb!]
\centering
\includegraphics[width=0.85\textwidth]{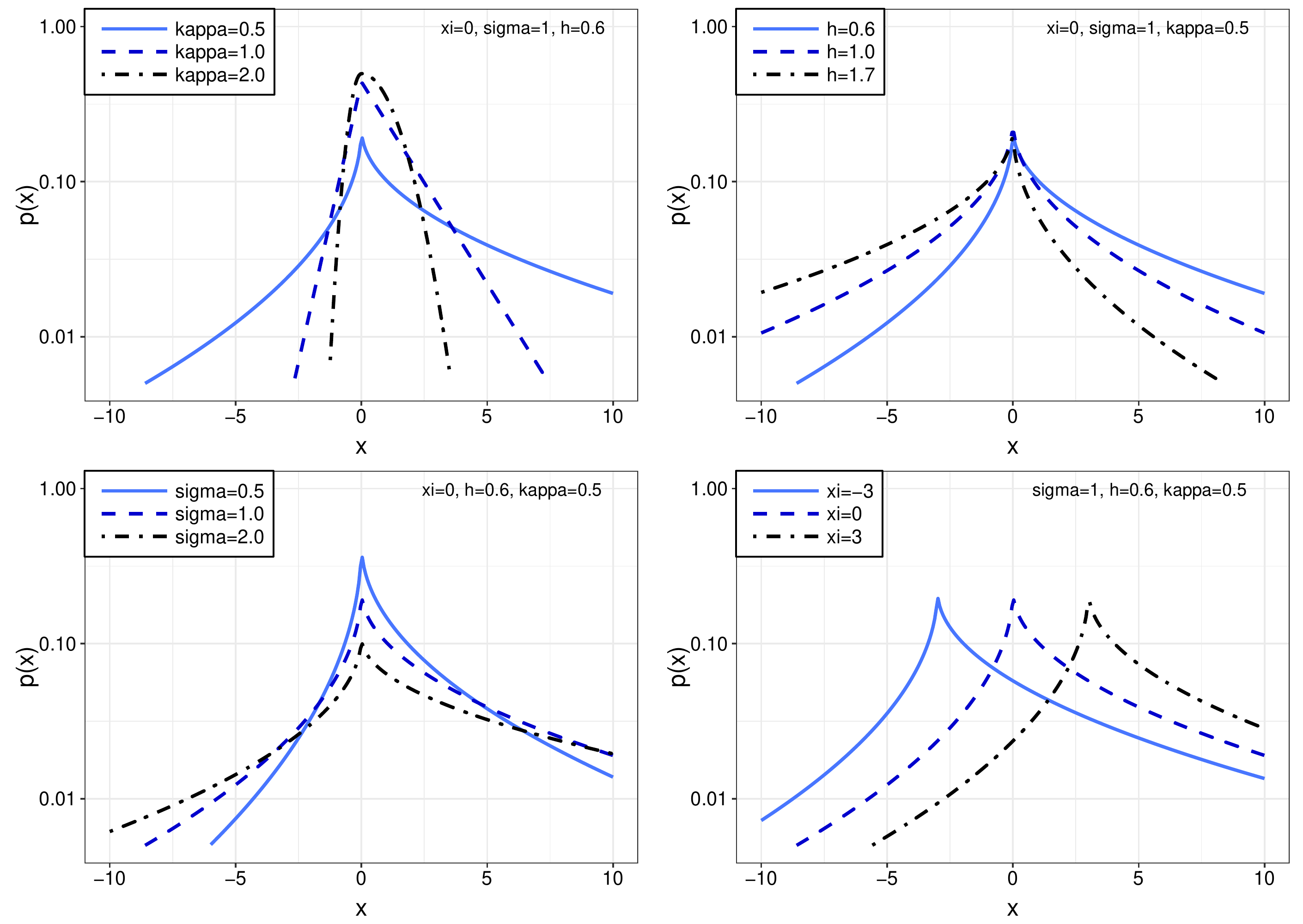}
\caption{Density of the Asymmetric Exponential Power (AEP) distribution for different parameter settings. Upper left: Variation of tail parameter $\kappa$. Upper right: Variation of skew parameter $h$. Lower left: Variation of scale parameter $\sigma$. Lower right: Variation of location parameter $\xi$.}
\label{fig:variations:AEP}
\end{figure}

\subsection{Asymmetric exponential power (AEP) distributions}
\label{app:dist:aep}

For the distribution of the growth rates at the firm-level, the model advanced by \citet{Bottazzi/Secchi06, Bottazzietal07, Bottazzi/Secchi11} is considered a strong candidate, the asymmetric exponential power (AEP) or Subbotin distribution. This model is of particular interest here as an alternative model for comparison, since growth rates must be expected to be related to productivities in general, and to the labor productivity in particular. Firms with high labor productivity will generally have good prospects for future growth, while firms with low labor productivity will likely write losses and be unable to grow or even sustain their present operations unless supported by an inflow of additional resources.

The AEP is a generalization of the symmetric Laplace distribution, the two-sided exponential distribution. \citet{Bottazzi/Secchi06} take a very similar approach to the one taken in this paper: They characterize the distribution of growth as constant in mean absolute differences, use this as entropy constraint $|x-\xi|=\sigma$ and compute the maximum entropy distribution.

\begin{equation}
\Lambda = -\int_{-\infty}^{\infty}p(x)\log\left(\frac{p(x)}{m(x)}\right)dx - \lambda_1\left(\int_{-\infty}^{\infty}p(x)dx -1\right) -\lambda_2(p(x)|x-\xi|-\sigma)
\label{eq:laplace:maxent}
\end{equation}

The first-order condition with respect to $p(x)$

$$\frac{\partial \Lambda}{\partial p(x)} = 0 = -\log\left(\frac{p(x)}{m(x)}\right)-1-\lambda_1-\lambda_2|x-\xi|$$

yields the functional form

\begin{equation}
p(x) = me^{-\lambda_1}e^{-\lambda_2|x-\xi|}=ke^{-\lambda_2|x-\mu|}.
\end{equation}

Solving the remaining first-order conditions (not given here) fixes parameters $k$ and $\lambda_2$ and results in the canonical form of the standard Laplace distribution, 

\begin{equation}
p(x) = \frac{1}{2\sigma}e^{-\frac{|x-\xi|}{\sigma}}.
\end{equation}

Relaxing the assumptions on symmetry\footnote{\citet{Bottazzi/Secchi06} find an empirical symmetry very close to $h=1$, the symmetric Laplace case.} and tail behavior yields the more general functional form of the AEP,\footnote{Note that this reduces to a Laplace distribution for $\kappa=1$, $h=1$.}

\begin{equation}   
    p(x) = \frac{\kappa h}{\sigma (1+\kappa^2)\Gamma(1/h)}e^{[-(\kappa^{\operatorname {sgn}(x-\xi)}(|x-\xi|/\sigma))^h}
\label{eq:AEP:density}
\end{equation}

where $\Gamma$ is the Gamma function. The given parametrization is for the 4-parameter AEP that we will use as an alternative model and point of comparison in Section~\ref{sect:results}. The four parameters again stand for the tail behavior ($\kappa$), the skew ($h$), the scale ($\sigma$), and the location ($\xi$) and are visualized in the four panels of Figure~\ref{fig:variations:AEP} in direct comparison to the L\'{e}vy alpha-stable distribution in Figure~\ref{fig:variations:levy}. Again, the diagrams are in semi-log scale with the vertical axis being logarithmic. As expected, all AEP variants approach a linear shape towards both tails in the semi-log form, indicating that they belong to the family of exponential distribution forms (which are linear in semi-log). In contrast, the L\'{e}vy alpha-stable functions above bend in outward direction and clearly have tails that are heavier than exponential. 

As an alternative to the 4-parameter AEP, there is a 5-parameter variant, which assigns two different tail parameters for positive and negative tails. We choose to work with the 4-parameter version to allow a more direct comparison with the 4-parameter L\'{e}vy alpha-stable function under consideration here.

\subsection{Tail behavior}
\label{app:dist:tails}

While the AEP results as the maximum entropy distribution for specific conditions, it is not a stable distribution: Summing AEPs will yield a Gaussian aggregate. While the difference may seem academic when considering the very similar body part of the distributions in Figures~\ref{fig:variations:levy} and \ref{fig:variations:AEP}, the difference becomes important with the tail behavior and the finiteness of moments: In the case of L\'{e}vy alpha-stable distributions with $\alpha<2$, the variance is infinite. That is, each sample will have a specific variance, but the sample variance will diverge in sample size $N$ \citep{NOLAN2018,Emberchts97,Yangetal19} with 

\begin{equation}
Var(x) \sim N^{\frac{2-\alpha}{2\alpha}}.
\end{equation}

The mean of the distribution may or may not (if $\alpha<1$) be finite. But as the expected deviation from the mean is infinite, the information carried by each observation about the true mean is practically zero. The mean, albeit existent, may be difficult or impossible to infer from a sample.

This is not to say that nothing can be known for sure about L\'{e}vy alpha-stable distributed samples. On the contrary, both the quantiles of the distribution and its fitted parameters will converge and convey everything there is to know about the distribution. It is merely a matter of choosing the correct interpretation of the data and using adequate measures to characterize it.

\section{Prior specification of the regression model}
\label{app:prior}

The likelihood function and the priors of a Bayesian multi-level model in section \ref{sect:results:determinants} are written as follows:
\begin{eqnarray}
\operatorname{Parameter_{i}} &\sim& \operatorname{Normal}(\mu_{i}, \sigma)  \nonumber \\
\mu_{i} &=& \alpha + \alpha_{_{j[i]}} +  \alpha_{_{t[i]}} + \beta_1\operatorname{GDP\_Growth} + \beta2\operatorname{Firm\_Age} + \beta3\operatorname{Emp} + \beta4\operatorname{Cap\_Intensity}  \nonumber \\
\alpha & \sim& \operatorname{Student-t}(3, 1, 10)  \nonumber \\
\beta_1, \beta_2, \beta_3, \beta_4 &\sim& \operatorname{Normal}(0, 1)  \nonumber \\
\sigma &\sim& \operatorname{Student-t}^{+}(3,0,10)\nonumber \\
\alpha_{_{j}} &\sim&  \operatorname{Normal}(\mu_j, \sigma_j)  \nonumber \\
\alpha_{_{t}} &\sim&  \operatorname{Normal}(\mu_t, \sigma_t)  \nonumber \\
\mu_j, \mu_t &\sim&  \operatorname{Normal}(0, 1)  \nonumber \\
\sigma_j, \sigma_t&\sim& \operatorname{HalfCauchy}(0,1)\nonumber
\end{eqnarray}

From line 3, we define the prior distribution for each parameter of the model. The overall intercept (the grand mean), $\alpha$ is given a weakly informative prior in the form of the Student's-t distribution centered on 1 with 3 degrees of freedom and 10 standard deviation. The population effect coefficients, $\beta_1, \beta_2, \beta_3, \beta_4$, are given a Gaussian prior centered on 0 with 1 standard deviation. The standard deviation of the Gaussian likelihood function, $\sigma$, is given a weak prior in the form of the half Student's-t distribution centered on 0 with 3 degrees of freedom and 10 standard deviation. The varying intercepts, $\alpha_{j[i]}$ and $ \alpha_{t[i]}$ are given a Gaussian prior with a hierarchical structure. The hyperpriors on the mean, $\mu_j$ and $\mu_t$ are given a Gaussian prior centered on 0 with 1 standard deviation.  The hyperpriors on the standard deviation, $\sigma_j$ and $\sigma_t$ are given a Half-Cauchy prior centered on 0 with scale parameter 1.  Note that  $\sigma_{\alpha_{t}}$ and $\sigma_{\alpha_{j}}$ represent the estimated between-year variance and the between-province variance, respectively. For a detailed discussion on the choice of prior in Bayesian statistics, see \cite{gelman2017prior}. 

\section{Historical note on productivity growth in the PR China}
\label{app:historical}

What caused the period of rapid economic growth in the PR China? Since the economic reform was initiated in 1978, the PR China has undergone significant structural change \citep{brandt2008,wu2011total,bosworth2008accounting,chow2002china} - going from an agricultural to an industrial economy with growing importance of the service-sector - and achieving many important milestones. The early period was dominated by productivity improvements in agriculture (notably with the transformation from the \textit{production-team system} to \textit{household responsibility system (HRS)} \citep{lin1992rural,mcmillan1989impact}. The rapid productivity growth in the agricultural sector - 6.5\% annually on average while labor input in the primary sector declines 4.5-5.5 \% annually \citep{cao2013agricultural,borensztein1996accounting} - came to an end around 1984, because the slowing new labor participation and technology adoption after 1984, and the institutional change exhausted its catch-up potential \citep{lin1992rural}.\footnote{\citet{gong2018agricultural} applies a varying coefficient production function to capture the structural change in different agricultural segments. He finds that the agricultural TFP growth rate  fluctuated cyclically in the past forty years, which is typical for policy-driven sectors and demands more intensive technological investment.} In turn, the industrial and service sectors absorbed the labor freed in the primary sector, while also being boosted by an increasing work participation rate \citep{lin1992rural} and improved education and human capital accumulation \cite{au2006migration,gordon1995change}. The 1980s saw a fundamental reform of the economic organization (enterprise reform) followed by increasing international investment in the PR China in the 1990s, which probably boosted economic growth through technology transfer and spillovers \citep{hu2005r}. In 2001, the PR China was admitted to the WTO, allowing better integration in the global economy with again a significant effect on economic growth \citep{brandt2017wto}. By then, manufacturing was the workhorse of the Chinese economy, with productivity growth in manufacturing between 1998 and 2007 being estimated as 7.7\% annually \citet{brandt2012creative}, of which two-thirds came from the productivity differences between entering and existing firms.

Firms in the PR China are typically categorized into seven types \citep{Yuetal15}: (1) Traditional state-owned enterprises (SOE), (2) collective owned enterprises, in particular \textit{Township and village enterprises} (TVE), (3) shareholding firms, (4) private firms, (5) Hong Kong, Macao, and Taiwan owned companies, (6) foreign-owned companies, and (7) other domestic firms. Up to the enterprise reform, the economy was dominated by the first two categories (SOEs and TVEs) with the first phase of growth and rising productivity in the 1980s being carried to a significant part by TVEs, before private ownership became legal with the enterprise reform \citep{goodhart1996rise,ito2006economic,jefferson1994enterprise}, which introduced categories (3) and (4). Categories (5) and (6) would only become important with the increasing international integration of the economy of the PR China in the 1990s and 2000s. 

\newpage
\section{Additional results}
\label{app:results}

 \begin{figure}[h!]
 \centering
 \includegraphics[width=0.65\textwidth]{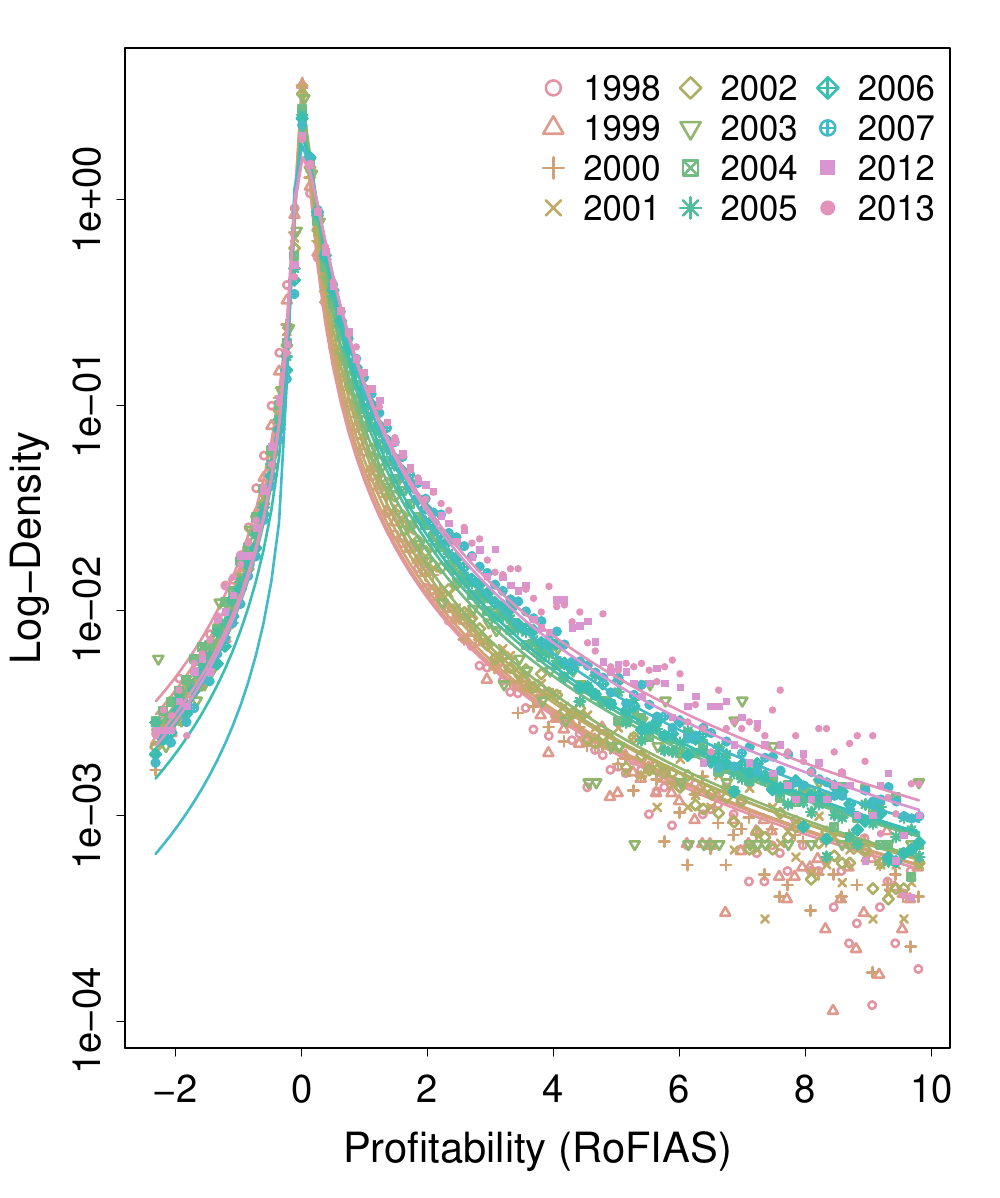}
 \caption{Density of the profitability (return on capital, $ROC$) distribution (full sample) by year in semi-log (vertical axis logarithmic). Solid lines indicate Levy alpha stable distribution fits as reported in Table~\ref{tab:patameterfits}.}
 \label{fig:density:year:roc}
 \end{figure}

 \begin{figure}[tb!]
 \centering
 \includegraphics[width=0.65\textwidth]{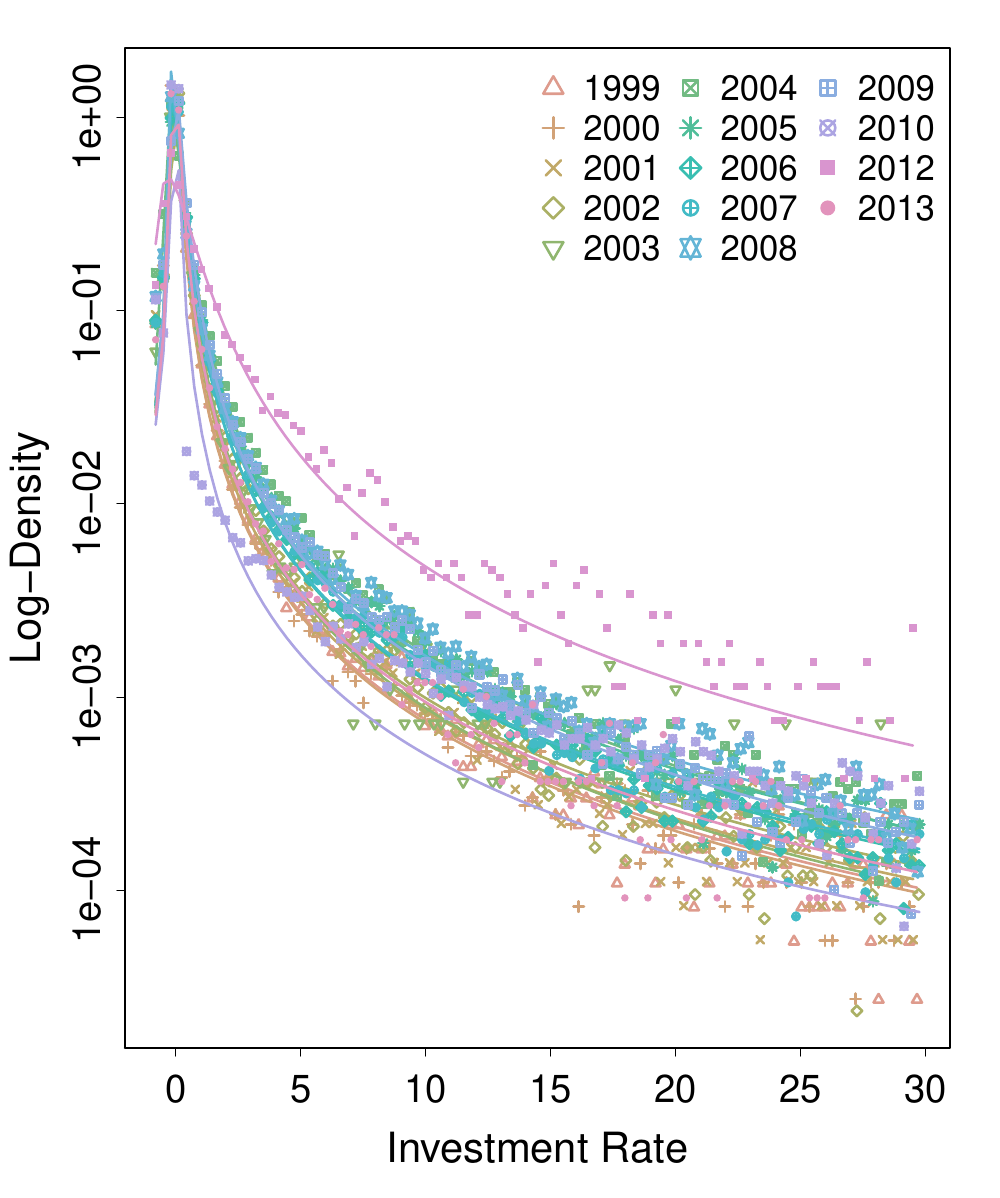}
 \caption{Density of the investment rate ($IR$) distribution (full sample) by year in semi-log (vertical axis logarithmic). Solid lines indicate Levy alpha stable distribution fits as reported in Table~\ref{tab:patameterfits}.}
 \label{fig:density:year:ir}
 \end{figure}

\begin{figure}[tb!]
\centering
\includegraphics[width=0.85\textwidth]{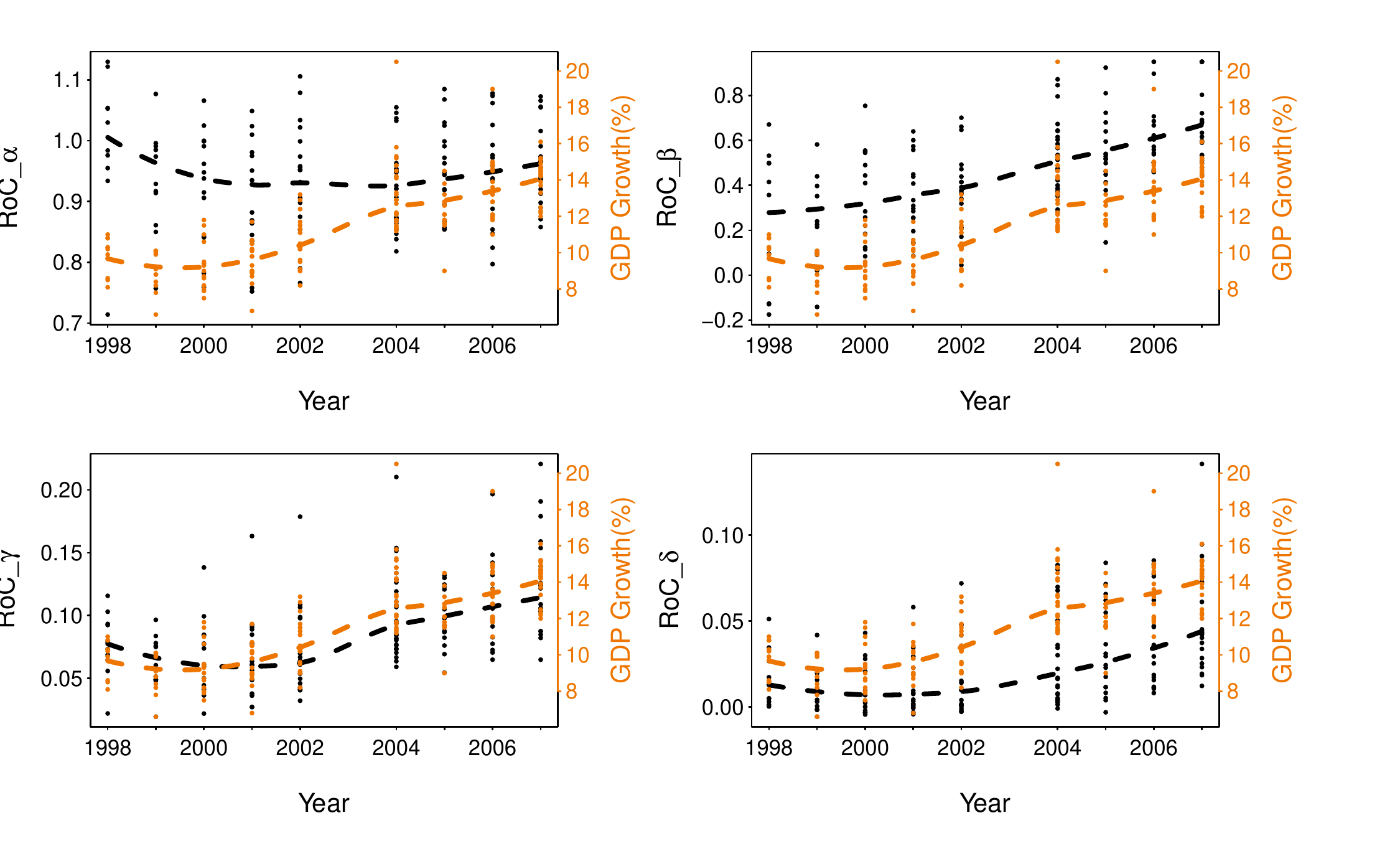}
\caption{\textit{Return on capital} by region and year (black) in comparison to GDP growth (orange).}
\label{fig:timedev:province:roc}
\end{figure}

\begin{figure}[tb!]
\centering
\includegraphics[width=0.85\textwidth]{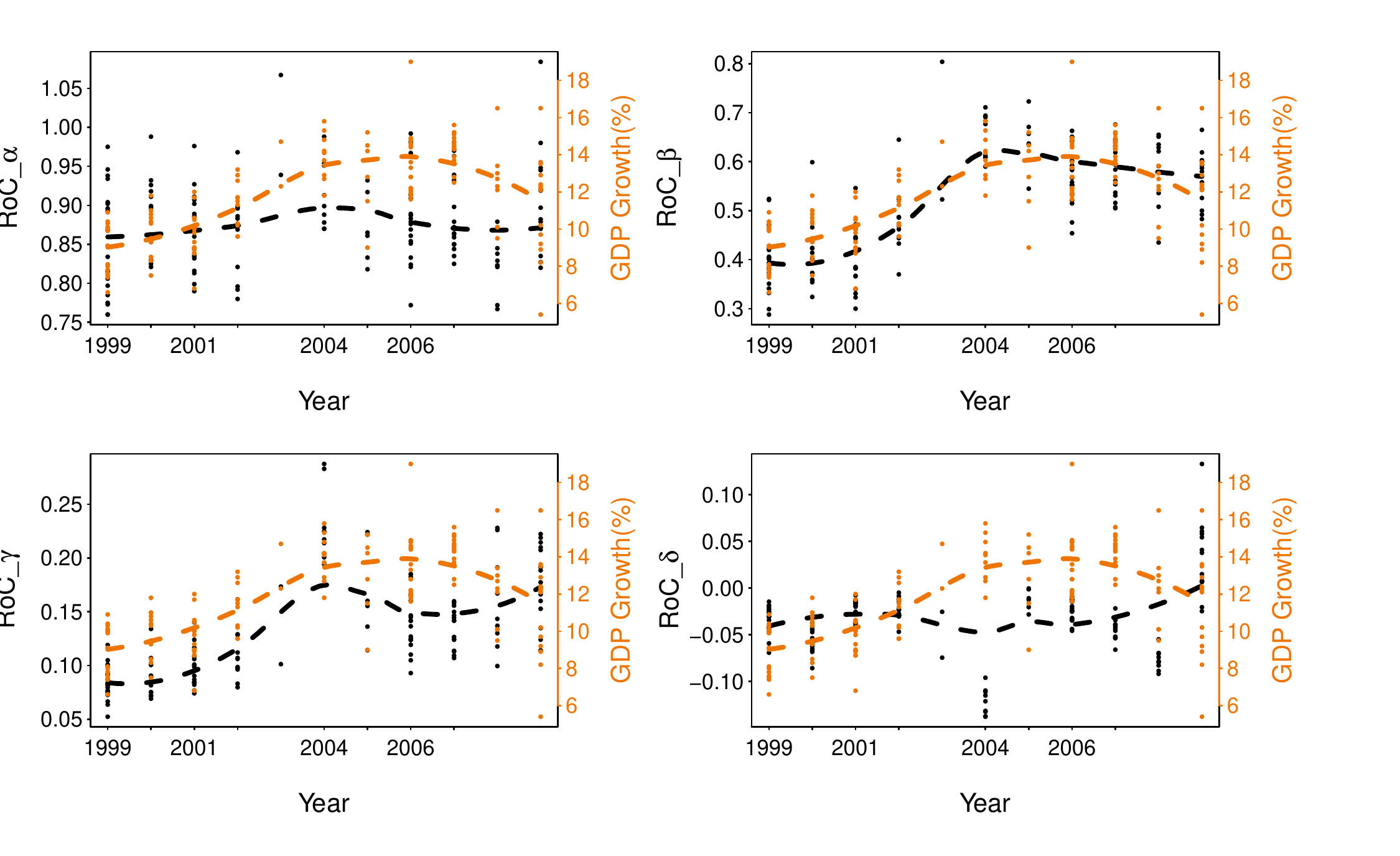}
\caption{\textit{Investment rate} by region and year (black) in comparison to GDP growth (orange).}
\label{fig:timedev:province:ir}
\end{figure}

\begin{figure}[hbt!]
\centering
\subfloat[$LP$]{\includegraphics[width=0.5\textwidth]{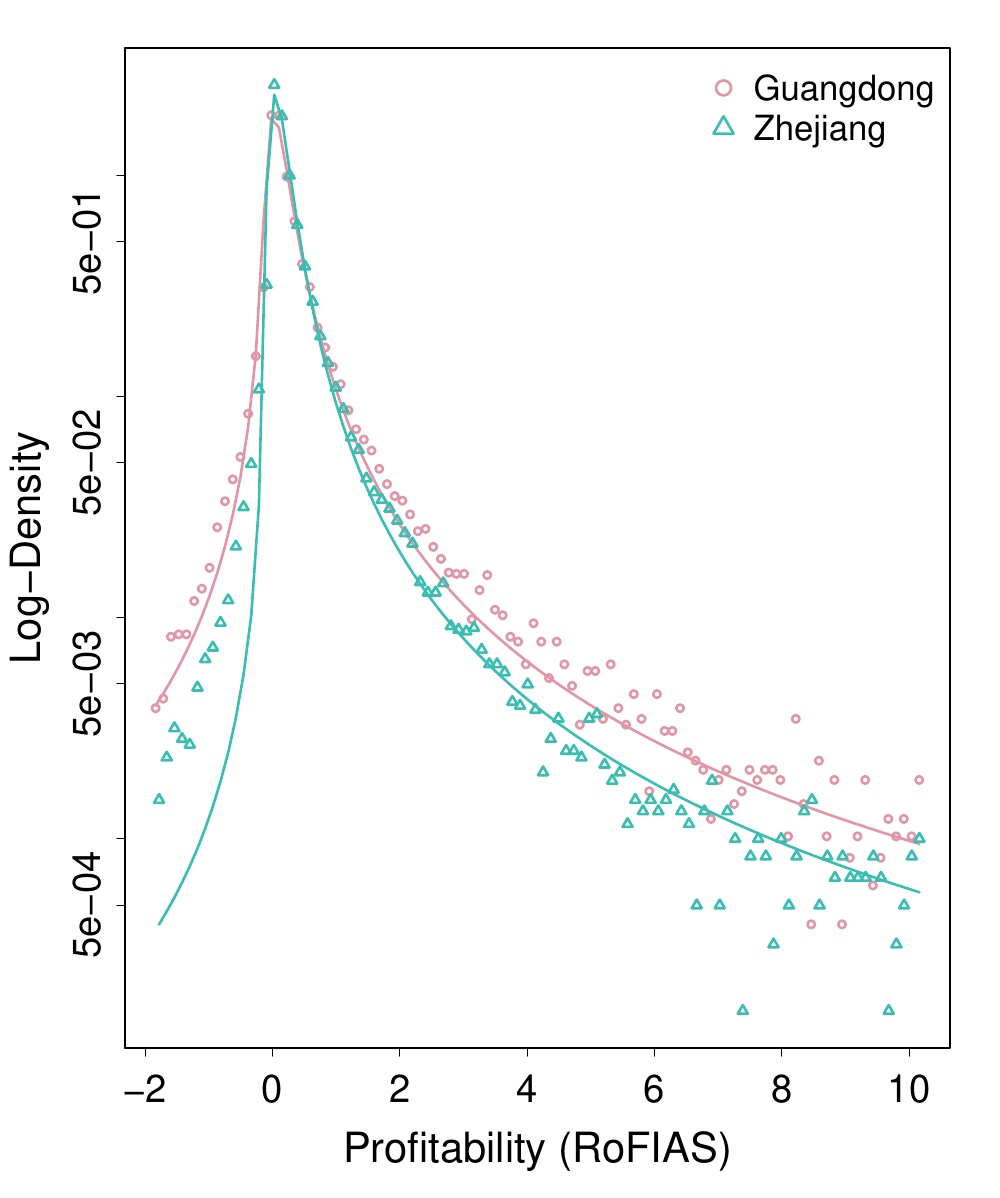}}
\subfloat[$\Delta LP$]{\includegraphics[width=0.5\textwidth]{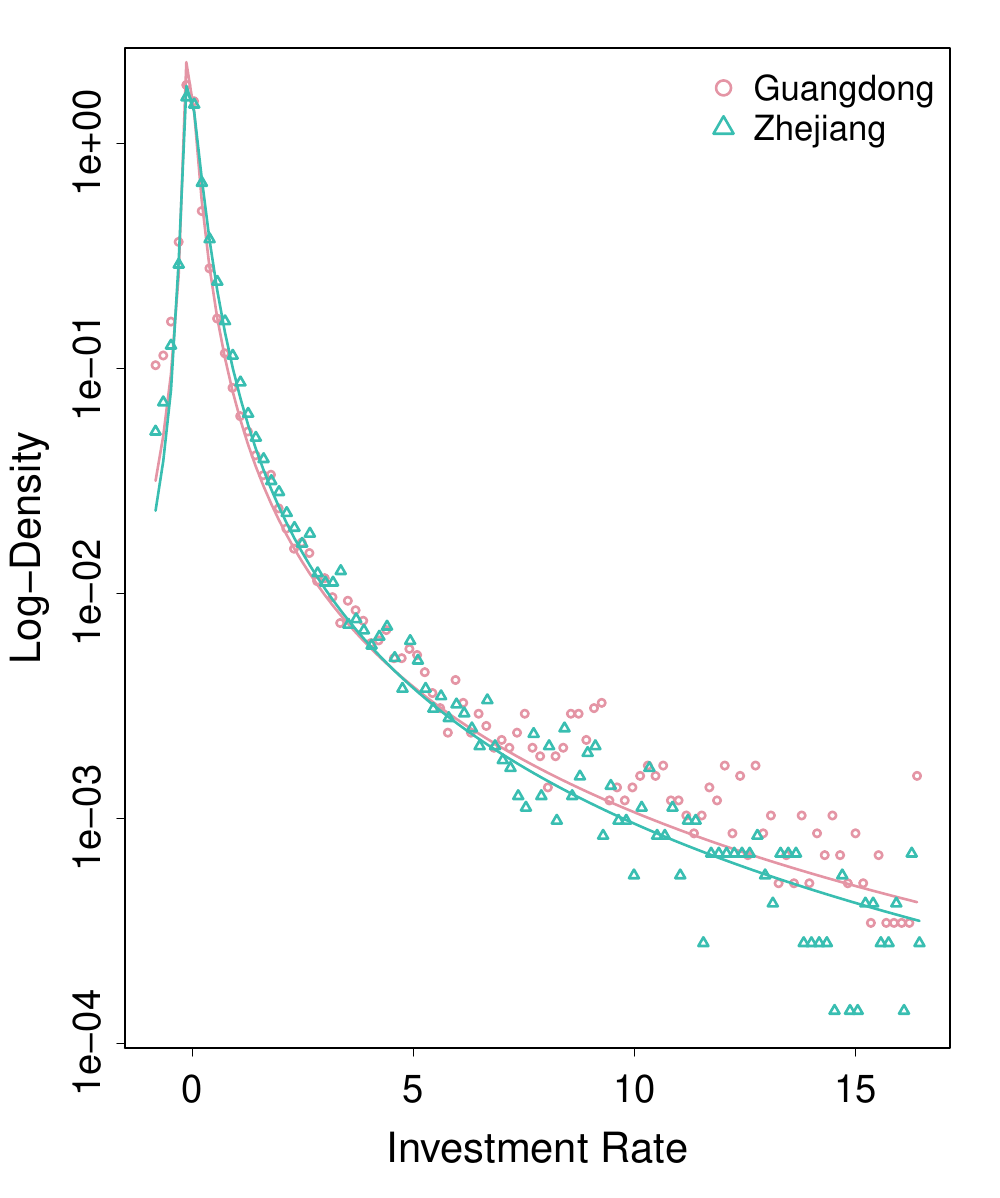}}\\
\caption{Density of $ROC$ and $IR$ for regions Guangdong and Zhejiang in 2007}
\label{fig:2provinces:RoC-IR}
\end{figure}

\begin{figure}[hbtp!]
\centering
\subfloat[1998]{\includegraphics[width=0.9\textwidth]{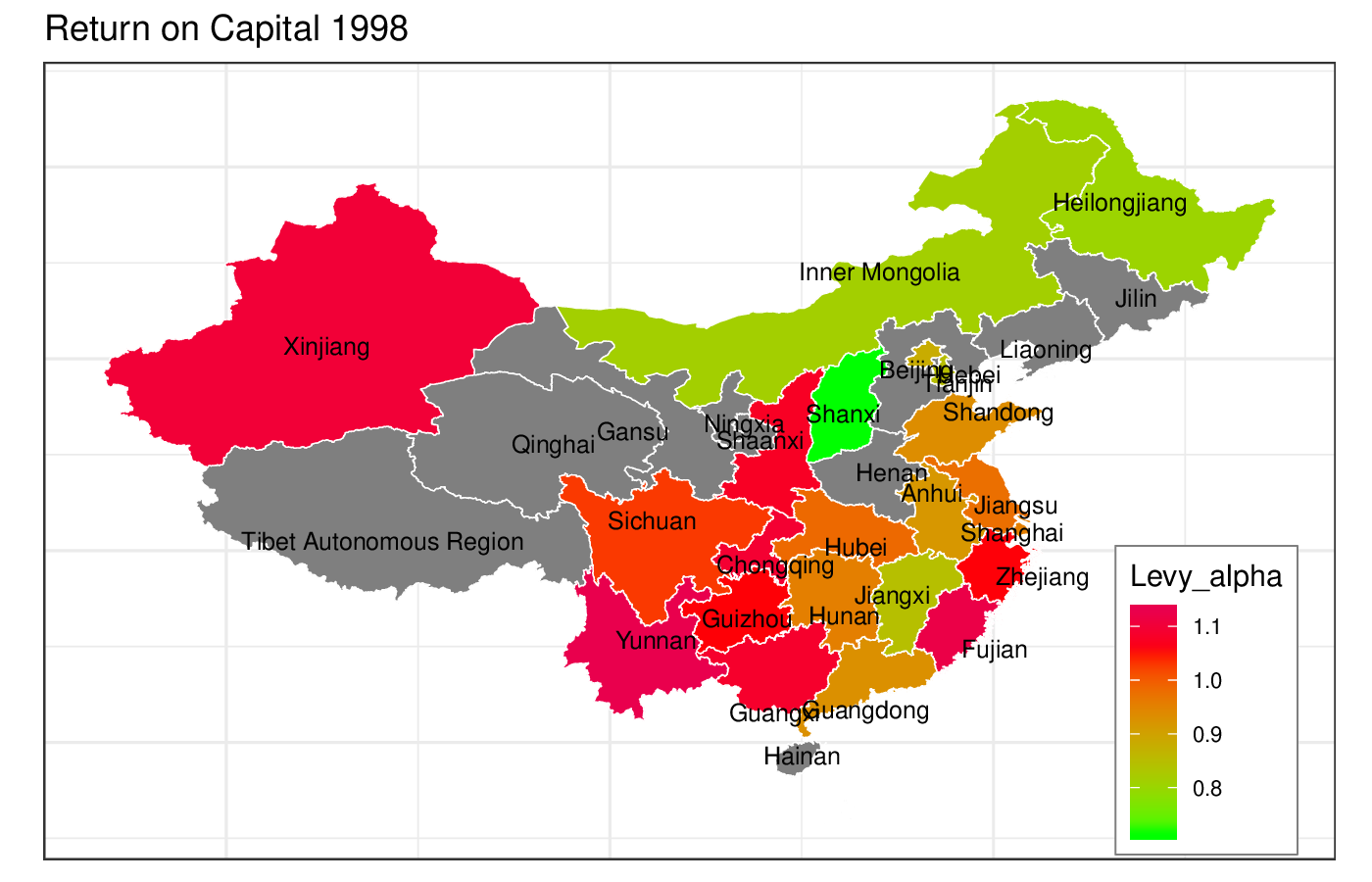}}\\
\subfloat[1999]{\includegraphics[width=0.33\textwidth]{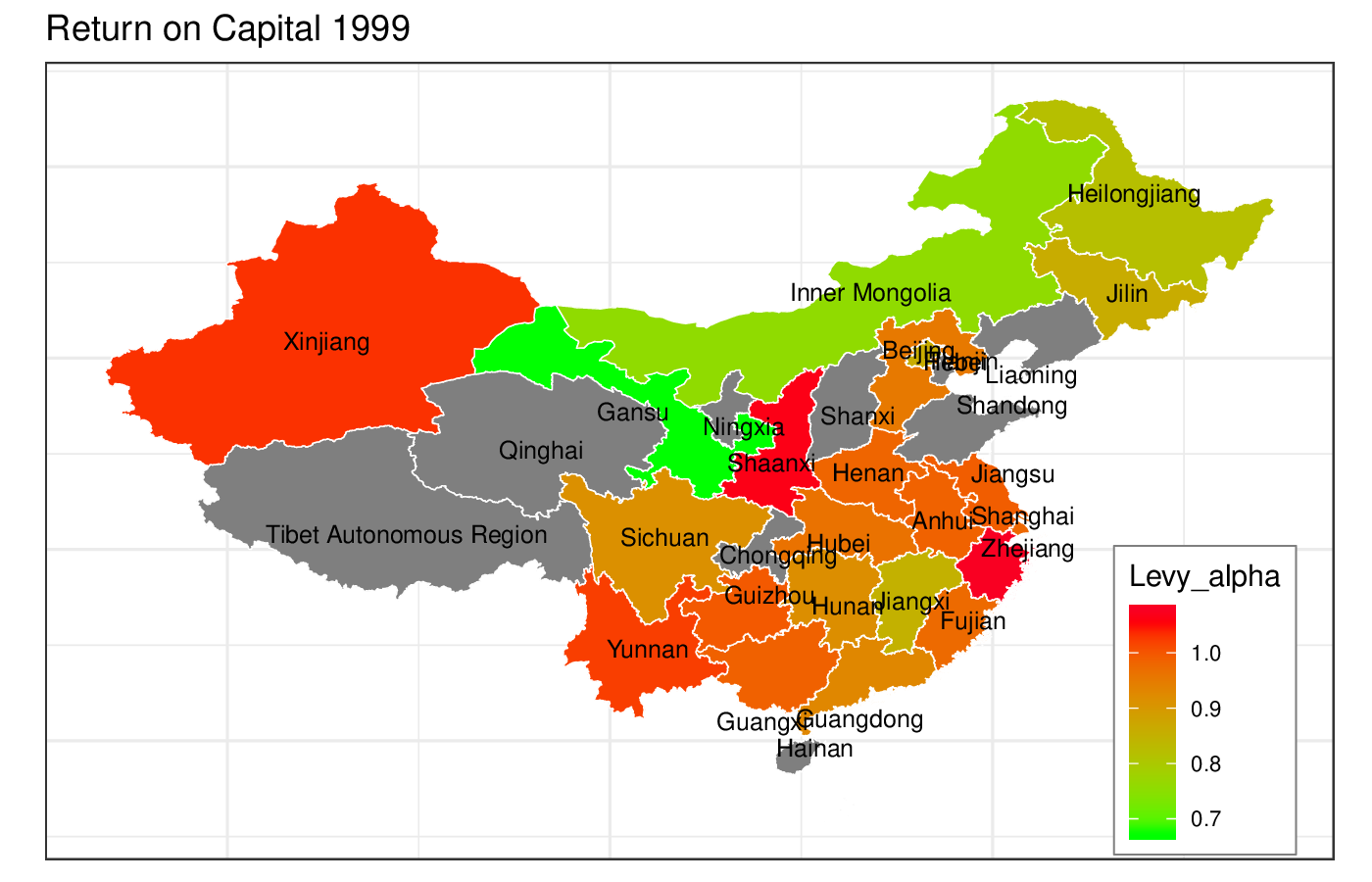}}
\subfloat[2000]{\includegraphics[width=0.33\textwidth]{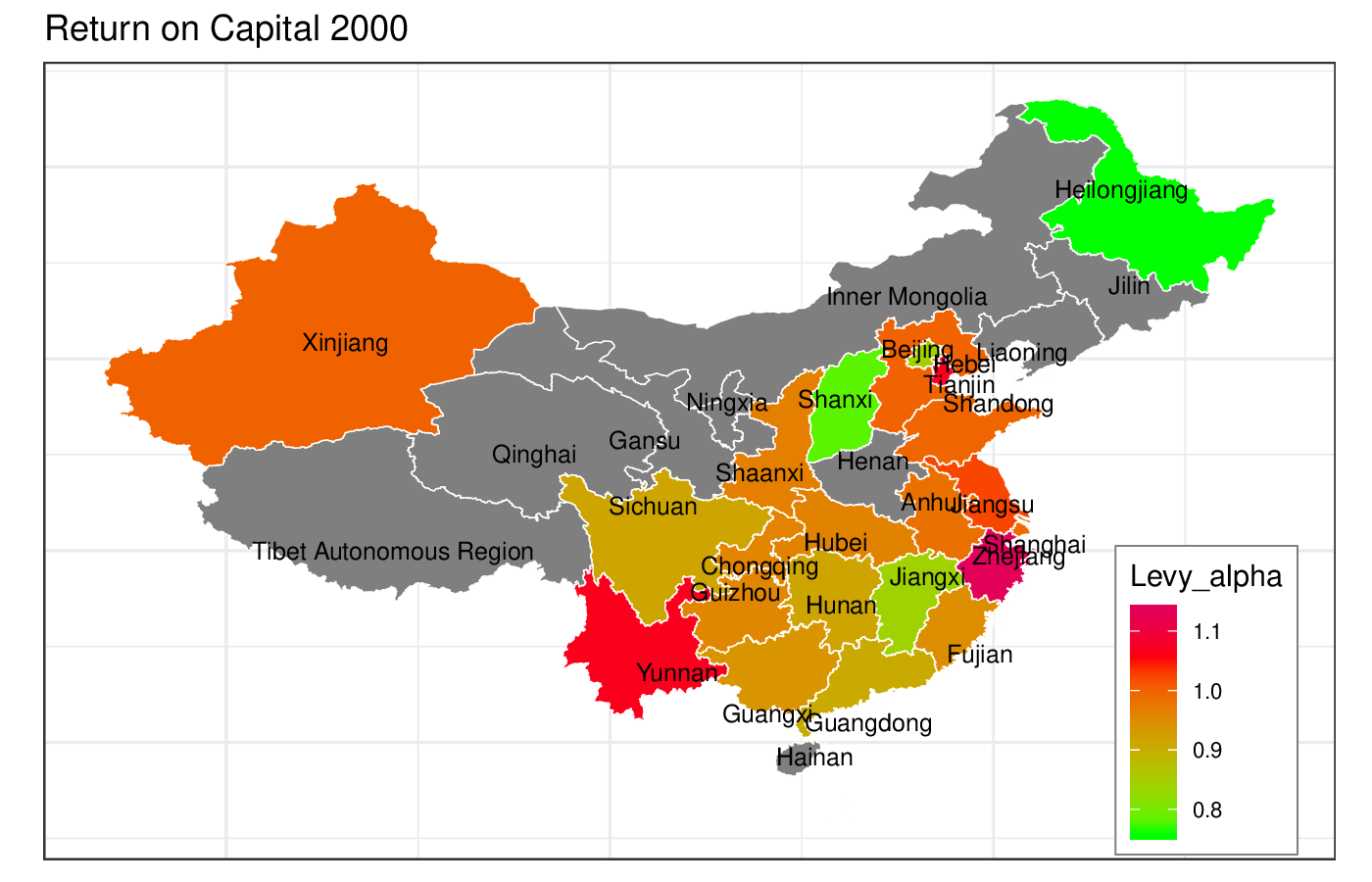}}
\subfloat[2001]{\includegraphics[width=0.33\textwidth]{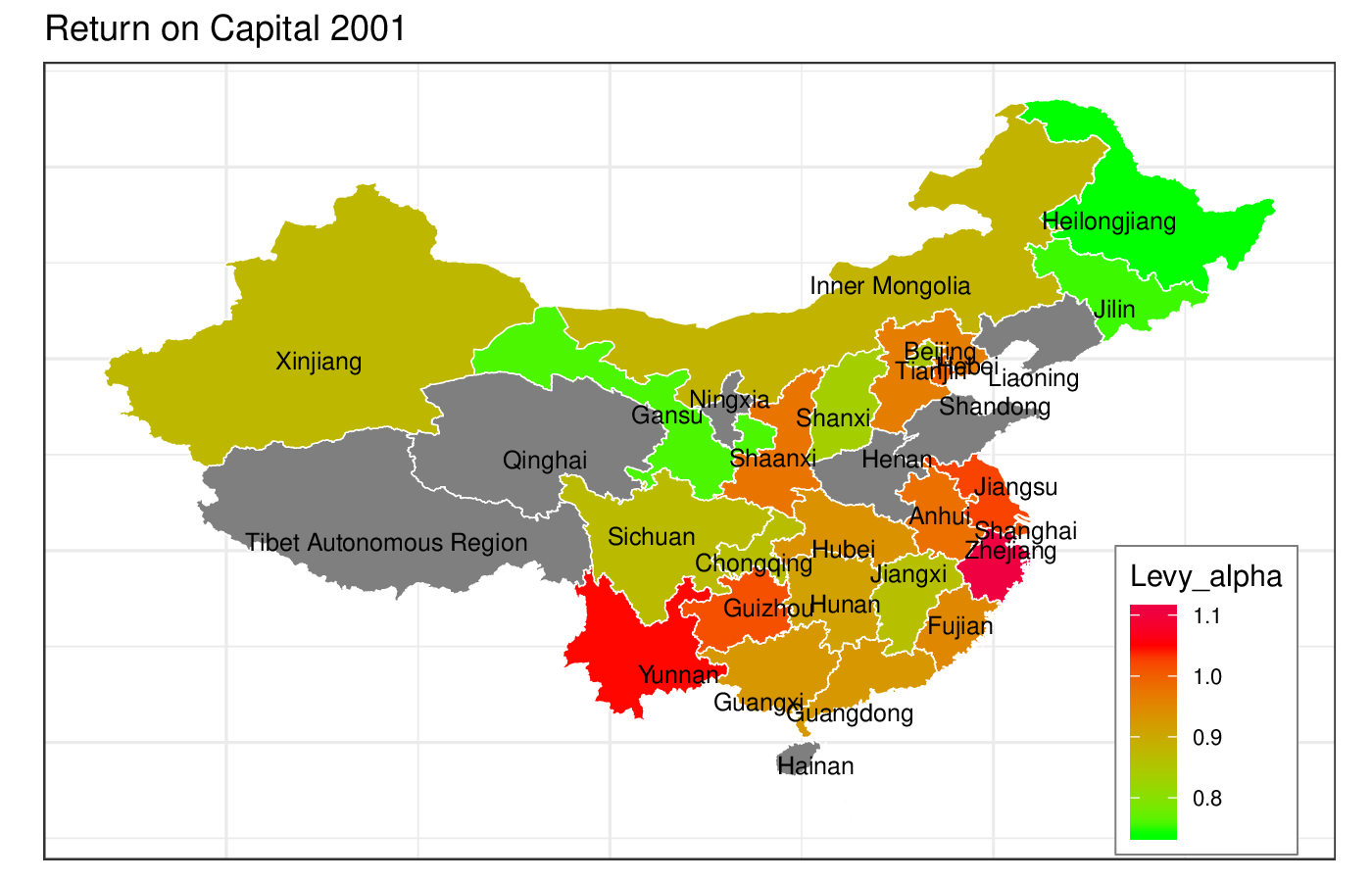}}\\
\subfloat[2002]{\includegraphics[width=0.33\textwidth]{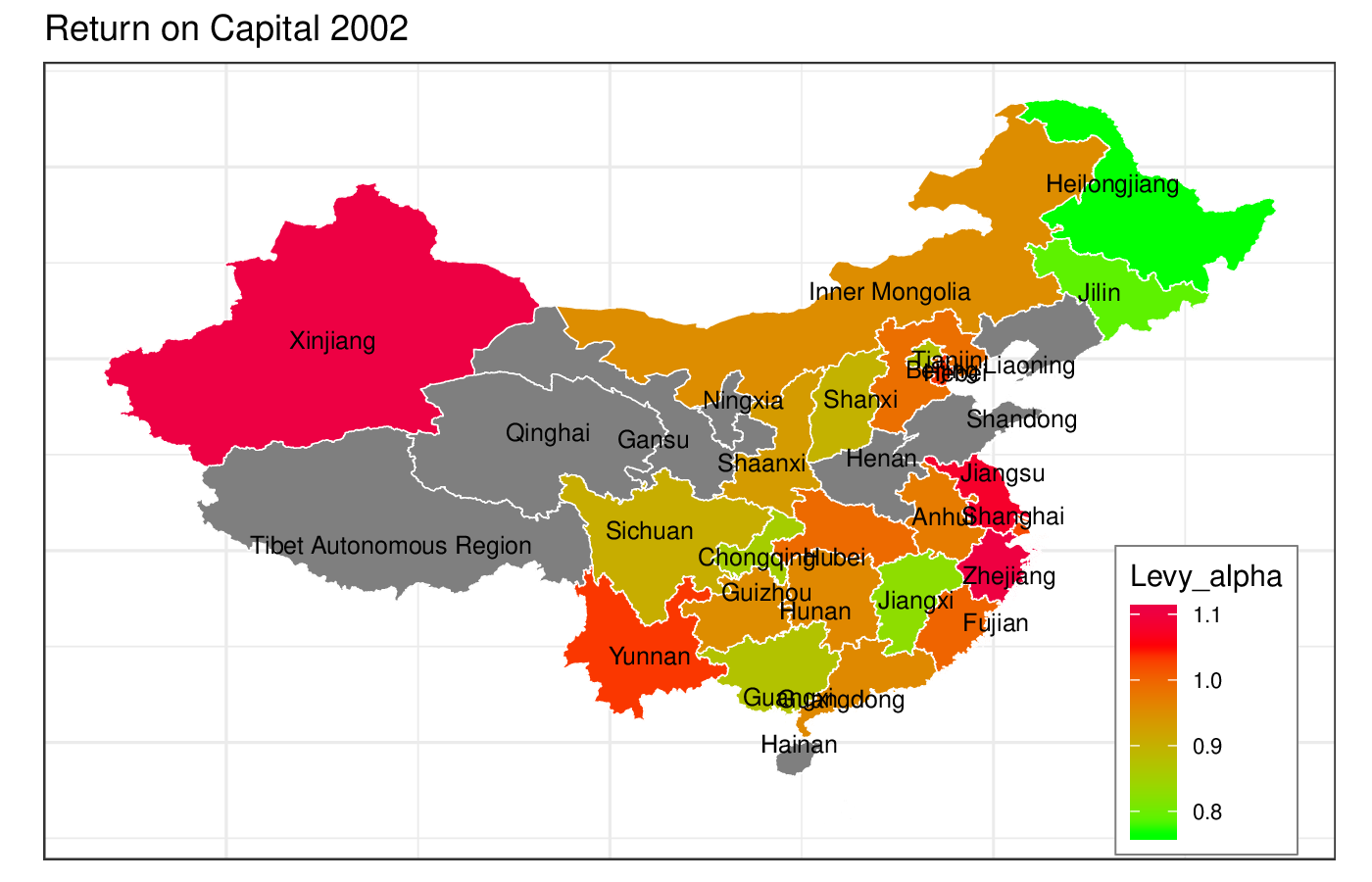}}
\subfloat[2004]{\includegraphics[width=0.33\textwidth]{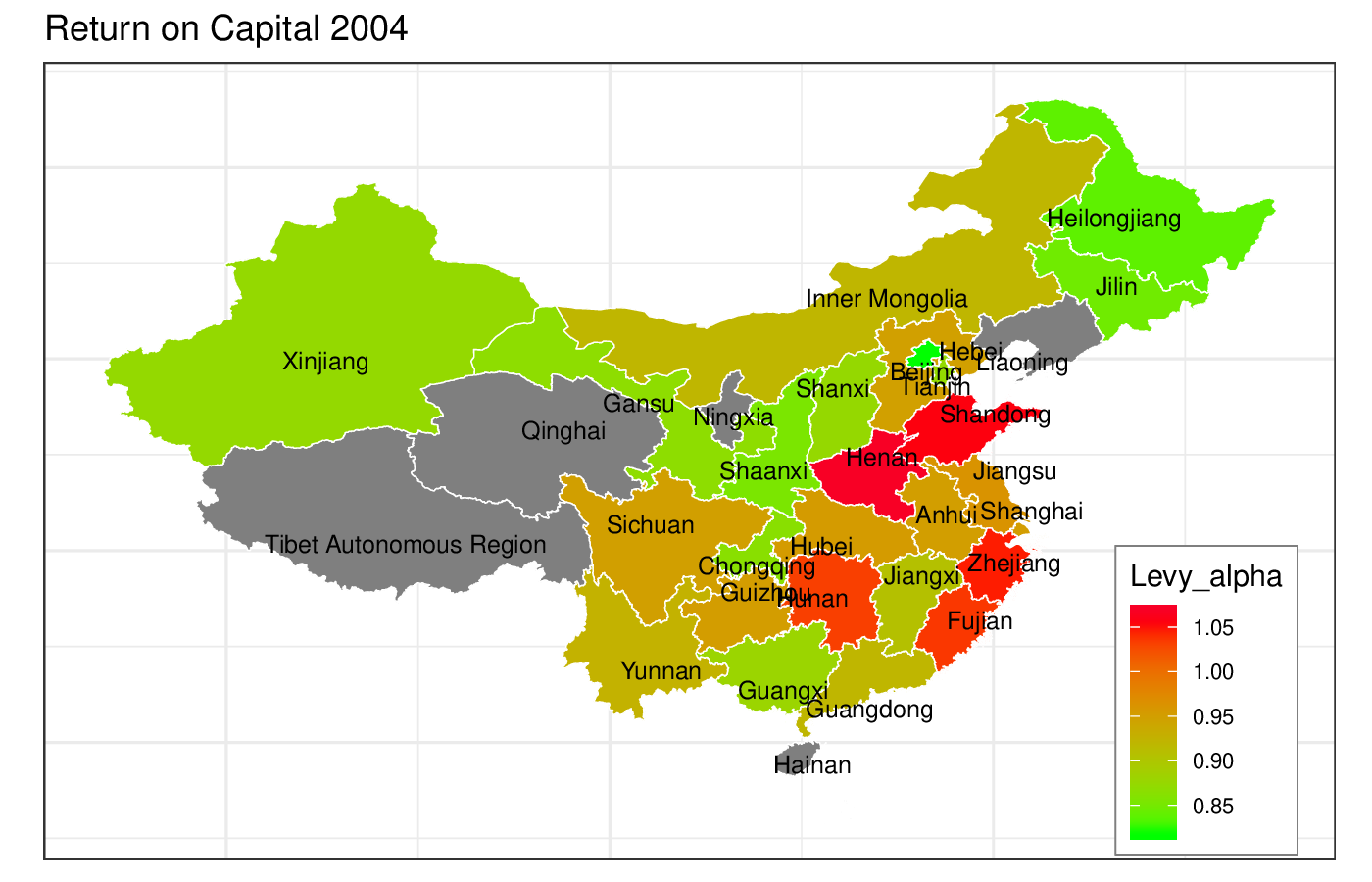}}
\subfloat[2005]{\includegraphics[width=0.33\textwidth]{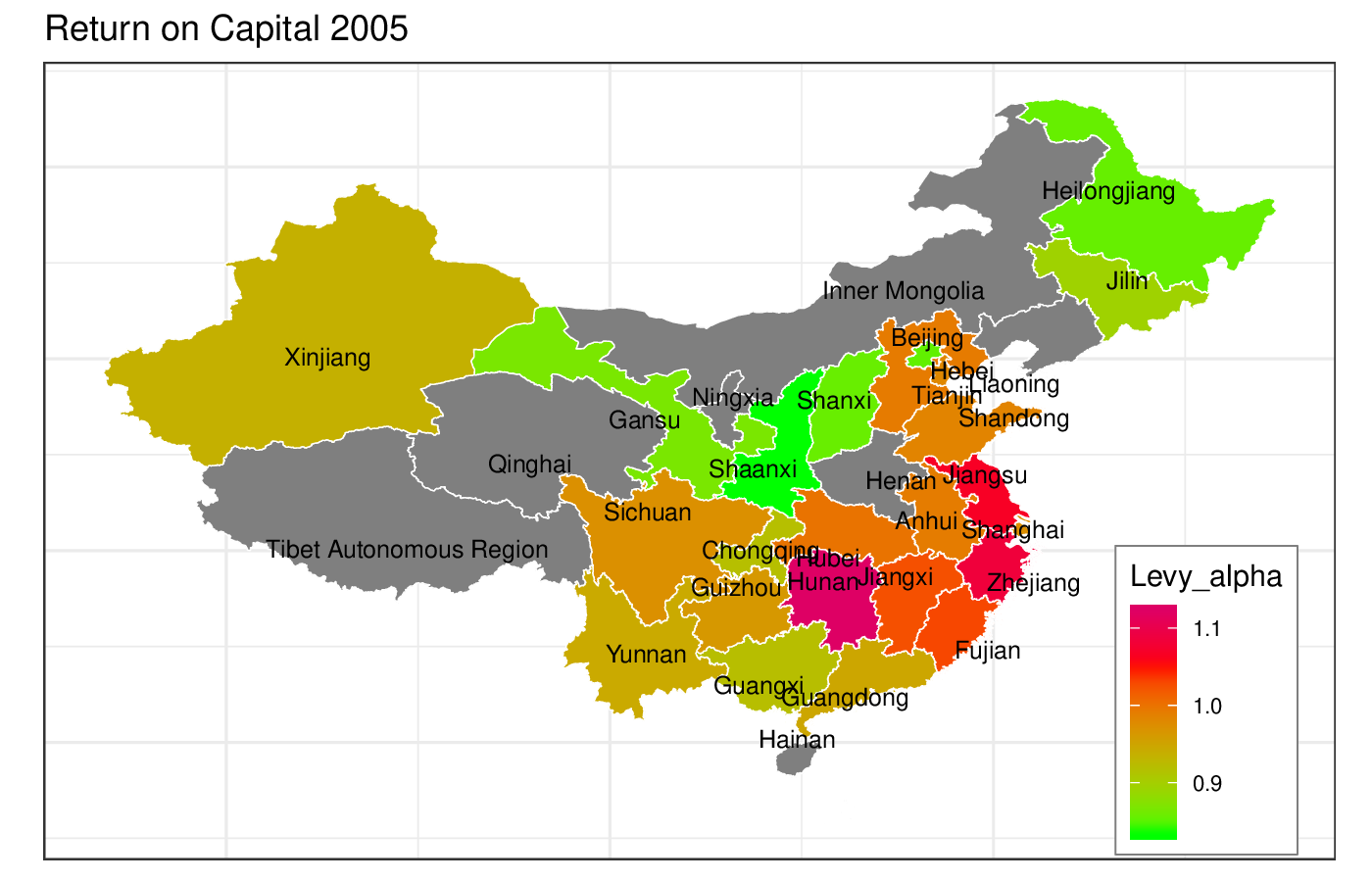}}\\
\subfloat[2006]{\includegraphics[width=0.33\textwidth]{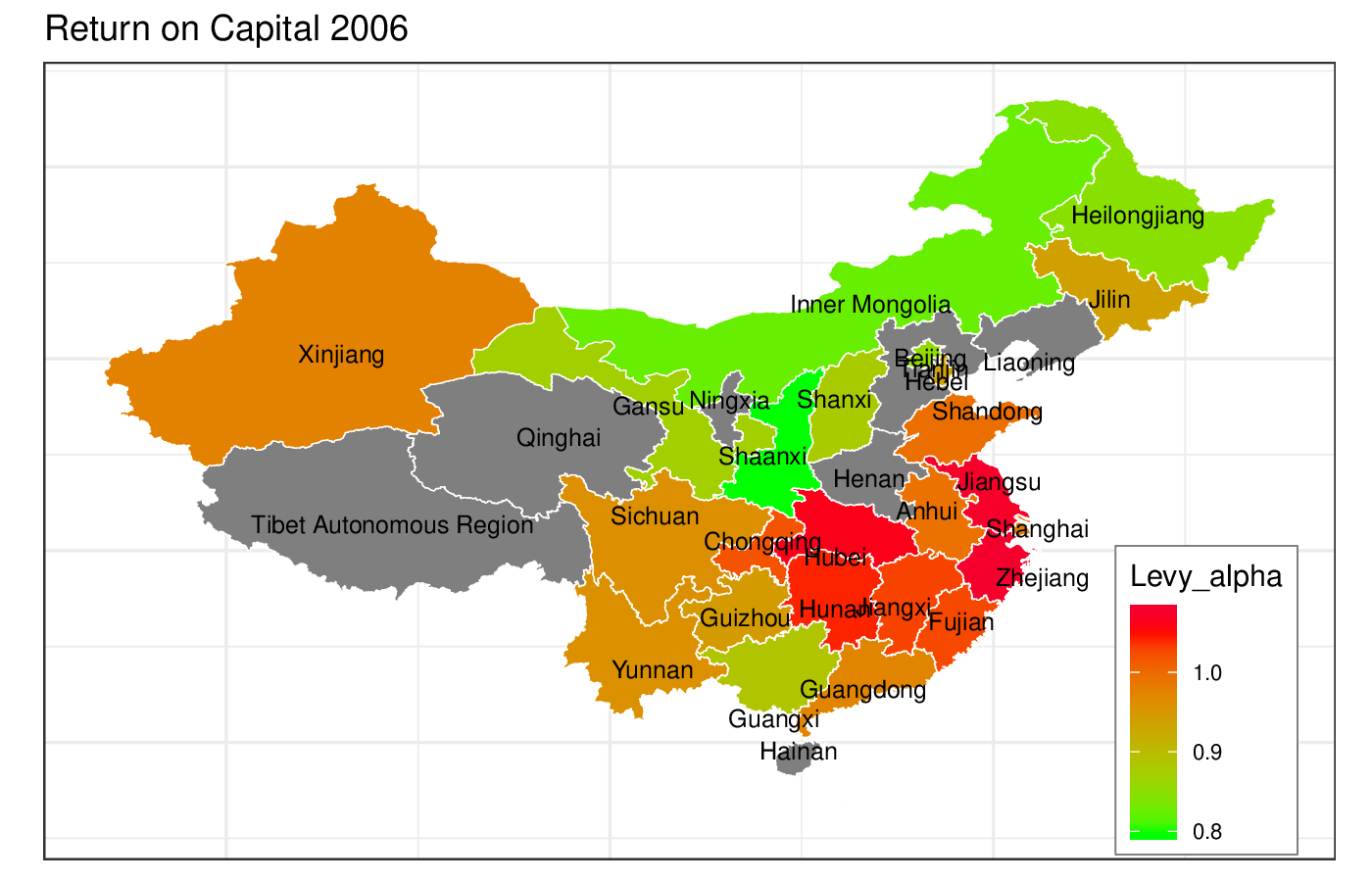}}
\subfloat[2007]{\includegraphics[width=0.33\textwidth]{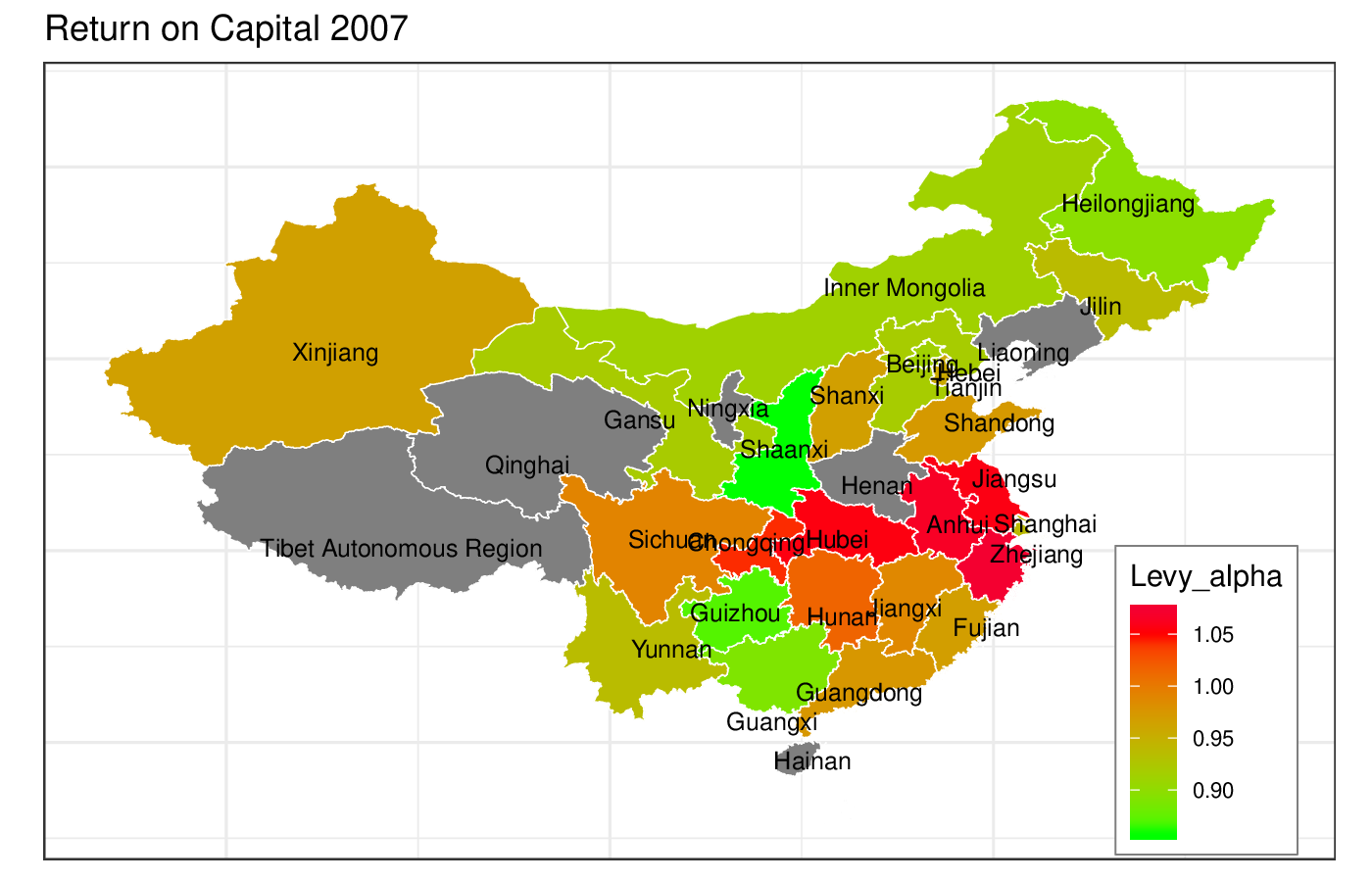}}
\caption{Levy $\alpha$ parameter fits for $ROC$ (profitability) by Region}
\label{fig:maps:roc}
\end{figure}

\begin{figure}[hbtp!]
\centering
\subfloat[1999]{\includegraphics[width=0.33\textwidth]{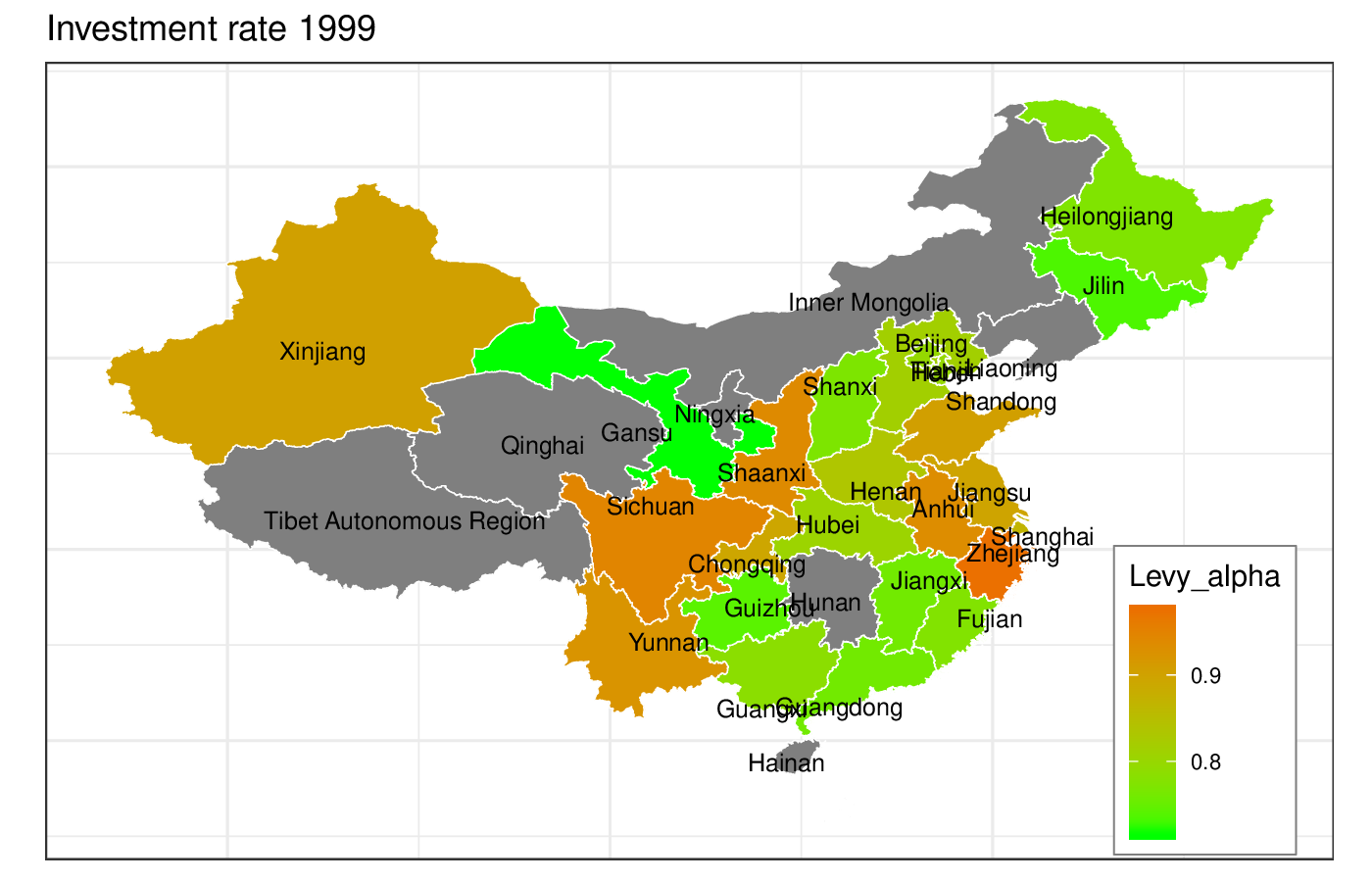}}
\subfloat[2000]{\includegraphics[width=0.33\textwidth]{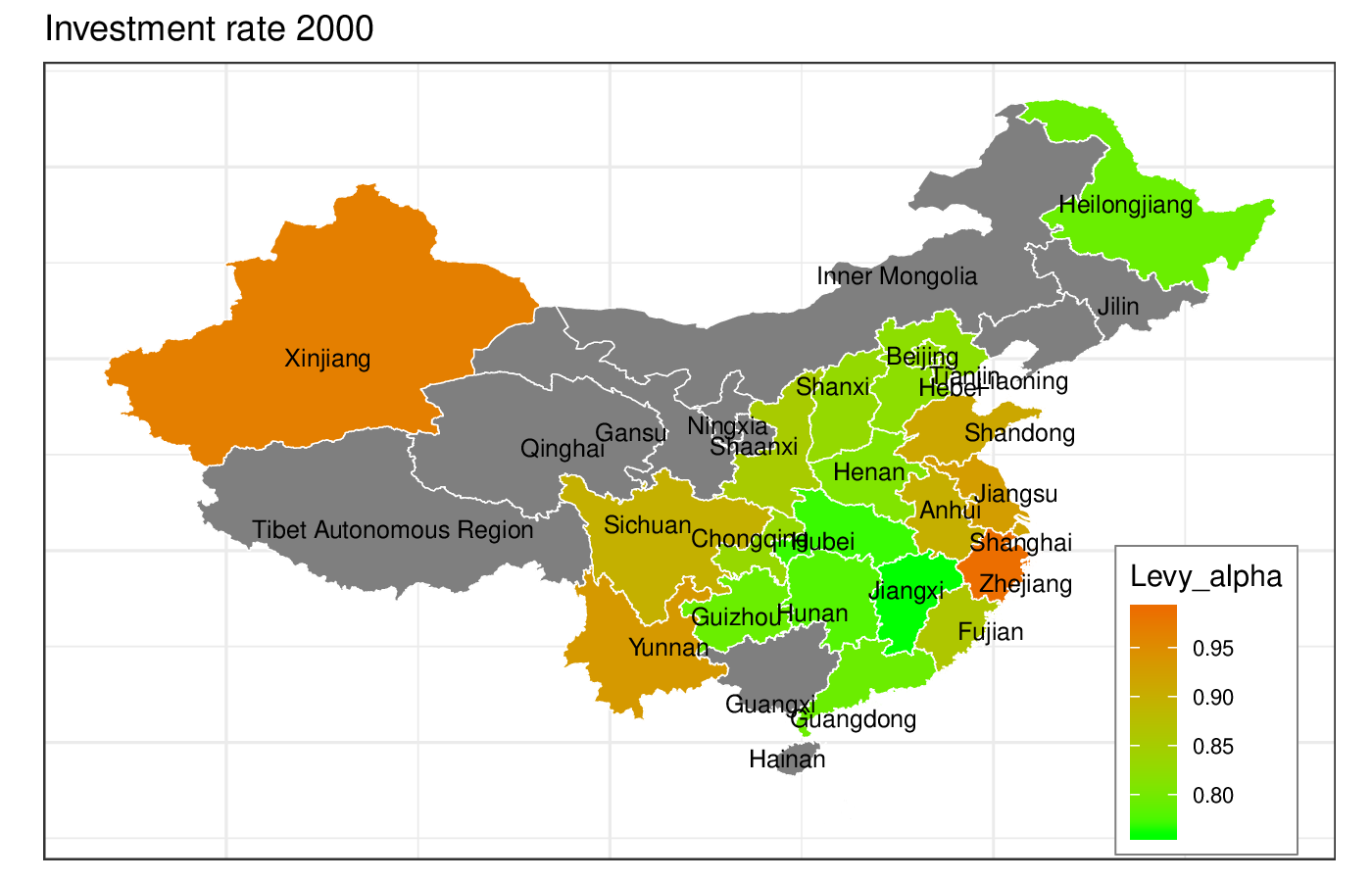}}
\subfloat[2001]{\includegraphics[width=0.33\textwidth]{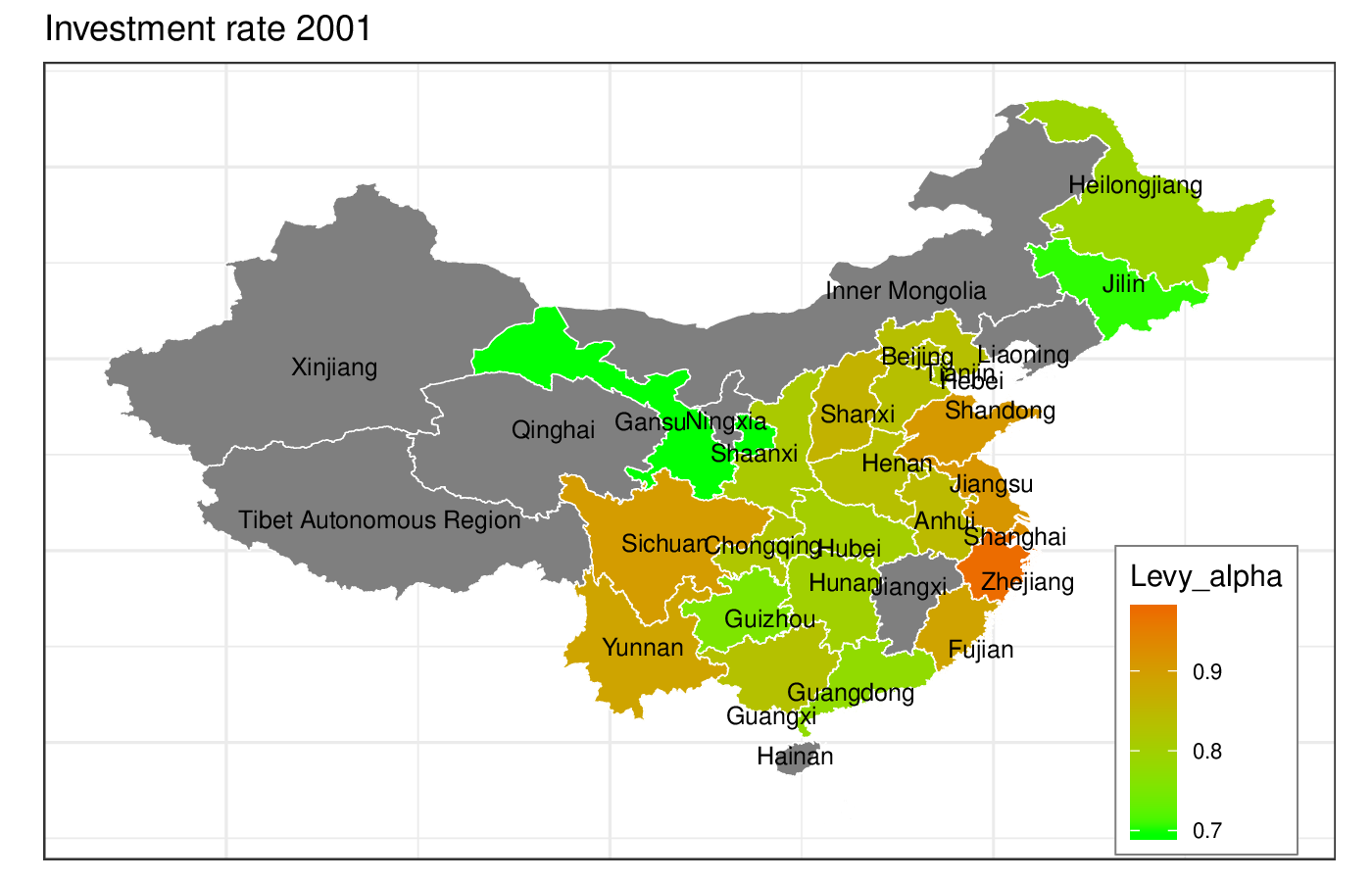}}\\
\subfloat[2002]{\includegraphics[width=0.33\textwidth]{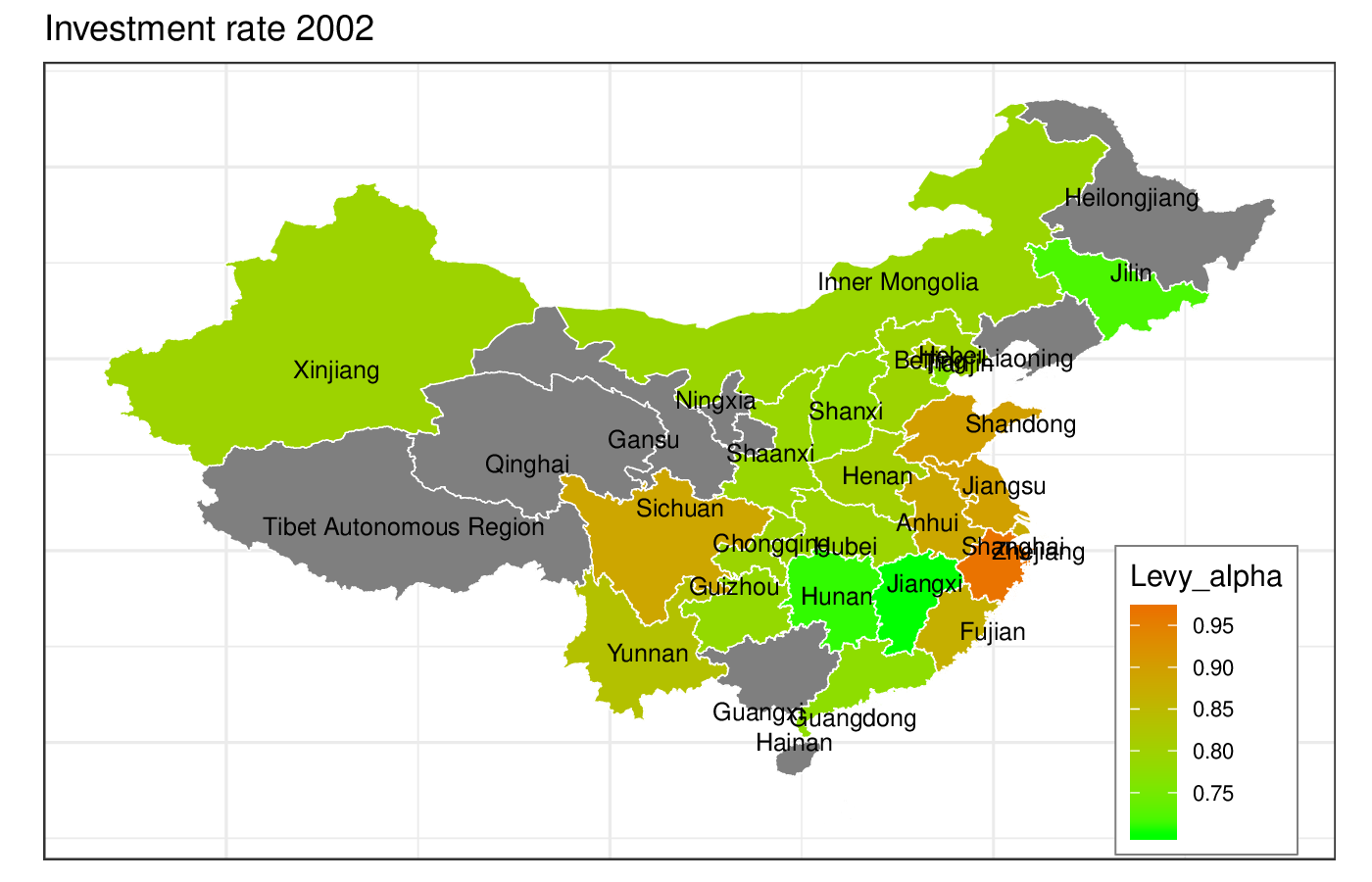}}
\subfloat[2004]{\includegraphics[width=0.33\textwidth]{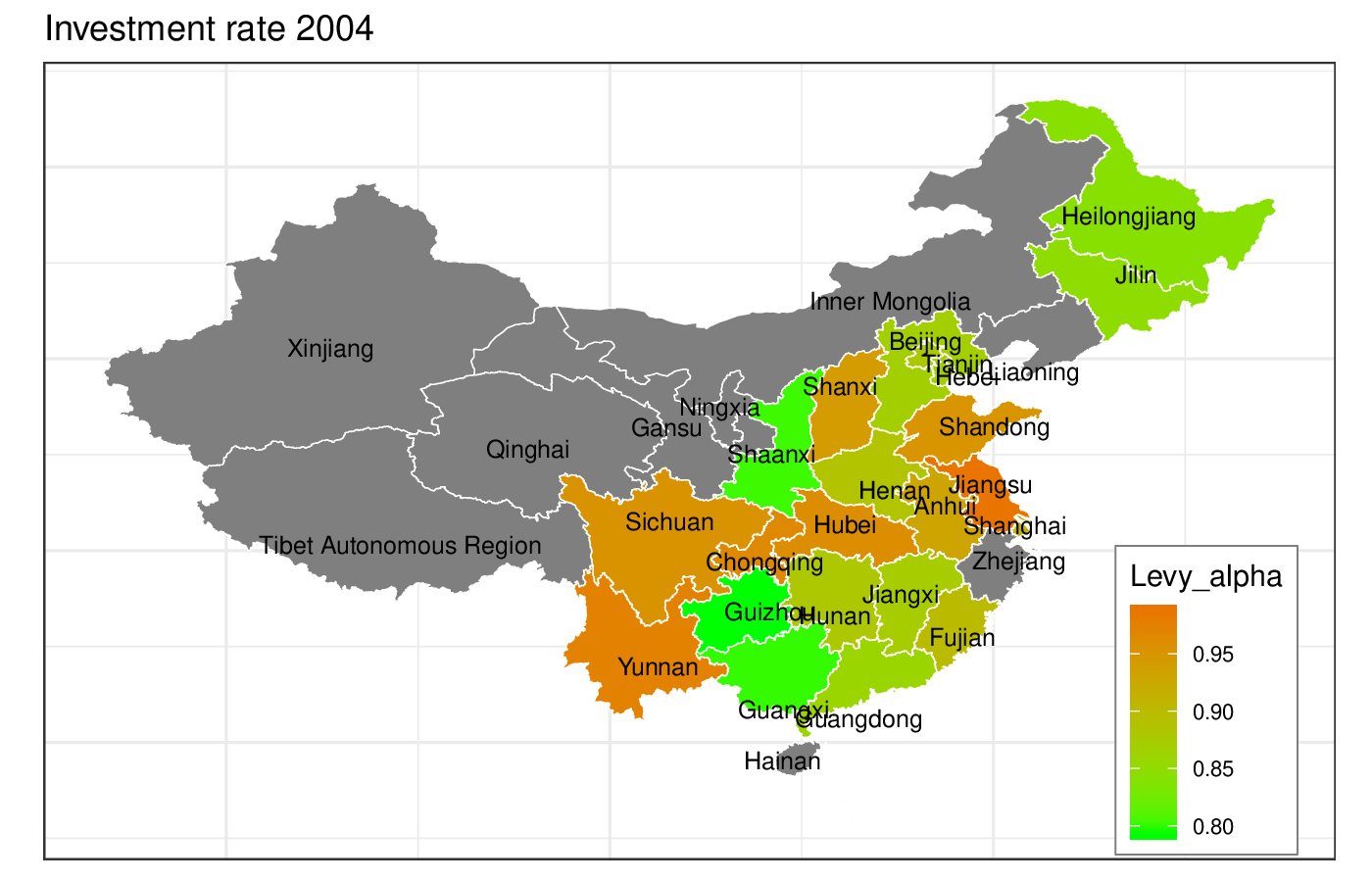}}
\subfloat[2005]{\includegraphics[width=0.33\textwidth]{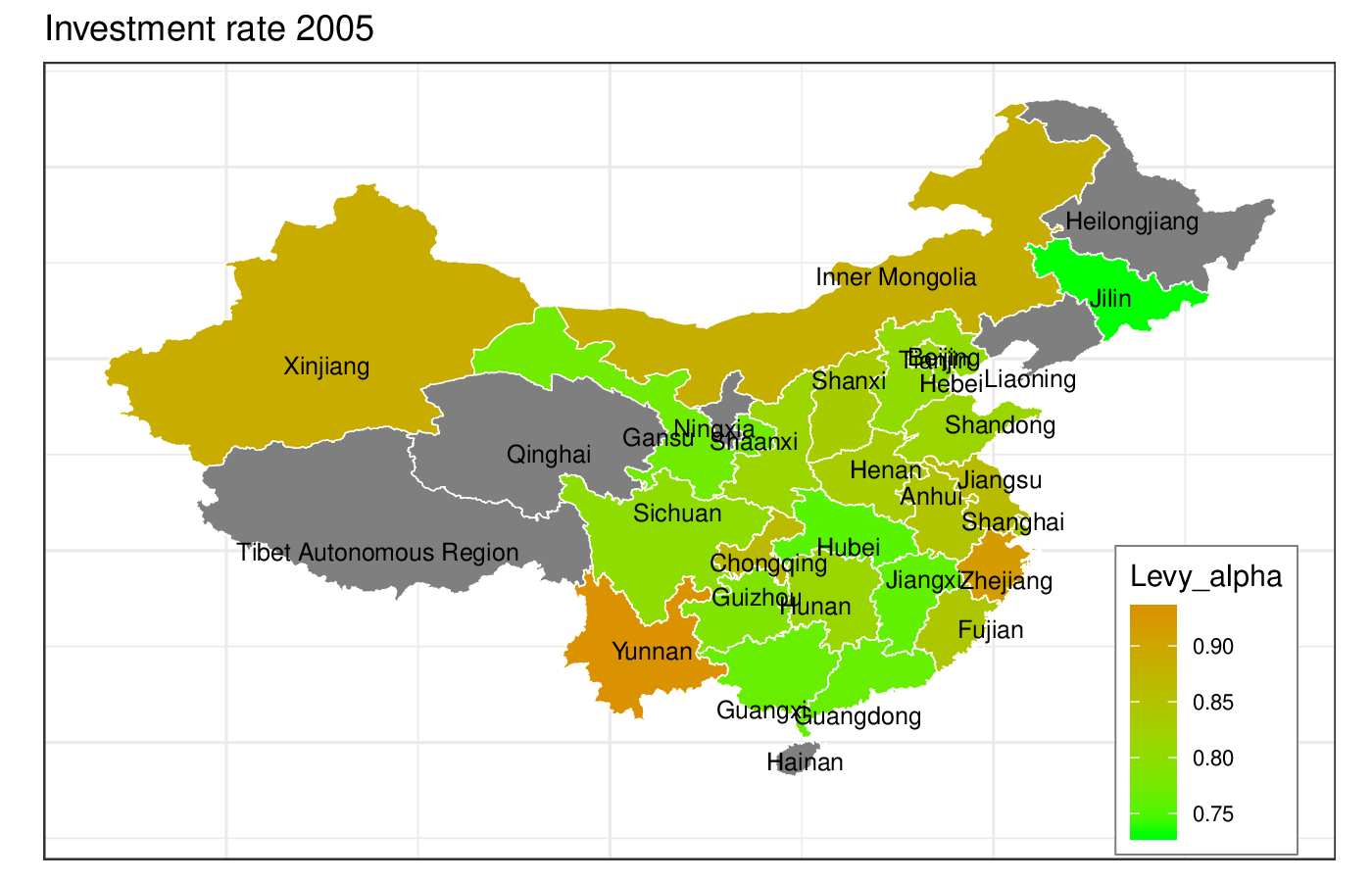}}\\
\subfloat[2006]{\includegraphics[width=0.5\textwidth]{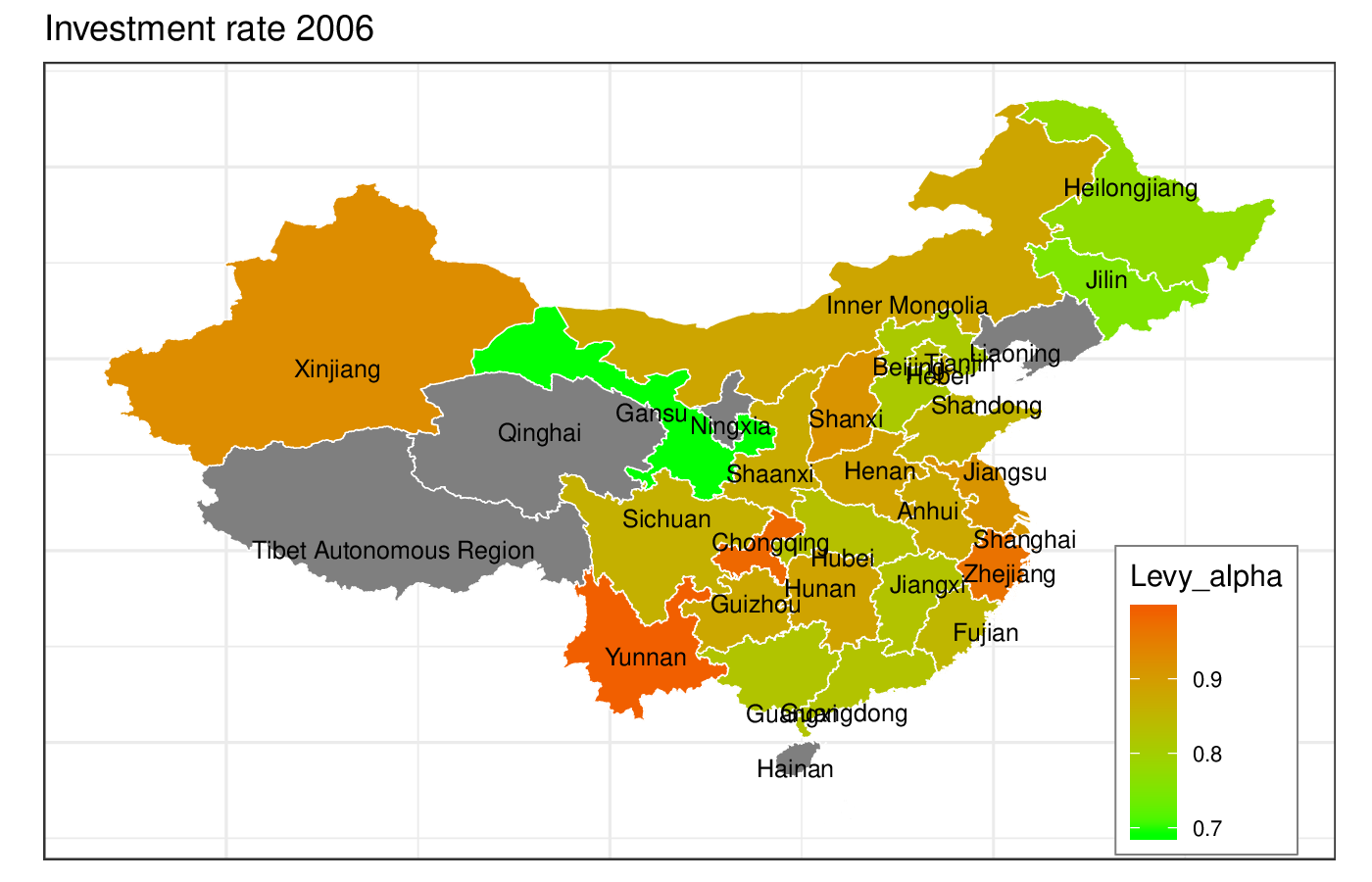}}
\subfloat[2007]{\includegraphics[width=0.5\textwidth]{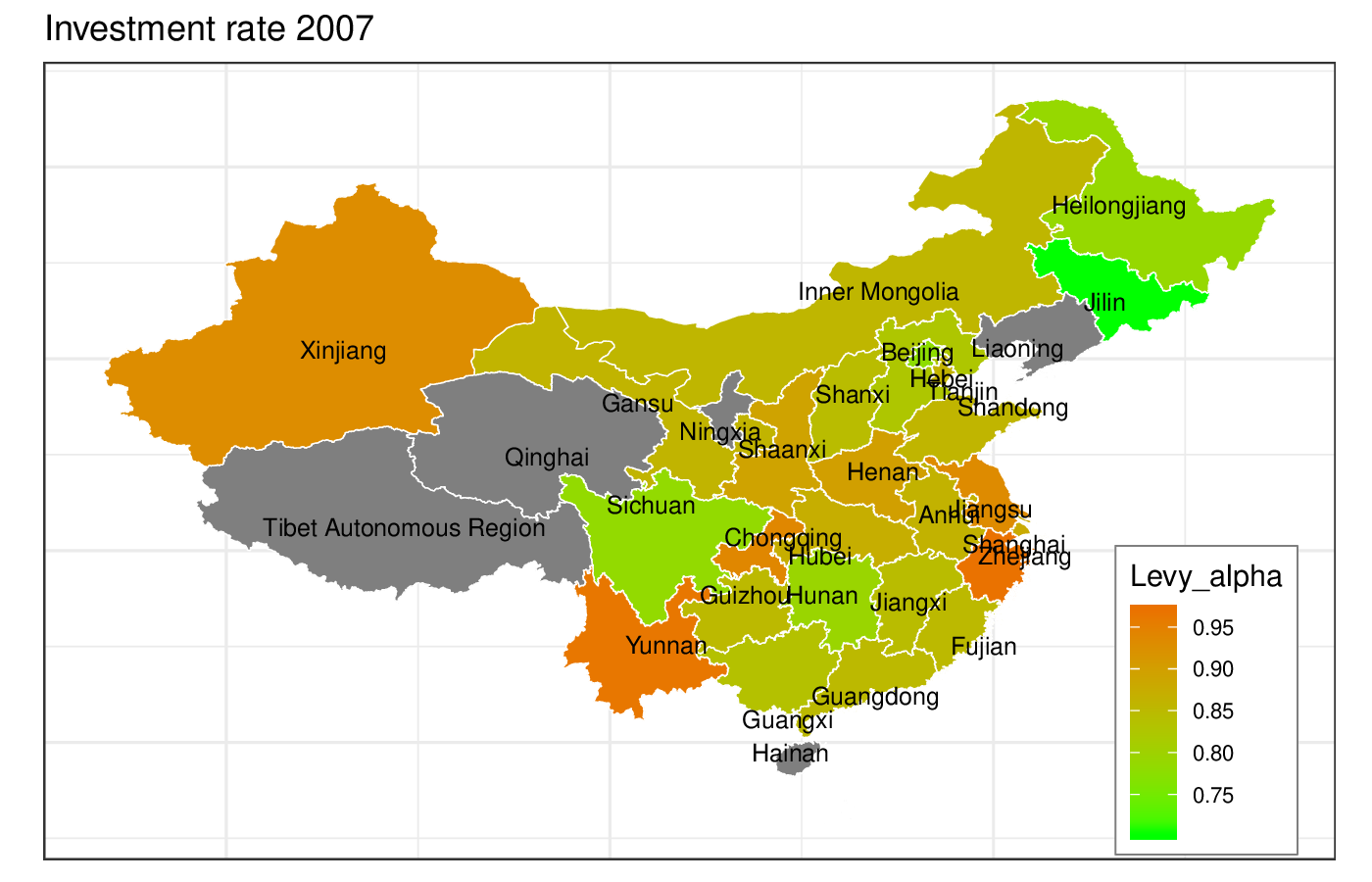}}
\caption{Levy $\alpha$ parameter fits for $IR$ (investment return) by Region}
\label{fig:maps:fias_g}
\end{figure}

\label{app:regression}

\begin{sidewaystable}[ht]
	\centering
	\scriptsize
	\begin{tabular}{llcccccccccccccccc}
		\toprule
		\multicolumn{1}{c}{Variable} & \multicolumn{1}{c}{Coefficient}  &\multicolumn{4}{c}{$\alpha$} & \multicolumn{4}{c}{$\beta$}    &\multicolumn{4}{c}{$\gamma$} & \multicolumn{4}{c}{$\delta$}  \\
		\cmidrule(lr){1-1}    \cmidrule(lr){2-2}  \cmidrule(lr){3-6} \cmidrule(lr){7-10}  \cmidrule(lr){11-14}  \cmidrule(lr){15-18}  
		
		\multicolumn{1}{c}{} & \multicolumn{1}{c}{}& \multicolumn{1}{c}{Est.} &  \multicolumn{1}{c}{SE} & \multicolumn{1}{c}{CI 5\%}  & \multicolumn{1}{c}{CI 95\%} & \multicolumn{1}{c}{Est.} &  \multicolumn{1}{c}{SE} & \multicolumn{1}{c}{CI 5\%}  & \multicolumn{1}{c}{CI 95\%} & \multicolumn{1}{c}{Est.} &  \multicolumn{1}{c}{SE} & \multicolumn{1}{c}{CI 5\%}  & \multicolumn{1}{c}{CI 95\%} & \multicolumn{1}{c}{Est.} &  \multicolumn{1}{c}{SE} & \multicolumn{1}{c}{CI 5\%}  & \multicolumn{1}{c}{CI 95\%}  \\ 
		\cmidrule(lr){3-6} \cmidrule(lr){7-10}  \cmidrule(lr){11-14}  \cmidrule(lr){15-18} 
		
		\\
		\multirow{8}{*}{LP Change}   &
Intercept & 0.984 & 0.046 & 0.894 & 1.075 & -0.056 & 0.102 & -0.26 & 0.151 & 0.055 & 0.047 & -0.033 & 0.143 & -0.011 & 0.024 & -0.056 & 0.037 \\ 
  & GDP Growth & 0.008 & 0.003 & 0.003 & 0.013 & 0.026 & 0.007 & 0.013 & 0.038 & 0.009 & 0.003 & 0.004 & 0.014 & 0.005 & 0.001 & 0.003 & 0.008 \\ 
  &Firm.Age & -0.001 & 0.001 & -0.002 & 0 & -0.003 & 0.001 & -0.006 & -0.001 & 0 & 0 & -0.001 & 0.001 & 0 & 0 & -0.001 & 0.001 \\ 
  & Employment & 0 & 0 & 0 & 0 & 0 & 0 & 0 & 0 & 0 & 0 & 0 & 0 & 0 & 0 & 0 & 0 \\ 
  & Cap Intensity & 0 & 0 & 0 & 0 & 0 & 0 & 0 & 0 & 0 & 0 & 0 & 0.001 & 0 & 0 & 0 & 0 \\ 
  & sd(Class) & 0.043 & 0.008 & 0.029 & 0.06 & 0.065 & 0.017 & 0.037 & 0.1 & 0.038 & 0.007 & 0.027 & 0.054 & 0.019 & 0.004 & 0.012 & 0.028 \\ 
  & sd(Year) & 0.009 & 0.007 & 0 & 0.027 & 0.08 & 0.031 & 0.039 & 0.156 & 0.041 & 0.014 & 0.021 & 0.076 & 0.017 & 0.006 & 0.009 & 0.033 \\ 
  & WAIC & -586.565 & 30.229 &  &  & -314.187 & 19.377 &  &  & -681.084 & 30.577 &  &  & -872.515 & 54.299 &  &  \\ 
	\midrule
		\multirow{8}{*}{LP}  & Intercept & 1.082 & 0.051 & 0.981 & 1.178 & 0.917 & 0.017 & 0.883 & 0.948 & 0.23 & 0.063 & 0.112 & 0.355 & 0.257 & 0.068 & 0.123 & 0.396 \\ 
  &GDP Growth & 0.001 & 0.003 & -0.005 & 0.007 & 0.002 & 0.001 & -0.001 & 0.004 & 0.005 & 0.003 & -0.001 & 0.012 & 0.009 & 0.004 & 0.002 & 0.016 \\ 
  &Firm.Age & 0 & 0 & 0 & 0 & 0 & 0 & 0 & 0 & 0 & 0 & 0 & 0 & 0 & 0 & 0 & 0 \\ 
  &Employment & 0 & 0 & 0 & 0 & 0 & 0 & 0 & 0 & 0 & 0 & -0.001 & 0 & 0 & 0 & -0.001 & 0 \\ 
  &Cap Intensity & 0 & 0 & 0 & 0 & 0 & 0 & 0 & 0 & 0 & 0 & 0 & 0.001 & 0 & 0 & 0 & 0 \\ 
  &sd(Class) & 0.051 & 0.009 & 0.036 & 0.07 & 0.008 & 0.004 & 0.001 & 0.016 & 0.08 & 0.014 & 0.057 & 0.112 & 0.086 & 0.014 & 0.062 & 0.117 \\ 
  &sd(Year) & 0.029 & 0.011 & 0.014 & 0.057 & 0.003 & 0.003 & 0 & 0.01 & 0.057 & 0.02 & 0.032 & 0.102 & 0.083 & 0.027 & 0.048 & 0.156 \\ 
  &WAIC & -577.335 & 20.28 &  &  & -763.403 & 92.099 &  &  & -626.065 & 39.083 &  &  & -581.903 & 39.123 &  &  \\ 	
		\midrule
		\multirow{8}{*}{ROC} & Intercept & 1.124 & 0.056 & 1.009 & 1.227 & 0.167 & 0.134 & -0.103 & 0.431 & 0.071 & 0.021 & 0.031 & 0.114 & 0.034 & 0.018 & -0.002 & 0.068 \\ 
  &GDP Growth & -0.003 & 0.003 & -0.01 & 0.003 & 0.033 & 0.008 & 0.017 & 0.049 & 0.003 & 0.001 & 0.001 & 0.006 & 0.002 & 0.001 & 0 & 0.004 \\ 
  &Firm.Age & -0.001 & 0 & -0.001 & 0 & 0 & 0 & -0.001 & 0.001 & 0 & 0 & 0 & 0 & 0 & 0 & 0 & 0 \\ 
  &Employment & 0 & 0 & 0 & 0 & 0 & 0 & -0.001 & 0 & 0 & 0 & 0 & 0 & 0 & 0 & 0 & 0 \\ 
  &Cap Intensity & 0 & 0 & 0 & 0 & 0 & 0 & 0 & 0 & 0 & 0 & 0 & 0 & 0 & 0 & 0 & 0 \\ 
  &sd(Class) & 0.046 & 0.009 & 0.031 & 0.066 & 0.164 & 0.028 & 0.118 & 0.232 & 0.029 & 0.005 & 0.022 & 0.04 & 0.017 & 0.003 & 0.012 & 0.024 \\ 
  &sd(Year) & 0.024 & 0.011 & 0.009 & 0.053 & 0.127 & 0.046 & 0.063 & 0.24 & 0.02 & 0.007 & 0.01 & 0.038 & 0.016 & 0.006 & 0.008 & 0.031 \\ 
  &WAIC & -439.61 & 18.542 &  &  & -275.578 & 20.391 &  &  & -861.659 & 23.479 &  &  & -869.361 & 32.042 &  &  \\ 		
		\midrule
		\multirow{8}{*}{Inv Rate}& Intercept & 0.795 & 0.049 & 0.703 & 0.89 & 0.42 & 0.096 & 0.234 & 0.604 & 0.122 & 0.029 & 0.066 & 0.179 & -0.069 & 0.019 & -0.107 & -0.031 \\ 
  &GDP Growth & 0 & 0.003 & -0.005 & 0.006 & 0.005 & 0.004 & -0.004 & 0.013 & 0.003 & 0.002 & 0 & 0.006 & 0 & 0.001 & -0.002 & 0.002 \\ 
  &Firm.Age & -0.001 & 0.001 & -0.003 & 0 & 0.001 & 0.002 & -0.002 & 0.004 & -0.001 & 0 & -0.002 & 0 & 0 & 0 & -0.001 & 0 \\ 
 & Employment & 0 & 0 & 0 & 0 & 0 & 0 & 0 & 0 & 0 & 0 & 0 & 0 & 0 & 0 & 0 & 0 \\ 
  &Cap Intensity & 0 & 0 & 0 & 0 & 0 & 0 & 0 & 0 & 0 & 0 & 0 & 0 & 0 & 0 & 0 & 0 \\ 
  &sd(Class) & 0.058 & 0.011 & 0.04 & 0.083 & 0.06 & 0.016 & 0.034 & 0.094 & 0.02 & 0.004 & 0.013 & 0.029 & 0.011 & 0.002 & 0.007 & 0.016 \\ 
  &sd(Year) & 0.032 & 0.01 & 0.018 & 0.057 & 0.193 & 0.052 & 0.117 & 0.322 & 0.046 & 0.013 & 0.029 & 0.078 & 0.04 & 0.01 & 0.026 & 0.065 \\ 
  &WAIC & -486.833 & 20.609 &  &  & -326.147 & 37.232 &  &  & -617.939 & 20.353 &  &  & -754.725 & 31.133 &  &  \\ 
		\bottomrule
	\end{tabular}
	\caption{Detailed regression results} 
\end{sidewaystable}
\end{document}